\documentclass[aps,pra,twocolumn,amsmath,superscriptaddress,longbibliography]{revtex4-2}

\newcommand{\bea}{\begin{eqnarray}}
\newcommand{\eea}{\end{eqnarray}}
\newcommand{\beq}{\begin{equation}}
\newcommand{\eeq}{\end{equation}}

\newcommand{\Si}{\Sigma}

\usepackage{amsmath}
\usepackage[urlcolor=blue,colorlinks=true,citecolor=blue,linkcolor=blue,pdfstartview={FitH},bookmarks=false]{hyperref}
\usepackage{appendix}
\usepackage{graphicx}
\usepackage{longtable}
\usepackage{epsfig}
\usepackage{dcolumn}
\usepackage{bm}
\usepackage{bbm}
\usepackage{amssymb}
\usepackage{multirow}
\usepackage{times,color}
\usepackage{hyperref}
\usepackage{amsmath}
\usepackage{color}
\usepackage{subfigure}
\usepackage{float}
\usepackage{epstopdf}

\begin{document}
\title{Quantum many-body scars in spin-1 Kitaev chain with uniaxial single-ion anisotropy}

\author{Wen-Yi Zhang}
\affiliation{College of Physics, Nanjing University of Aeronautics and Astronautics, Nanjing, 211106, China}
\affiliation{Key Laboratory of Aerospace Information Materials and Physics (NUAA), MIIT, Nanjing 211106, China}

\author{Ya-Nan Wang}
\affiliation{College of Physics, Nanjing University of Aeronautics and Astronautics, Nanjing, 211106, China}
\affiliation{Key Laboratory of Aerospace Information Materials and Physics (NUAA), MIIT, Nanjing 211106, China}

\author{Dongchang Liu}
\affiliation{Mathematical Sciences Institute, The Australian National University, Canberra ACT 2601, Australia}

\author{Jie Ren}
\affiliation{Department of Physics, Changshu Institute of Technology, Changshu 215500, China}

\author{Jia Li}
\affiliation{College of Engineering Physics, Shenzhen Technology
University, Shenzhen 518118, China}
\affiliation{Key Laboratory of Aerospace Information Materials and Physics (NUAA), MIIT, Nanjing 211106, China}

\author{\\ Ning Wu}
\email{wunwyz@gmail.com}
\affiliation{Center for Quantum Technology Research, School of Physics, Beijing Institute of Technology, Beijing 100081, China}

\author{Andrzej M. Ole\'s}
\email{a.m.oles@fkf.mpi.de}
\affiliation{\mbox{Max Planck Institute for Solid State Research,
             Heisenbergstrasse 1, D-70569 Stuttgart, Germany} }
\affiliation{\mbox{Institute of Theoretical Physics, Jagiellonian University,
             Prof. Stanis\l{}awa \L{}ojasiewicza 11, PL-30348 Krak\'ow, Poland}}

\author{Wen-Long You}
\email{wlyou@nuaa.edu.cn}
\affiliation{College of Physics, Nanjing University of Aeronautics and Astronautics, Nanjing, 211106, China}
\affiliation{Key Laboratory of Aerospace Information Materials and Physics (NUAA), MIIT, Nanjing 211106, China}

\begin{abstract}
To establish a solid-state-based framework for the coexistence of quantum
many-body scars and quantum criticality, we investigate the spin-1 Kitaev
chain with uniaxial single-ion anisotropy (SIA). In the subspace with
uniform $\mathbb{Z}_2$ gauge fields, this model can be exactly mapped to
the spin-1/2 effective detuned PXP Hamiltonian, where the SIA plays a
role of the static detuning term. The quench dynamics starting from the
product states is symmetric between positive and negative values
of the SIA, while a quantum phase transition from the Kitaev spin liquid
to the dimer phase only occurs at the critical point with a negative
$D_c$, implying the spontaneous breaking of the translational symmetry.
We find that the coherent oscillations of quantum fidelity and certain
local observables are sustained against small SIA perturbations in a
quantum quench from special initial states. While the oscillation
amplitudes of these observables decay with time as the SIA strength is
increased, the system completely thermalizes upon approaching the
critical point. In contrast, the initial polarized state, which shows
an absence of revivals of quantum fidelity, will exhibit long revivals
for $D<D_c$. Finally, we investigate the evolution of phase boundaries of the Kitaev spin liquid and  dimer phase by introducing Heisenberg interactions, which spoil the $\mathbb{Z}_2$ gauge fields. A complete
phase diagram is given by the infinite time-evolving block decimation
method and the ground state properties of each phase are accurately
captured by various spin correlations. Our work opens the door
to understanding exotic connections between many-body scars and
quantum criticality in systems with higher spins.
\end{abstract}

\date{\today}

\maketitle

\section{Introduction}
\label{intro}

In the past decade, there has been significant progress in understanding out-of-equilibrium dynamics of isolated quantum systems~\cite{Polkovnikov2011Nonequilibrium}. The eigenstate thermalization hypothesis (ETH)~\cite{Deutsch1991,Srednicki1994,Rigol2008,Rigol2012,Kim2014,Deutsch_2018} has been regarded as a cornerstone of contemporary statistical mechanics, which states that in a thermalizing system,  the expectation value of a generic local observable in individual eigenstates should be equivalent to its microcanonical average.  Despite the significant success of ETH in explaining thermalization of chaotic systems,
instances of ergodicity breaking are continually being discovered.
The integrable systems~\cite{Anderson1958,Rigol2007,Biroli2010,Stefan2009,Gamayun2014,Lionel2015} and many-body localization~\cite{Vidmar_2016,Gornyi2005,BASKO20061126,Pal2010,Lazarides2015,Altman2018,Schreiber2015} are the most noteworthy exceptions. The strong ergodicity breaking phenomena in counter-examples, where most of the eigenstates violate the ETH,  can be ascribed to the presence of conserved quantities~\cite{Pakrouski2020}. In an integrable system, the number of conserved quantities is equal to the number of degrees of freedom~\cite{Rigol2009}. On the other hand, many-body localization occurring in systems where disorder and interactions prevent the system from thermalizing can be also described by the emergence of an extensive set of quasi-local integrals of motions~\cite{Abanin2019}. Recently, a Rydberg-atom quantum simulator~\cite{Bernien2017} revealed the emergence of a new
type of ETH-violating eigenstates in certain nonintegrable quantum many-body systems, dubbed quantum many-body scar (QMBS) states
~\cite{Choi2019,Iadecola2019,Lin2020,Iadecola2020,Bull2020,Turner2021,Ljubotina2022,Windt2022,Ren2022,Dooley2023,Zhang2023}. Some specific low-entanglement states in a many-body quantum system are exceptional in that they violate the ETH and can retain quantum coherence for long times, even when the system is chaotic and thermalizing~\cite{Geraedts2016}.

To be specific, the number of QMBSs is exponentially smaller than the
Hilbert space dimension.
The discovery of QMBSs has opened up a new paradigm for studying unusual nonequilibrium phenomena including many-body revivals and nonthermal stationary states~\cite{M.D.Lukin2019,Moudgalya_2022}. Soon the scarred states have been observed in a variety of physical systems, including {\it inter alia,} interacting spin chains~\cite{ZheXuanGong2013,Neyenhuis2017}, cold atom systems~\cite{Tang2018,W.Kao2021,Kinoshita2006},
superconducting qubits~\cite{XuKai2018,Guo2021}, etc.
In parallel with exciting experimental advances,  theoretical studies have shown that QMBSs are not related to the usual symmetries~\cite{RenJie2021}. Known systems that host QMBS states also include the Affleck-Kennedy-Lieb-Tasaki (AKLT) model~\cite{Affleck1987,Moudgalya2018}, the spin-1 XY model~\cite{Schecter2019}, and the generalized Fermi-Hubbard model~\cite{Desaules2021}. The associated weak
ergodicity breaking not only challenges the validity of ETH but also
poses a different scenario of nonthermal dynamics.

Later it was pointed out theoretically that the Rydberg experiment can
be described by the one-dimensional (1D) chain of spin-1/2 degrees of freedom~\cite{Jaksch2000,Turner2018,Khemani2019,Lin2019ExactQuantum,Mark2020,Mukherjee2020},
where the spin-up state $\vert 1 \rangle$ corresponds to a Rydberg
atom occupying an excited state and
the spin-down state $\vert 0 \rangle$  denotes an atom in the
ground state.
Such a spin-1/2 spin chain, known as the PXP Hamiltonian and
resulting from the first-order Schrieffer-Wolff transformation applied to a tilted Ising chain,
is described by
\begin{equation}\label{eq:H_PXP}
\hat{H}_{\rm PXP} = \sum_{i=1}^N P_{i-1} X_i P_{i+1},
\end{equation}
where $N$ is the number of sites,  $X=\vert 0 \rangle \langle 1 \vert +\vert 1 \rangle \langle 0 \vert $ and $P  = \vert 0 \rangle \langle 0 \vert $ is the projector onto the ground state,
ensuring that the
nearby atoms are not simultaneously in the excited state.

Such Rydberg blockade induced kinetic constraint is responsible for
the atypical dynamics of QMBS states.
When
the system is initialized at time $t = 0$ in the product state $\vert \psi(0) \rangle  \equiv \vert \mathbb{Z}_k \rangle \, (k=1, 2,3,4)$, namely,
\begin{eqnarray}
\label{eq:Zk}
\vert \mathbb{Z}_1 \rangle &=& \vert 0000 \cdots 00 \rangle,  \quad~\,
\vert \mathbb{Z}_2 \rangle = \vert 010101 \cdots 01 \rangle,  \nonumber\\
\vert \mathbb{Z}_3 \rangle &=&\vert 001001 \cdots 001 \rangle,
\vert \mathbb{Z}_4 \rangle = \vert 00010001 \cdots 0001 \rangle, \quad \quad
\end{eqnarray}
the system then follows the evolution
governed by
the PXP Hamiltonian, $\vert \psi(t) \rangle = \exp(-i\hat{H}_{\rm PXP}  t) \vert \psi(0)\rangle$.  It was noted that the quantum quench from either $\vert \mathbb{Z}_2 \rangle$ or $\vert \mathbb{Z}_3 \rangle$ exhibits periodic revivals in the quantum fidelity
\begin{eqnarray}
F(t) = \vert \langle \psi(0) \vert   \psi(t) \rangle \vert^2,
\end{eqnarray}
while $\vert \mathbb{Z}_1 \rangle$  or  $\vert \mathbb{Z}_4 \rangle$  thermalize under time evolution. The observed oscillations and apparent nonergodic dynamics are due to the existence of  equal spacing of the QMBS eigenstates~\cite{Serbyn2021}.

Considering the experimental realization and
the important role of the emergence of QMBS in the PXP model,
the intensive study of the PXP model has been the subject of a separate thread of investigation of much current interest~\cite{Roux2010,Sierant_2018,ZhaoHongzheng2020,Mukherjee2020_2,Halimeh2020}. In fact, this effective
model has a long history dating at least as far back as an effective Hamiltonian for the tilted Bose-Hubbard model~\cite{Fendley2004}.
The PXP model has been studied in various other contexts including Fibonacci anyon chains~\cite{Trebst2008,Lesanovsky2012,Chandran2020},
Ising models on dimer ladders~\cite{Moessner2001,Laumann2012},
U(1) lattice gauge theory in its quantum link~\cite{Surace2020,Chen2021,Desaules12023,Desaules22023}
dipole-conserving Hamiltonians~\cite{ Sala2020},
the quantum Hall effect on a thin torus at filling $\nu$= 1/3~\cite{Moudgalya2020}, etc.
 Meanwhile, the PXP model
was extended to Floquet Hamiltonians~\cite{Mizuta2020}, higher spins~\cite{Mukherjee2021}, and
higher dimensions~\cite{Michailidis2020}.
The PXP model can be deduced from the biaxial Ising model with both transverse and longitudinal
fields at zero detuning~\cite{Turner2018PhysRevB}
and the Bose-Hubbard model at resonance~\cite{JianWeiPan2023observation}.
It was also claimed that there is an intimate relation between QMBS and quantum criticality~\cite{ZhaiHui2022} or quantum integrability~\cite{CuiXiaoling2022}.

Remarkably, the 1D PXP chain is shown to be embedded in the spin-1 Kitaev model~\cite{You2020,You2022prr}, highlighting
a solid-state-based realization of the PXP model. The celebrated Kitaev model is renowned as a prototype model of quantum spin liquid (QSL), which hosts massive long-range entanglement and fractional quasiparticles from localized spins described by bosonic/fermionic spinons and $\mathbb{Z}_2$ gauge fields~\cite{KITAEV20062}.
Solid-state material realizations of the bond-dependent Kitaev interactions with $S$=1/2 local moments have vitalized the research in QSLs~\cite{Jackeli2009,Liu2020}, where strong spin-orbit coupling in a strongly correlated Mott
insulator plays an essential role. This poses  $4d$ and $5d$ transition-metal compounds are proposed to be candidate materials, such as
triangular lattice  $\rm YbMgGaO_4$~\cite{LiYuesheng2015}, $\rm 1T$-$\rm TaSe_2$~\cite{Ruan2021} and $\rm NaYbS_2 $~\cite{Wu2022}, kagome lattice $\rm ZnCu_3(OH)_6Cl_2$~\cite{Khuntia2020} and $\rm Na_4Ir_3O_{8}$~\cite{Shockley2015},
honeycomb lattice $\rm \alpha$-$\rm RuCl_3$~\cite{A.Banerjee2017}, $\rm H_3LiIr_2O_6$~\cite{Y.Ravi2018}, $\rm Cu_2IrO_3$~\cite{Srishti2021}, $\rm RuBr_3$~\cite{Yoshinori2022} and $\rm BaCo_2(AsO_4)_2$~\cite{Halloran2023},
pyrochlore lattice $\rm Ce_2Zr_2O_7$~\cite{Gao2019} and $\rm Ba_3Yb_2Zn_5O_{11}$~\cite{Chern2022}.
After the groundbreaking proposal for realizing the higher-spin analogs of the Kitaev interactions~\cite{Stavropoulos.P.Peter2019}, a number of materials with strong
Hund’s coupling among two electrons in $e_g$-orbitals of transition metal ions and  strong spin-orbit coupling of anions have emerged as potential candidates for the $S = 1$ Kitaev model. Recently the importance of studying the higher-spin Kitaev physics has attracted a lot of attention.

Both experimental and numerical analyses have been indispensably
carried out to explore the higher-spin Kitaev physics, such as $S=1$~\cite{Koga2018,Lee2020,Chen2022,Pohle2023,Taddei2023,Mohapatra2023},
$S=3/2$~\cite{Xu2020,Jin2022,natori2023quantum}, and even $S=2$ systems~\cite{Fukui2022}.
It is noteworthy that non-Kitaev interactions widely exist in candidate
materials, which is a chief  obstacle of keeping the system away from
the pure Kitaev limit. The ferromagnetic Heisenberg interactions are
generated from superexchange paths together with Kitaev interactions,
in parallel with  the antiferromagnetic Heisenberg term from direct-exchange paths.

For Mott insulators with two or more atoms per site, the direct on-site
interactions can give rise to a nonlinear term $\propto D\sum_j(S_j^z)^2$
for $S\ge 1$, where $D$ is the so-called uniaxial single-ion anisotropy
(SIA) constant.
Recently, theoretical~\cite{Xu2018,Bradley.Owen.2022,Erik2023}
and experimental~\cite{Fishman.Randy.S.2021} studies on the Kitaev model
with additional SIA have attracted increasing attention.
In this work, we will show that the static detuning in the PXP model,
which describes the static frequency difference between the ground and excited states, can be mimicked by the additional SIA in the
spin-1 Kitaev model, which normally stems from zero-field splitting due
to a crystal-field anisotropy. Upon varying the strength of the SIA,
a corresponding second-order phase transition will occur with
a~translational symmetry breaking.
A comprehensive study of the phase diagram has been conducted by incorporating significant Heisenberg interactions.
In the numerical calculation, we employ the exact diagonalization (ED) method, the time-evolving a matrix product state (MPS) with matrix product operators (MPOs)~\cite{Zaletel2015} based on ITensor~\cite{ITensor} and the infinite time-evolving block decimation (iTEBD) algorithm~\cite{VidaliTEBD2007}.

The remainder of this paper is organized as follows:
In Sec. \ref{sec:KD}, we present the spin-1 Kitaev model with SIA
(the KD model), and deduce the effective spin-1/2 detuned PXP model in the
ground-state manifold. The quantum many-body scars in the spin-1 KD
model are studied in detail. In Sec. \ref{sec:QPTKD}, we investigate the quantum criticality
in the KD model, and find the characteristics of the dimer phase. Under the cooperative effects of the single-ion anisotropy and Heisenberg
interactions (the KHD model), we reveal the rich quantum phase diagram of
KHD model in Sec. \ref{sec:Heisenberg}. The summary and conclusion are
given in Sec. \ref{Summary and conclusions}.

\section{Spin-1 Kitaev chain with uniaxial single-ion anisotropy}
\label{sec:KD}
In this work, we consider a spin-1 Hamiltonian composed of the Kitaev
interaction and SIA, given by
\begin{eqnarray}
\label{equ:KD}
\hat{H}_{\rm KD}\! = \!
\sum_{j=1}^{N/2} \left(K_{2j-1} S_{2j-1}^x S_{2j}^x\!+\!K_{2j}S_{2j}^y S_{2j+1}^y\right)\!+\!\sum_{j=1}^{N}\!D_j (S_{j}^z)^2\!,\nonumber \\
\end{eqnarray}
where $K_j$ parameterizes the strength of the bond-dependent Kitaev exchange coupling between two neighbouring sites  $\langle j,j{+}1\rangle$, and $D_j$ denotes the amplitude of the SIA at the $j$-th site. $S_j^{a}$
($a = x,y,z$) is the $a$-component of the spin-1 operator at the $j$-th site among total $N$ sites, obeying the SU(2) algebra, i.e.,
$[S_i^{a},S_j^{b}] = i \delta_{ij} \epsilon_{abc} S_j^c $, with the
antisymmetric tensor $\epsilon_{abc}$ and $(\boldsymbol{S}_j)^2= S(S+1)= 2$.

We will work with a special spin-1 representation, i.e.,
\begin{eqnarray}
\vert x\rangle\!&=&\!\frac{1}{\sqrt{2}}(\vert -1\rangle\!-\!\vert1\rangle),
\quad
\vert y\rangle\!=\!\frac{i}{\sqrt{2}}(\vert -1\rangle\!+\!\vert1\rangle),
\quad
\vert z\rangle\!=\!\vert 0\rangle, \nonumber \\
\end{eqnarray}
where $\vert m\rangle$ is the eigenstate of the spin operator $S^z$ with eigenvalues $m$ =$-1$, $0$, $1$. In such a representation, we have \mbox{$S_{bc}^a= -i\epsilon_{abc}$} and $\{S^x, S^y, S^z \}$ are given by
\begin{eqnarray}
 \left(
                 \begin{array}{ccc}
                   0 & 0 & 0 \\
                    0& 0 & -i\\
                   0 & i & 0 \\
                 \end{array}
               \right),
 \left(
                 \begin{array}{ccc}
                   0 & 0 & i \\
                    0& 0 & 0\\
                   -i & 0 & 0 \\
                 \end{array}
               \right),
 \left(
                 \begin{array}{ccc}
                   0 & -i & 0 \\
                    i& 0 & 0\\
                   0 & 0 & 0 \\
                 \end{array}
               \right).
\end{eqnarray}
The corresponding site parity matrices are defined as
$\Sigma_j^a \equiv e^{i \pi S_j^a}$$ = 1-2 (S_j^a)^2$ and become diagonal,

\begin{eqnarray}
\!\left(
                 \begin{array}{ccc}
                   1 & 0 & 0 \\
                    0& -1 & 0\\
                   0 & 0 & -1 \\
                 \end{array}
               \right),
   \left(
                 \begin{array}{ccc}
                   -1 & 0 & 0 \\
                    0& 1 & 0\\
                   0 & 0 & -1 \\
                 \end{array}
               \right), \left(                 \begin{array}{ccc}
                   -1 & 0 & 0 \\
                    0& -1 & 0\\
                   0 & 0 & 1 \\
                 \end{array}
               \right). \label{Sigmaa}
\end{eqnarray}

It has been revealed that different Ising interactions on odd and even
bonds in Eq. (\ref{equ:KD}) can be rewritten into a similar form through a
unitary transformation on the even sites~\cite{Sen2010,You2022prr}:
\begin{eqnarray}
U=\prod_j  \exp(i \pi S_{2j}^x) \exp\left(i \frac{\pi}{2} S_{2j}^z\right),
\label{e_rot}
\end{eqnarray}
which gives $U S_{2j}^x U^\dagger= S_{2j}^y$,
$U S_{2j}^y  U^\dagger= S_{2j}^x$, and
$U S_{2j}^z  U^\dagger = -S_{2j}^z$, as well as
$U \vert x\rangle= \vert y\rangle$, $U \vert y\rangle= \vert x\rangle$,
$ U \vert z\rangle= -\vert z\rangle$. Note that the order of rotations
about $x$- and $z$- axes in Eq. (\ref{e_rot}) is essential as they do not commute. After the unitary transformation, the Kitaev exchange
couplings in Eq. (\ref{equ:KD}) take a translation-invariant form
\begin{eqnarray}
\tilde{H}_{\rm K} =  \sum_{j=1}^N K_j S_{j}^xS_{j+1}^y.
\label{equ:tilde Hk}
\end{eqnarray}
It is easy to see that the SIA term remains in its original form and
the Hamiltonian (\ref{equ:KD}) can be rewritten
\begin{eqnarray}
\tilde{H}_{\rm KD} = \sum_{j=1}^N K_j S_{j}^xS_{j+1}^y + D_j (S_{j}^z)^2. \label{equ:tilde HkD}
\end{eqnarray}
Note that the sign of the Kitaev interactions is still under debate with conflicting results from theoretical and experimental studies~\cite{Sugita.Yusuke2020,Sears2020}. Hereafter the uniform couplings with $K_j=1$ and $D_j=D$ ($\forall j$) are assumed unless otherwise specified.

Under the rotation (\ref{e_rot}),
the local bond parity operators are defined by
\begin{eqnarray}
\hat{W}_j = \Si_{j}^y\,\Si_{j+1}^x. \label{equ:tilde W}
\end{eqnarray}
One can readily find that $\hat{W}_j$ is invariant by inspecting $[\hat{W}_j,\tilde{H}_{\rm KD}]=0$.
As the eigenvalues of $\Si_j^a$ in Eq. (\ref{Sigmaa}) are $\pm 1$,
the eigenvalues of $\hat{W}_j$ are related to $\mathbb{Z}_2$-valued invariants, i.e., $w_j=\pm1$. It is straightforward to deduce from Eq. (\ref{Sigmaa}) that
for a pair of nearest neighbor sites $\langle j,j{+}1\rangle$,  total
 $3{\times}3=9$ allowed states can be distinguished into the $w_j=1$
sector spanned by $\vert xy\rangle$, $\vert xz\rangle$,
$\vert yx\rangle$, $\vert zy\rangle$, $\vert zz\rangle$ and the $w_j=-1$
sector spanned by $\vert xx\rangle$, $\vert yy\rangle$,
$\vert yz\rangle$, $\vert zx\rangle$. Hence, the whole Hilbert space
$\mathcal{H}$ can be decomposed into $2^N$ dynamically
disconnected Krylov subspaces of unequal sizes characterized by  $\vec{w}=\{w_1,w_n,\cdots,w_N\}$ as
 \begin{eqnarray}
\mathcal{H}=\bigoplus_{n=1}^{2^N} \mathcal{K}_n.
\end{eqnarray}
The Krylov subspace $\mathcal{K}_n$ is spanned by
\begin{eqnarray}
\{\mathcal{K}_{n}\}\equiv {\rm Span}\{\vert \psi_n \rangle,
\tilde{H}_{\rm KD} \vert \psi_n\rangle,
\tilde{H}_{\rm KD}^2 \vert \psi_n \rangle, \cdots\},
\end{eqnarray}
where $\vert \psi_n \rangle$ is the so-called root state, which
is a product state having explicit $\mathbb{Z}_2$ symmetries.

We have identified the ground state of spin-1 Kitaev chain lies within
the flux-free sector, i.e., $\vec{w}=\{1,1,\cdots,1\}$~\cite{You2020}.
In such a constrained Hilbert space, there is one-to-one mapping between base configurations $\{\tilde{\mathcal{K}}_{S=1}\}$ of Eq. (\ref{equ:tilde Hk})
within the flux-free sector and the configurations $\{{\mathcal{K}}_{S=1/2}\}$
of Eq. (\ref{eq:H_PXP}) with nearest neighbor exclusion. The rule for
constructing the mapping is simple.
The one-to-one mapping between the 5 allowed two-site configurations for a pair of nearest neighbor sites $\langle j,j{+}1\rangle$  and spin-$1/2$ degree of freedom for the bond center $j+1/2$  is given by~\cite{Moudgalya2020_2}
\begin{eqnarray}
    \label{statesmap}
        & &| \cdots zz \cdots\rangle_{j,j+1}   \leftrightarrow | \cdots \downarrow \downarrow \downarrow  \cdots \rangle_{j-\frac{1}{2}, j+\frac{1}{2}, j+\frac{3}{2}}, \nonumber \\
        & &| \cdots yx  \cdots \rangle_{j,j+1}  \leftrightarrow  | \cdots \downarrow \uparrow \downarrow \cdots\rangle_{j-\frac{1}{2}, j+\frac{1}{2}, j+\frac{3}{2}},\nonumber\\
        & & |  \cdots zy \cdots \rangle_{j,j+1}  \leftrightarrow |\cdots \downarrow \downarrow \uparrow \cdots \rangle_{j-\frac{1}{2}, j+\frac{1}{2}, j+\frac{3}{2}},\nonumber \\
        & & | \cdots xz  \cdots  \rangle_{j,j+1} \leftrightarrow |\cdots \uparrow \downarrow \downarrow \cdots\rangle_{j-\frac{1}{2}, j+\frac{1}{2}, j+\frac{3}{2}}, \nonumber \\
        & &  | \cdots xy  \cdots  \rangle_{j,j+1}   \leftrightarrow |\cdots \uparrow \downarrow \uparrow \cdots\rangle_{j-\frac{1}{2}, j+\frac{1}{2}, j+\frac{3}{2}}.
\end{eqnarray}
It is worthy noting that the prime lattice of the spin-1 Kitaev chain is
defined on the sites $\{j\}$, while the dual lattice of spin-1/2 PXP
model lives on the linking bonds at sites $\{j+1/2\}$. This
mapping from sites to bonds includes links to the two surrounding
sites and vice verse, which becomes subtle for open boundary conditions.
As an example, the four product states given by Eq. (\ref{eq:Zk}) can
be mapped to the following states in $\{\tilde{\mathcal{K}}_{S=1}\}$:
\begin{eqnarray}
 \vert \widetilde{\mathbb{Z}}_1 \rangle &=& \vert zzzz \cdots zz \rangle,
\quad\;\;\,
\vert\widetilde{\mathbb{Z}}_2 \rangle =  \vert xyxyxy \cdots xy \rangle , \nonumber \\
    \vert \widetilde{\mathbb{Z}}_3 \rangle &=& \vert yxzyxz \cdots yxz \rangle ,
    \vert \widetilde{\mathbb{Z}}_4 \rangle = \vert yxzzyxzz \cdots yxzz \rangle.\quad \quad
\end{eqnarray}

\begin{figure}[t!]
\centering
\includegraphics[width=\columnwidth]{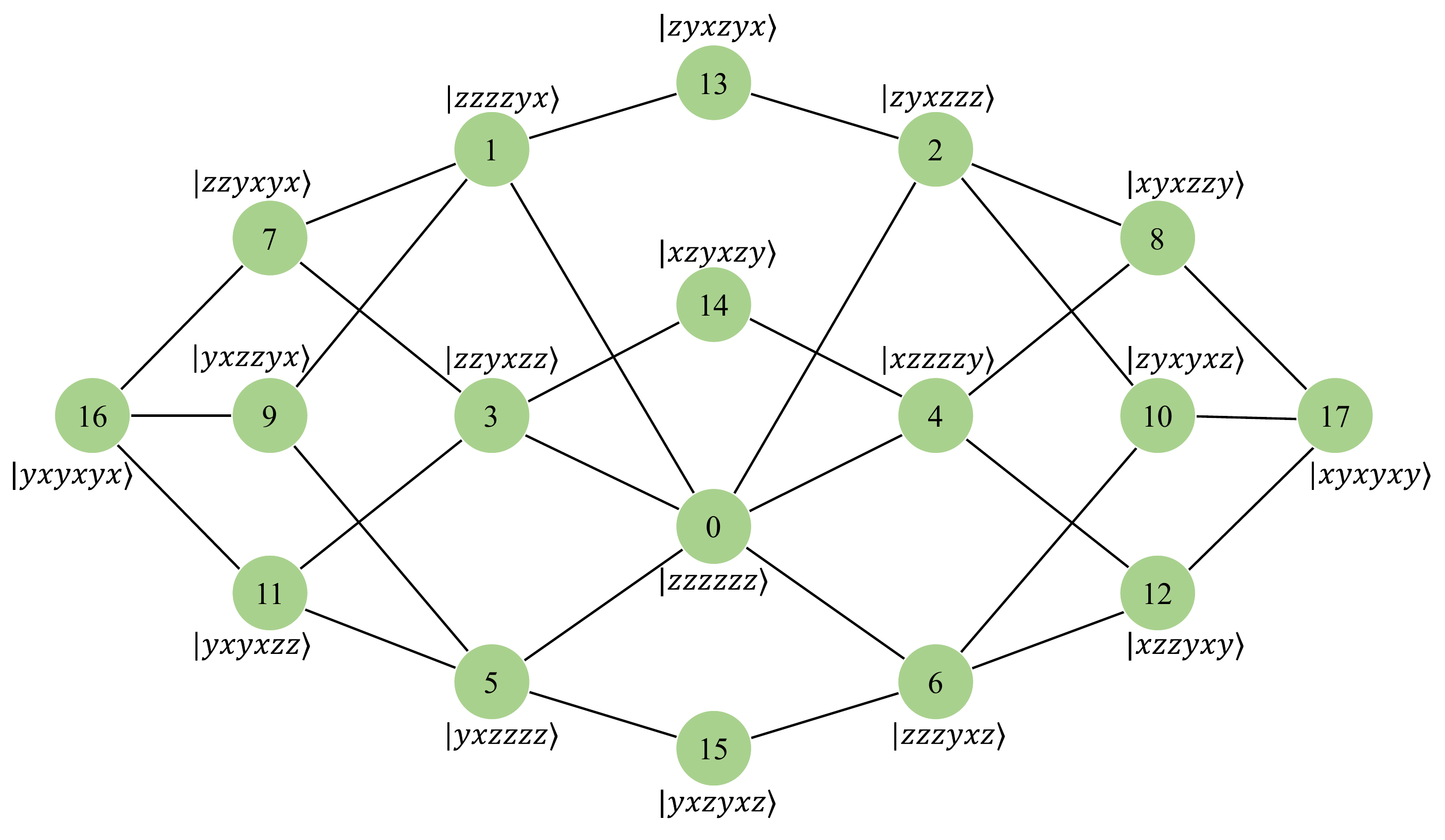}
\caption{The Hilbert space graph of the Kitaev Hamiltonian in Eq. (\ref{equ:tilde Hk}) within the $\vec{w}=\{1,1,1,1,1,1\}$ subspace for $N=6$ sites with periodic boundary conditions.
The nodes of the graph  $\vert m \rangle$ ($m=0, 1, 2, \ldots, 17)$  label the allowed product states,
and the edges connect product state configurations that differ by an excitation $\vert\cdots zz \cdots\rangle \leftrightarrow\vert\cdots y x\cdots\rangle$ due to the action of the Hamiltonian.
}
\label{fig:N_6_zzzzzz}
\end{figure}

The simplest root configuration in $\{\tilde{\mathcal{K}}_{S=1}\}$ is the
product state $\vert\widetilde{\mathbb{Z}}_1\rangle$, which is the ground
state in the $D\to\infty$ limit, and the Hilbert space of this sector
can be constructed by successively applying the Hamiltonian on this root
state, i.e.,
\begin{eqnarray}
 \label{Haction}
   \{\tilde{\mathcal{K}}_{S=1}\} \equiv
   {\rm Span}\{\vert\widetilde{\mathbb{Z}}_1 \rangle, \tilde{H}_{\rm K}
   \vert \widetilde{\mathbb{Z}}_1 \rangle, \tilde{H}_{\rm K}^2 \vert \widetilde{\mathbb{Z}}_1 \rangle, \cdots\}.
\end{eqnarray}
The corresponding dimension $d$ of the flux-free sector is proven to be a
Lucas number~\cite{You2022prr}, i.e., $d=F_{N-1}+F_{N+1}$, where $F_\ell$
is the $\ell$th Fibonacci number. More precisely, $d=g^N+g^{-N}$ with
$g=(1+\sqrt{5})/2 $ being the golden ratio. This exponentially large
subspace belongs to the largest Krylov subspace among the exponential
number of Krylov subspaces, implying strong fragmentation of the Hilbert
space. The graphical representation of the constrained Hilbert space in the  $\vec{w}{=}\{1,1,{\cdots},1\}$  subspace is schematically shown in Fig. \ref{fig:N_6_zzzzzz} for $N=6$.  The vertices in the 18-dimensional
hypercube is uniquely labeled by the connected configurations (\ref{Haction}),
which have been arranged by the action of the Kitaev Hamiltonian $\tilde{H}_{\rm K}$ on the product state $\vert\cdots zzzz\cdots\rangle$.

The process of bond converting
$\vert\cdots zz \cdots\rangle_{j,j+1}\leftrightarrow\vert\cdots y x\cdots\rangle_{j,j+1}$ under the action of $\tilde{H}_K$
corresponds to the spin flip $\vert\cdots 0\cdots\rangle_{j+1/2}
\leftrightarrow \vert \cdots 1 \cdots \rangle_{j+1/2}$ in
$\{\mathcal{K}_{S=1/2}\}$. In this regard, the spin-1 Kitaev chain with
periodic boundary conditions can be exactly mapped to the a single
qubit-flip model represented by the effective spin-1/2 PXP model in
Eq. (\ref{eq:H_PXP}). Remarkably, we find the ground state remains in
the flux-free sector even in the presence of the SIA.
The action of the SIA term on the active bases yields,
\begin{eqnarray}
D\!\left[(S_{j}^z)^2 +\!(S_{j+1}^z)^2\right]\!\vert \cdots yx \cdots\rangle_{j,j+1}
\!&=&2D \vert \cdots  y x \cdots\rangle_{j,j+1},\nonumber \\
D\!\left[(S_{j}^z)^2 +\!(S_{j+1}^z)^2\right]\!\vert \cdots  zz \cdots\rangle_{j,j+1}
\! &=& 0,
\end{eqnarray}
which results in an effective detuning term on the spin-1/2 degrees of
freedom, such that the effective Hamiltonian can be mapped to the
spin-1/2 detuned PXP model,
\begin{eqnarray}
\label{equ:dPXP}
\hat{H}_{\rm dPXP} \!=\sum_{i=1}^{N} P_{i-1}X_{i} P_{i+1}
+ 2D \sum_{i=1}^N P_{i-1}n_{i }P_{i+1},
\end{eqnarray}
where $n =1-P= \vert 1\rangle \langle 1 \vert$.
Note that in both Eq. (\ref{eq:H_PXP}) and Eq.~(\ref{equ:dPXP}),
$i$ labels the bonds between sites, while the index $j$ labels the
sites in Eq. (\ref{equ:tilde HkD}). The detailed derivation of Eq.
(\ref{equ:dPXP}) can be found in Appendix~\ref{APPENDIX A}.

\begin{figure}[t!]
\centering
\includegraphics[width=\columnwidth]{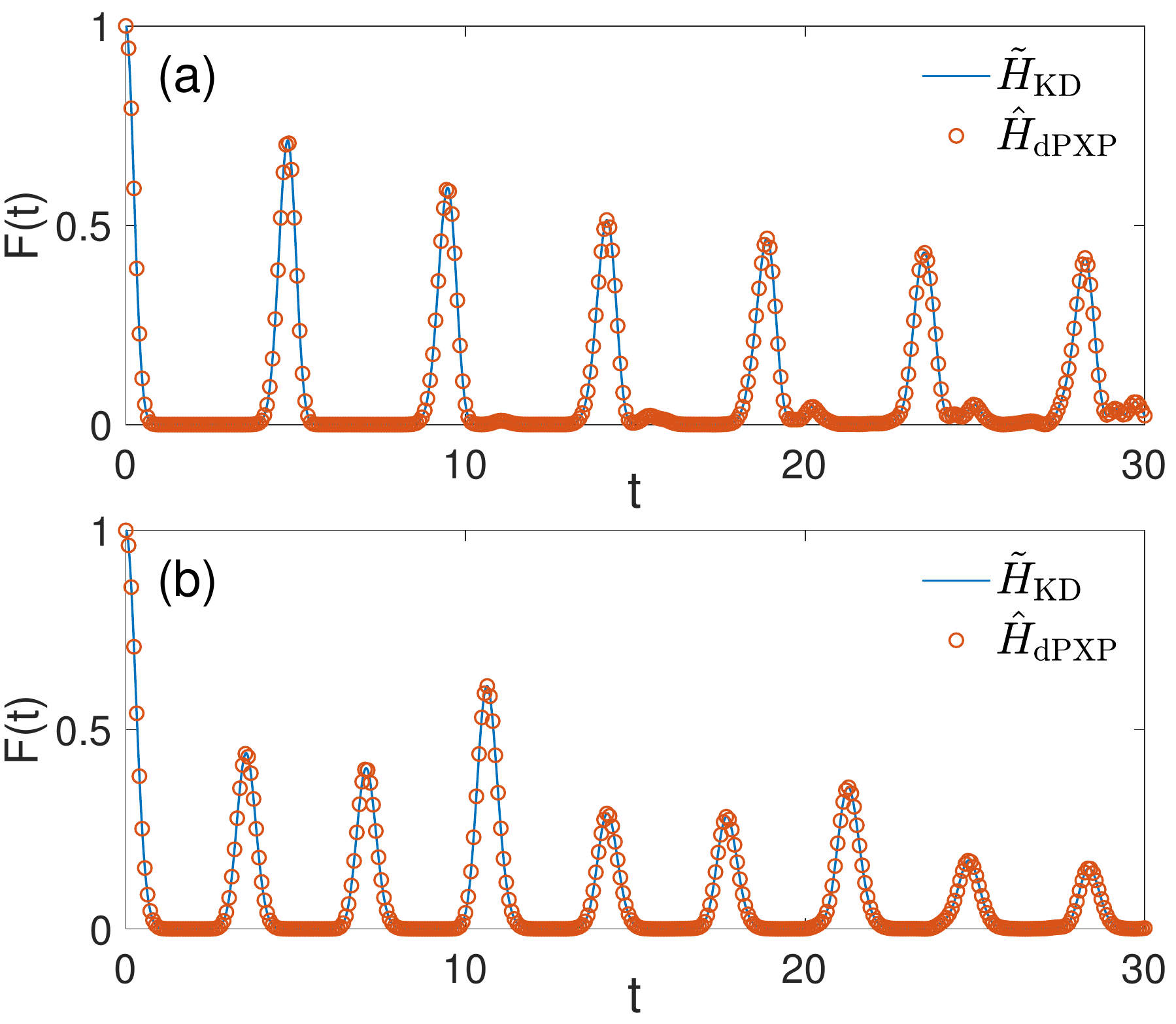}
\caption{ Characteristic quantum features of the spin-1 KD model and the
spin-1/2 detuned PXP model with $D = 0.1$.  Quantum fidelity $F(t)$ for
$\tilde{H}_{\rm KD}$ in Eq. (\ref{equ:tilde HkD}) [$\hat{H}_{\rm dPXP}$ in
Eq. (\ref{equ:dPXP})] starting from the initial states:
 \hfill\break
(a) $\vert\widetilde{\mathbb{Z}}_2\rangle$ ($\vert\mathbb{Z}_2\rangle$)
with $N = 18$, and (b) $\vert\widetilde{\mathbb{Z}}_3\rangle$
($\vert \mathbb{Z}_3 \rangle$) with $N = 18$. }
 \label{fig:Ft_KD_PXPPnP}
\end{figure}

The detuning term is commonly prevalent in practical experiments.
The static detuning (also called chemical potential~\cite{Daniel2023}) of the driving laser from the excited state can be finely tuned in the cold-atom platforms.
It has been noted that quantum quench from initial states $\vert \widetilde{\mathbb{Z}}_2 \rangle$ or $\vert \widetilde{\mathbb{Z}}_3 \rangle$ results in coherent oscillations, indicating the existence of ETH-violating QMBSs.  In our ED simulation~\cite{mynote}, the time-evolved  operator $\exp({-i\hat{H}  t})$
governed by either $\tilde{H}_{\rm KD}$  as defined in Eq. (\ref{equ:tilde HkD}) or $\hat{H}_{\rm dPXP}$ as given in Eq. (\ref{equ:dPXP})
is discretized using time steps of $dt = 0.01$, and the time-evolved state  $\vert \psi(t) \rangle$ is subsequently computed using the fourth-order Runge-Kutta method within the corresponding constraint Hilbert space.   Figure \ref{fig:Ft_KD_PXPPnP}  demonstrates  these oscillations in the dynamics of the quantum fidelity for $D=0.1$. The periodic revivals for the spin-1 KD model (\ref{equ:tilde HkD})  starting from the $\vert \widetilde{\mathbb{Z}}_2 \rangle$,  $\vert \widetilde{\mathbb{Z}}_3 \rangle$  initial states completely coincide with the ones observed for the spin-1/2 detuned PXP model, which starts from the corresponding  $\vert \mathbb{Z}_2 \rangle$,  $\vert \mathbb{Z}_3 \rangle$  initial states.

\begin{figure}[t!]
\centering
\includegraphics[width=1.07\columnwidth]{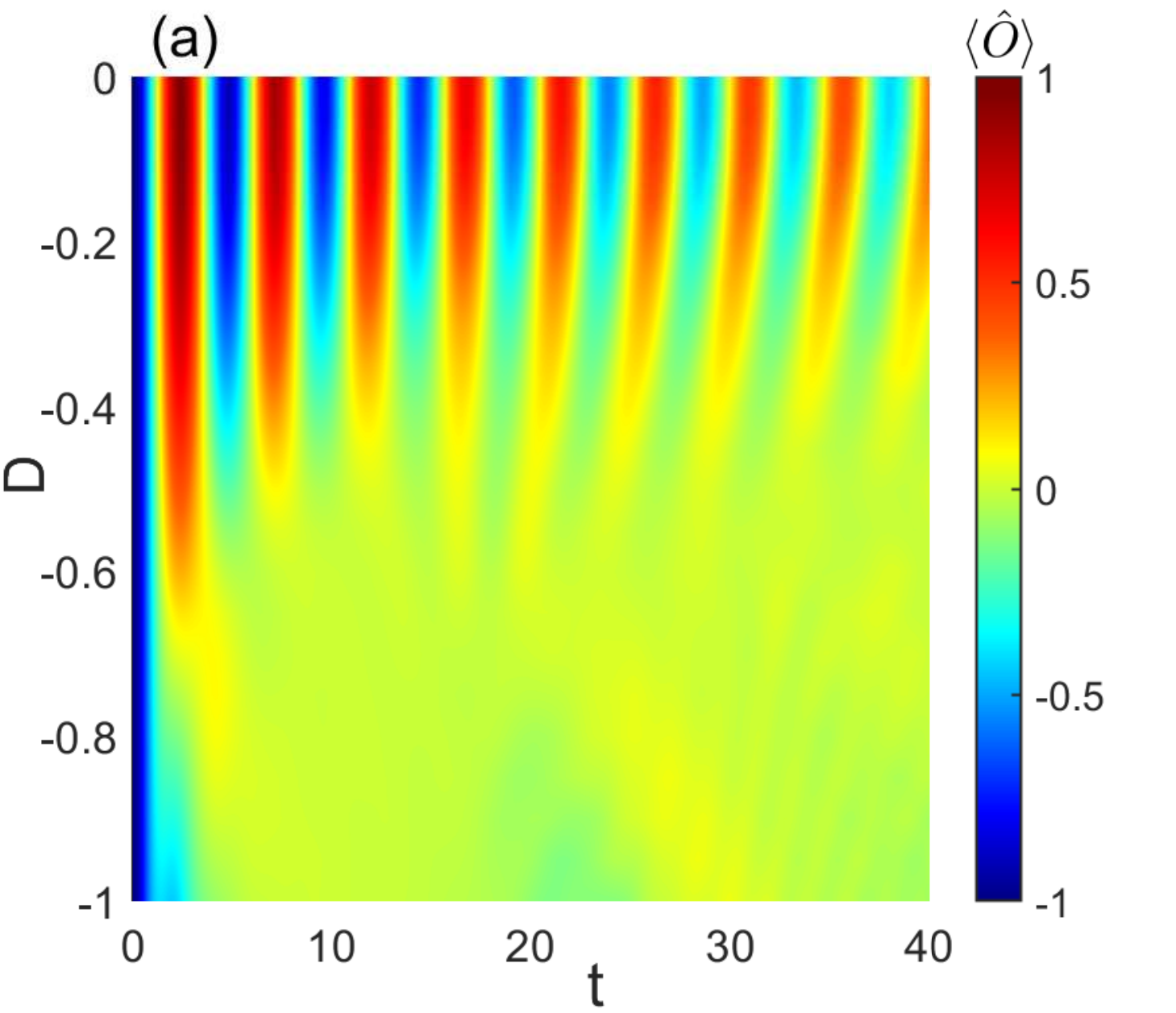}
\includegraphics[width=1.07\columnwidth]{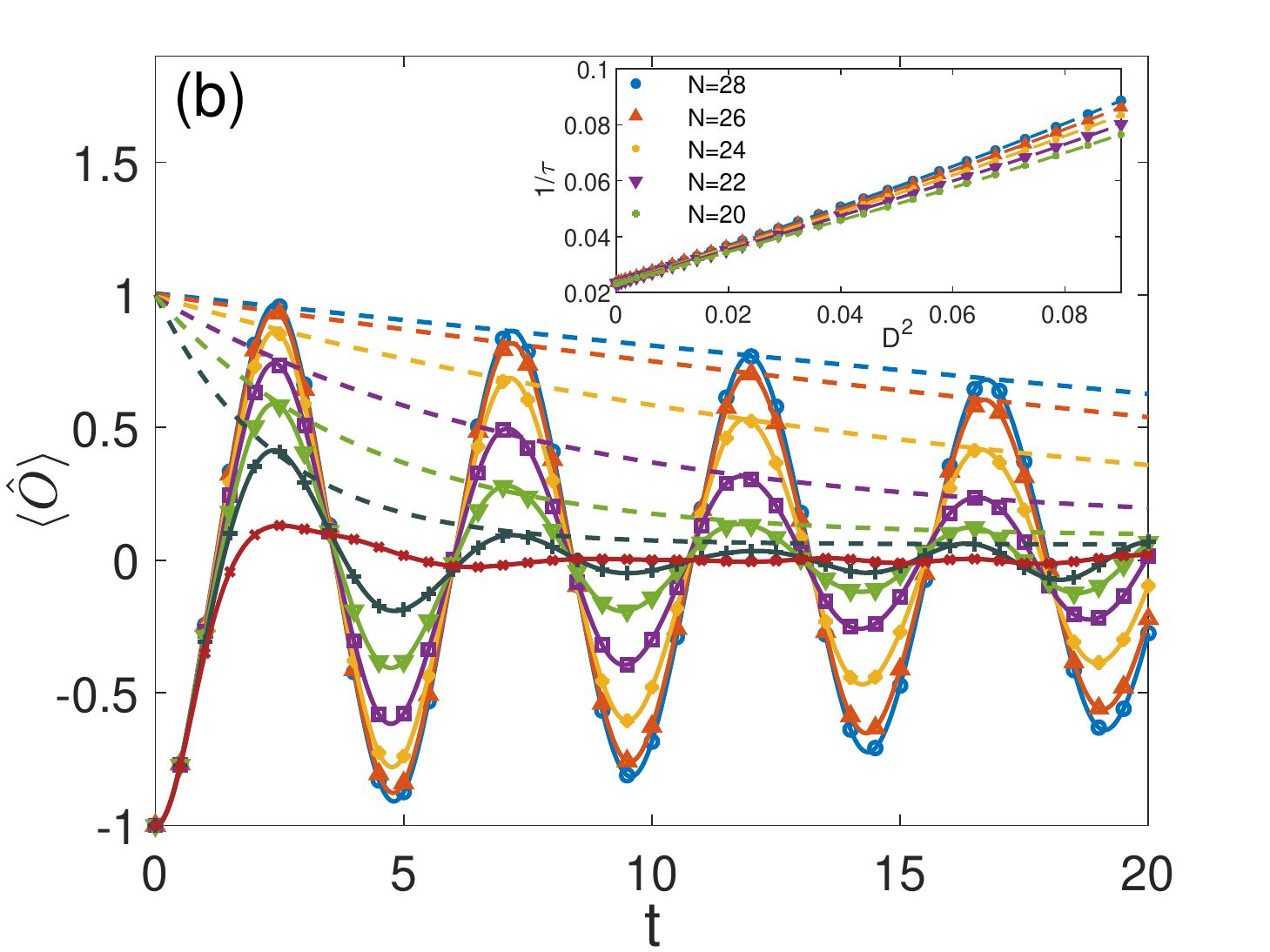}
\caption{ (a) The contour map of time evolution of $\langle \hat{O} \rangle$ defined in (\ref{eq:Ot}) of $\tilde{H}_{\rm KD}$ (\ref{equ:tilde HkD}) in a system of $N = 28$ spins prepared in $\vert \widetilde{\mathbb{Z}}_2 \rangle$ obtained by ED. (b)  The  time evolution of $\langle \hat{O} \rangle$ for different values of $D$. The curves correspond to $D=0.0$, $-0.1$, $-0.2$, $-0.3$, $-0.4$, $-0.5$, $-0.655$ (from bottom
to top at $t=5$). The dashed lines are fits capturing the amplitude decay.   
Inset shows inverse
lifetime of  $\langle \hat{O}(t)\rangle$  envelop with increasing $D^2$. We extract the decay time by fitting the data to the scaling ansatz $\langle \hat{O}(t) \rangle  =A e^{-t/\tau}\cos \omega t $ (see main text). }
 \label{fig:O_envelope_line}
\end{figure}

\begin{figure}[t!]
\centering
\includegraphics[width=1.07\columnwidth]{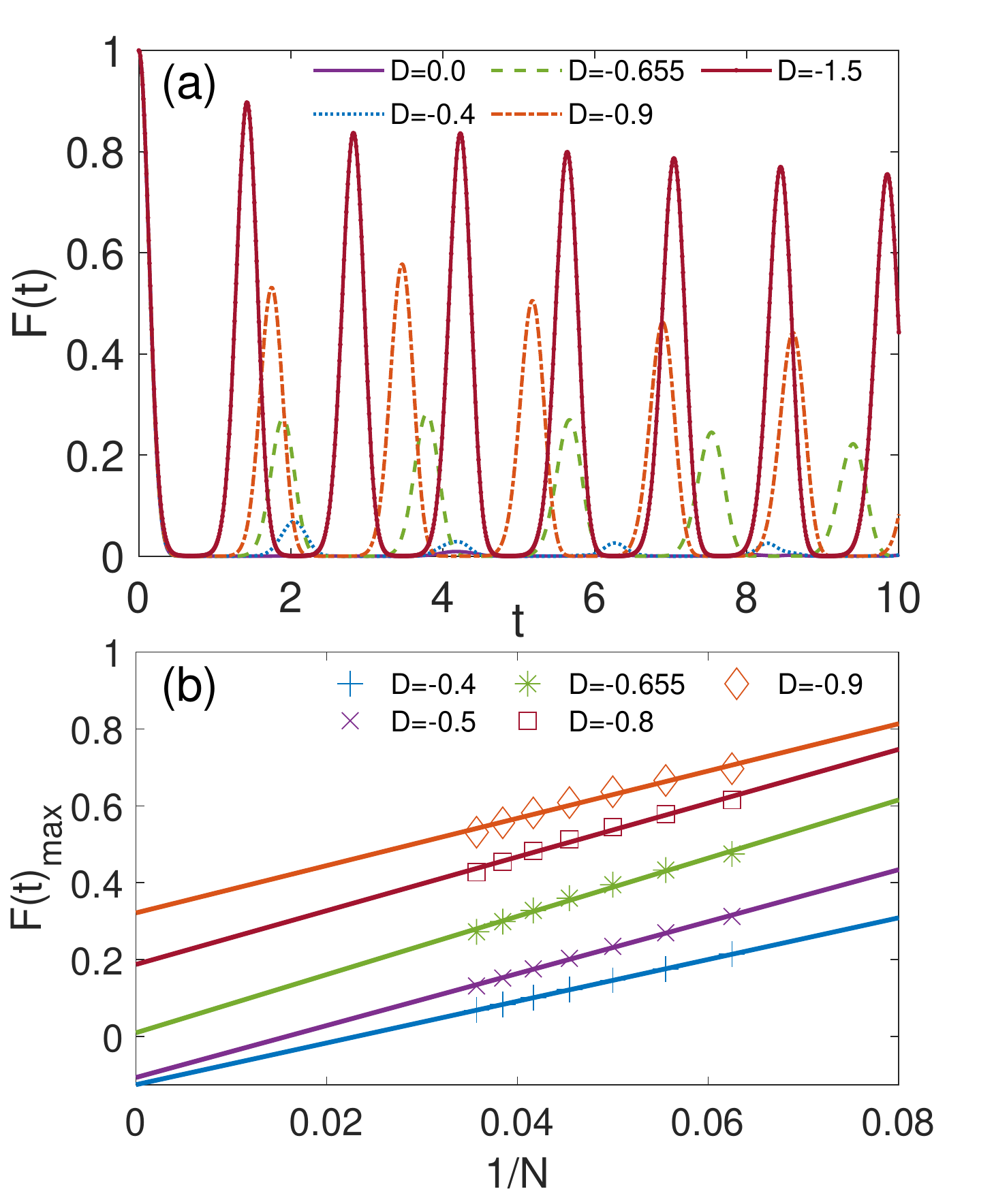}
\caption{ The dynamic evolution of the spin-1/2 detuned PXP model starting from the initial state $\vert \mathbb{Z}_1 \rangle$:
\hfill\break
(a) The quantum fidelity $F(t)$ with respect to different $D$ for $N\!=\!28$;
(b) Finite-size scaling of $F(t)$ for the first peak in panel (a).
}
\label{fig:Ft_a_L_28_b_scaling_L_16_28_polarised}
\end{figure}

Recent studies have signified an intimate relation between QMBS and
quantum criticality~\cite{CuiXiaoling2022,ZhaiHui2022}. As $D$ is tuned
to $D_c\approx -0.655 $, the ground state of the detuned PXP model
undergoes a Ising phase transition associated with a spontaneous
breaking of
$\mathbb{Z}_2$-symmetry~\cite{Byrnes2002,Rico2014,Yang2020,VanDamme2020}.
The non-thermalizing dynamics can be also captured by measuring the
expectation values of certain local observables ~\cite{LinChengJu2020}, e.g.,
\begin{eqnarray}
\label{eq:Ot}
\left\langle \hat{O}\right\rangle =
\frac{1}{2}\left\langle[\left(S_1^+\right)^2+\left(S_1^-\right)^2]\right\rangle.
\end{eqnarray}
Under the dual transformation (\ref{statesmap}), the correlator
$\langle\hat{O}\rangle$ of the KD model in Eq. (\ref{equ:tilde HkD}) is found
to be equivalent to the density imbalance,
$\langle n_2 \rangle - \langle n_1 \rangle$, an observable
corresponding to the staggered magnetization in the detuned PXP model
in Eq. (\ref{equ:dPXP}). Performing a quantum quench from an initial
state  $\vert \widetilde{\mathbb{Z}}_2 \rangle$ leads to nearly perfect
coherent dynamics. The coherence oscillations persist for long times for
$D=0$, as is shown in Fig.\ref{fig:O_envelope_line}. Note that the values of $F(t)$ and $\langle \hat{O}\rangle$ is
independent of the sign of $D$ when the system starts from the product
states, $\vert\widetilde{\mathbb{Z}}_k\rangle$ (see details in Appendix
\ref{APPENDIX B}).

As exhibited in Fig. \ref{fig:O_envelope_line}(a), these oscillations
are found to be remarkably robust to small SIA perturbations, while
moderate perturbations make the oscillations damp sharply until $D$
reaches a threshold value. One carefully observes from Fig. \ref{fig:O_envelope_line}(b) that the oscillations remain strong for
deviations up to $D \approx\pm D_c$, and there is barely oscillation at
$D=D_c$, upon which the thermalization completely sets in. Suppose that
the envelope of $\langle\hat{O}\rangle$ can be described by
exponentially decaying oscillations,
$\langle \hat{O}(t)\rangle=Ae^{-t/{\tau}}\cos\omega t$ over time $t$,
with the fitting parameters $A$, $\tau$, and $\omega$.
We observe that  the inverse lifetime approximately follows $\tau^{-1}\sim D^2$ at
small $D$, reminiscent of the Fermi’s golden rule~\cite{Langlett2022}.
Additionally, it is worth noting that the decay rate of oscillations $\tau^{-1}$  at $D=0$ remains small but finite, suggesting that the $\vert \widetilde{\mathbb{Z}}_2 \rangle$ initial state only approximates the near-perfect scar states in the standard PXP model.
The fact that the quantum critical point $D_c$ is negative and there is
no quantum phase transition for positive $D$, combined with the
significant difference between the ground states as $D$ tends
infinity, undermine the viewpoint that quantum many-body scars and
quantum criticality are directly bridged.

\begin{figure}[t!]
\centering
\includegraphics[width=1.07\columnwidth]{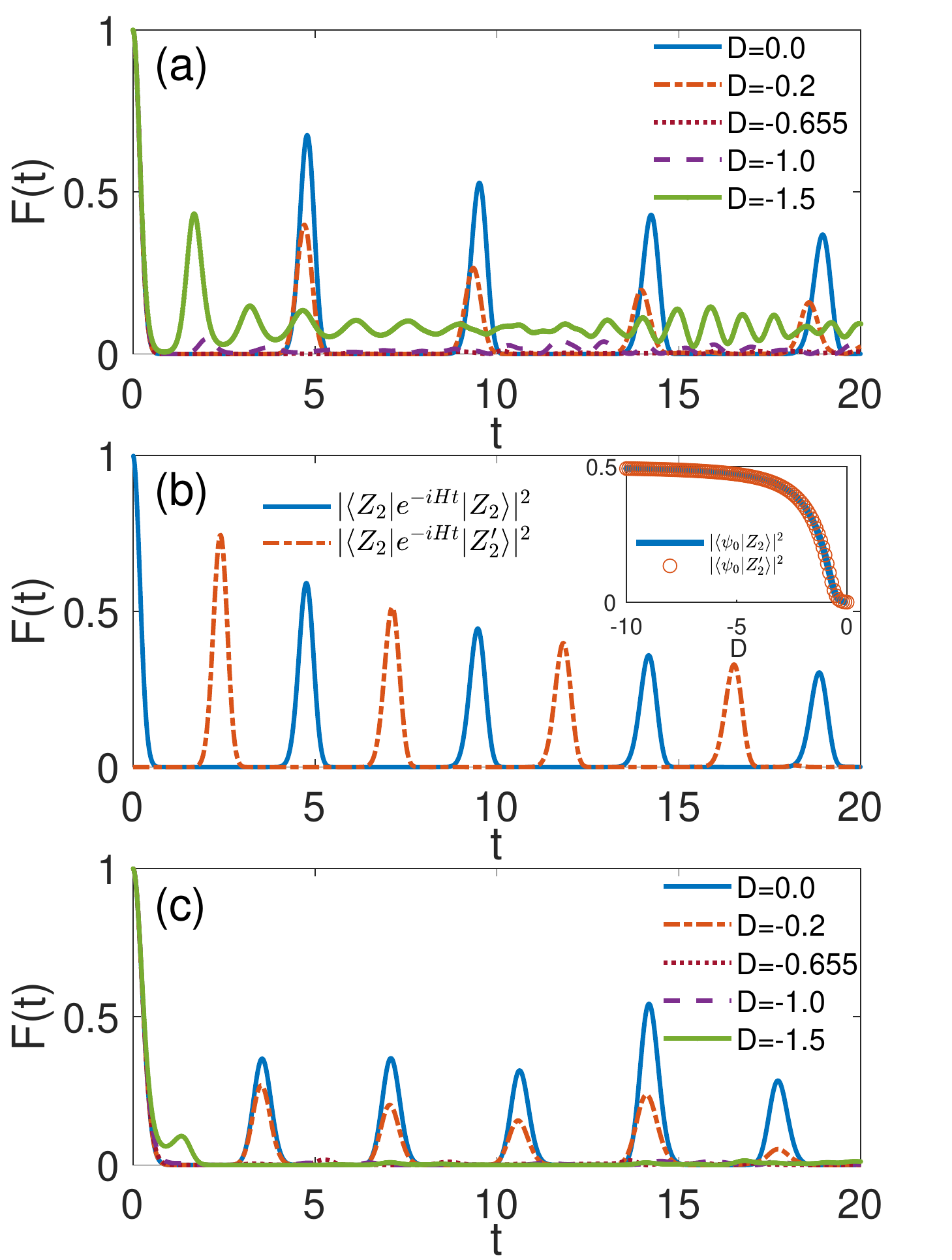}
\caption{ Dynamics of quantum fidelity for the detuned PXP model:
(a) Starting from initial state $\vert \mathbb{Z}_2 \rangle$  for $N=28$
sites;
(b) The overlaps between the product state $\vert\mathbb{Z}_2\rangle$
and the time-evolved state starting from  $\vert\mathbb{Z}_2\rangle$
(solid) and $\vert \mathbb{Z}_2^\prime \rangle$ (dashed) with $D=-0.1$
for $N=28$ sites. The inset shows the overlap of the prequench ground
state with $\vert\mathbb{Z}_2\rangle$ and
$\vert \mathbb{Z}_2^\prime \rangle$.
(c) Starting from initial $\vert \mathbb{Z}_3 \rangle$ state for $N=24$
sites.
   }
 \label{fig:Ft}
\end{figure}

For $D=0$, the quantum fidelity revivals do not occur for the initial
state $\vert \mathbb{Z}_1\rangle$. As $D$ decreases from
zero, surprisingly,  there will be a slight revival in fidelity for the
same initial state. As $D$ continues to decrease, the oscillation
becomes more clearly visible with smaller periods, as observed in Fig. \ref{fig:Ft_a_L_28_b_scaling_L_16_28_polarised}(a). When the value of
$D$ is smaller than the critical value $D_c$, the revivals become more
pronounced. 
Finite-size scaling in Fig. \ref{fig:Ft_a_L_28_b_scaling_L_16_28_polarised}(b) reveals that the first peak will disappear for large $N$ before the value of $D$ exceeds $D_c$. When $D > D_c$, the intercepts of the finite-size scaling curves become negative, which is an unphysical artifact and indicates that the linear fit is no longer applicable.
In contrast, the first peak will always have a finite value for $D<D_c$
in the thermodynamic limit, which may be related to asymptotic scars~\cite{gotta2023asymptotic}.

Figure \ref{fig:Ft}(a) shows the quantum fidelity of $\hat{H}_{\rm dPXP}$ 
with different values of
$D$ for $N=28$ using the initial $\vert \mathbb{Z}_2 \rangle$ state. Remarkably, when $D$ decreases from
zero, persistent
oscillations first decrease when $D>D_c$, then damp in the critical
regime $D\approx D_c$, and finally revive beyond the critical point for
$D<D_c$. Figure \ref{fig:Ft}(b) shows the overlaps between the
$\vert\mathbb{Z}_2\rangle$ state and the time evolved state starting
from $\vert\mathbb{Z}_2\rangle$ and $\vert\mathbb{Z}_2^\prime\rangle$,
where $\vert\mathbb{Z}_2^\prime\rangle\equiv\vert 101010\cdots 10\rangle$
is obtained by translating one lattice spacing on
$\vert\mathbb{Z}_2\rangle$.

The peaks of the oscillations of  $\vert \langle\mathbb{Z}_2\vert\exp(-i\hat{H}t)
\vert\mathbb{Z}_2 \rangle \vert^2$ and $\vert \langle\mathbb{Z}_2\vert
\exp(-i\hat{H}t)\vert\mathbb{Z}_2^\prime\rangle\vert^2$ are separated by half
a period.
We also show the fidelity between the $\vert \mathbb{Z}_2\rangle$ state
and the ground state $\vert\psi_0 \rangle$ at different values of $D$,
as shown in the inset of Fig. \ref{fig:Ft}(b). We observe that as $D$
approaches negative infinity, the fidelity between the ground state, and
$\vert\mathbb{Z}_2\rangle$ ($\vert\mathbb{Z}_2^\prime\rangle$) gradually
approaches $1/2$.
We remark that due to the Hilbert space constraint, at most half of the
atoms could be in the spin-up states. In fact, in the limit of
$D\to -\infty$, the ground state becomes an antiferromagnetic phase in
zero-momentum sector, i.e., $\vert\psi_0(D=-\infty)\rangle=
(\vert \mathbb{Z}_2 \rangle+\vert\mathbb{Z}_2^\prime\rangle)/\sqrt{2}$.
In contrast, Fig. \ref{fig:Ft}(c) demonstrates a complete absence of
revivals for $D<D_c$ in the case of an initial state of
$\vert\mathbb{Z}_3\rangle$, featuring approximate QMBS states vanish.

\begin{figure}[t!]
\centering
\includegraphics[width=0.49\columnwidth]{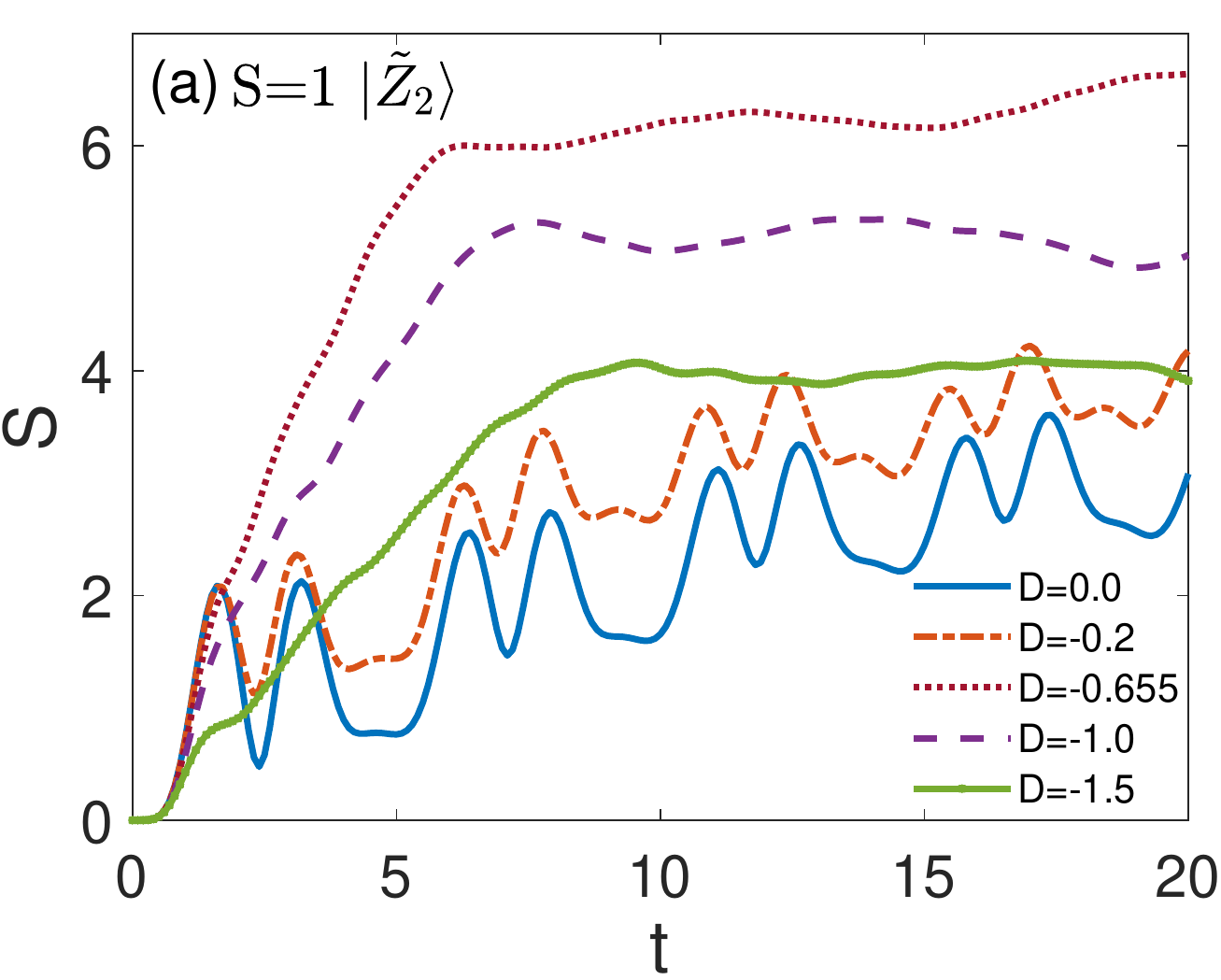}
\includegraphics[width=0.49\columnwidth]{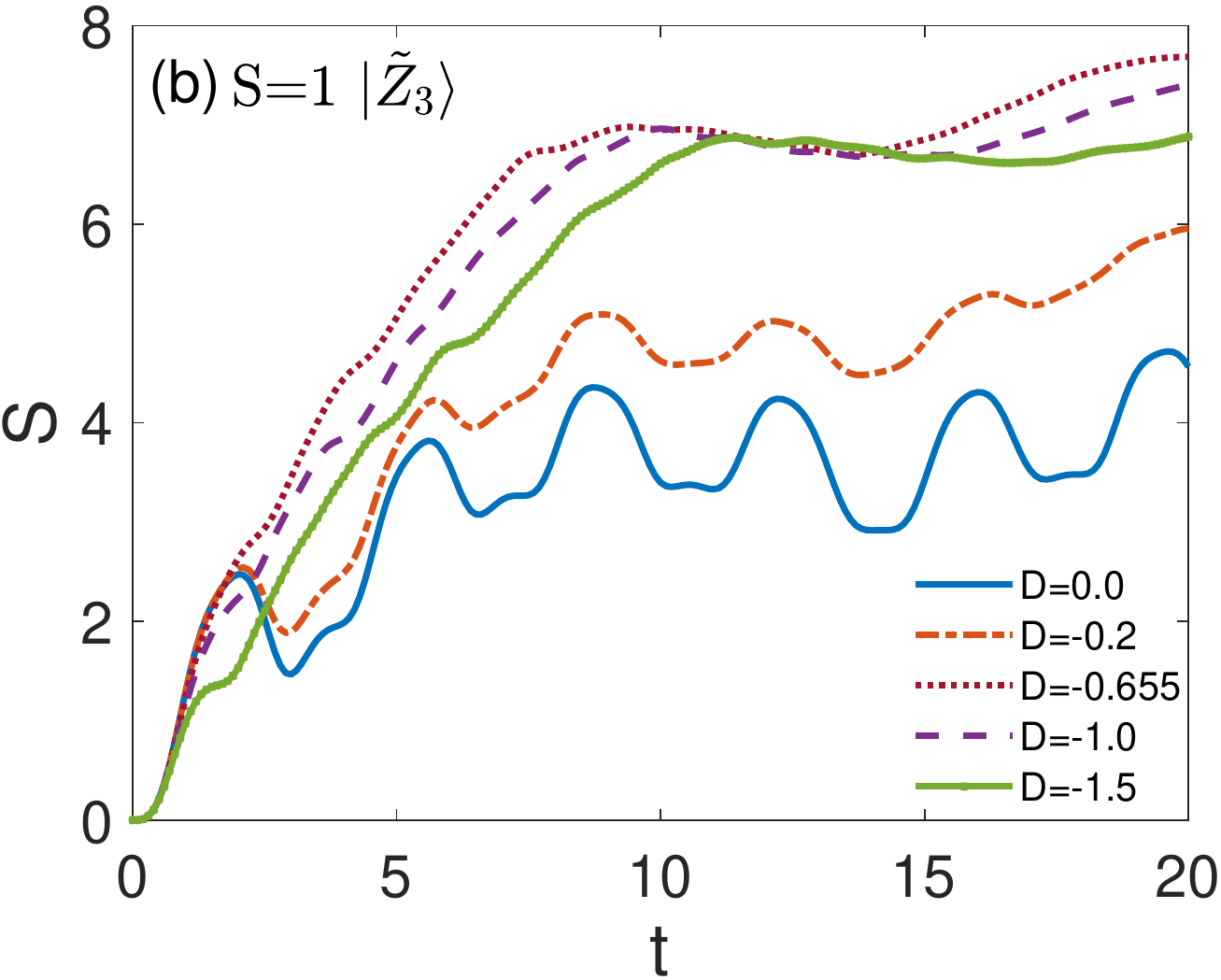}
\vskip .2cm
\includegraphics[width=0.49\columnwidth]{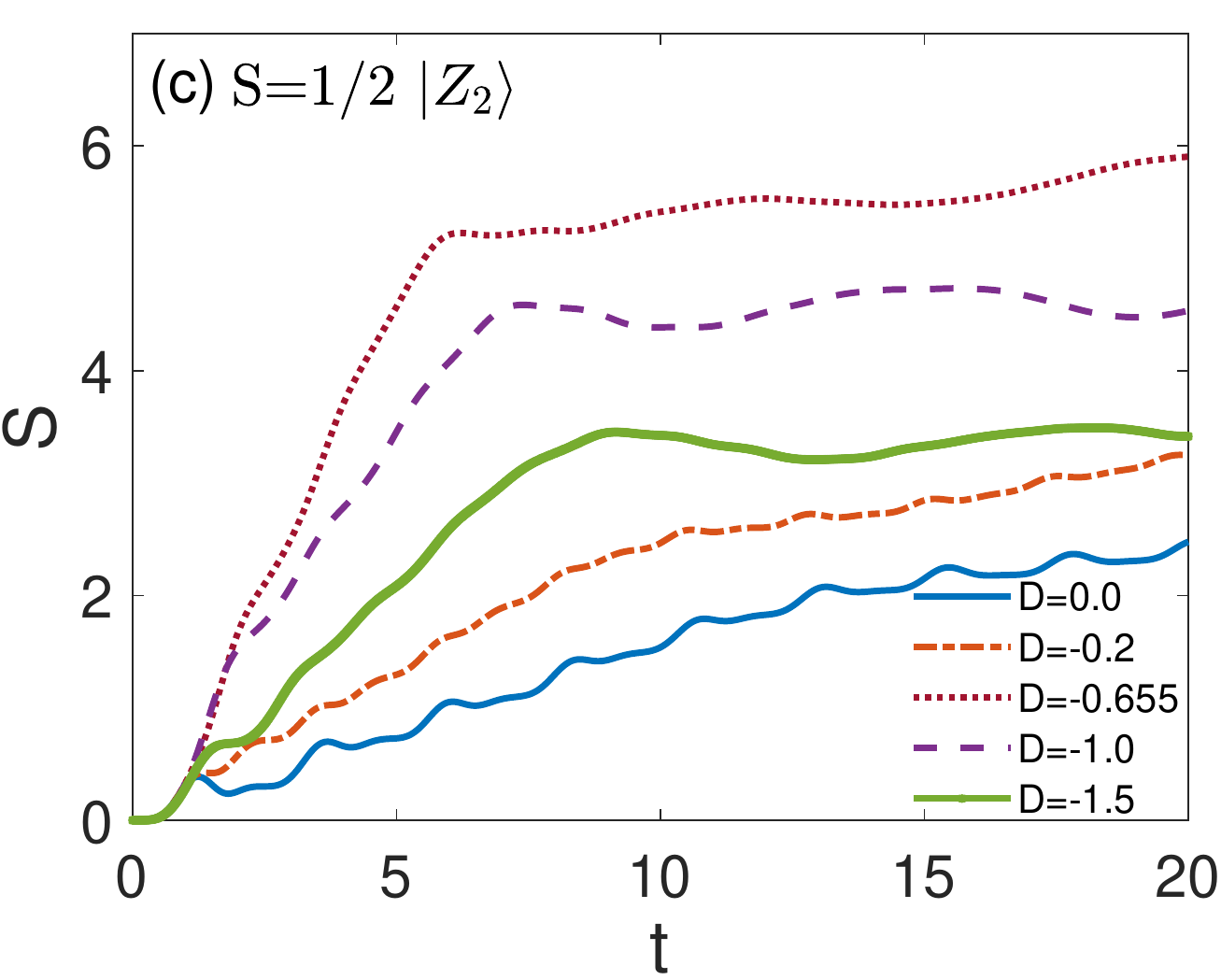}
\includegraphics[width=0.49\columnwidth]{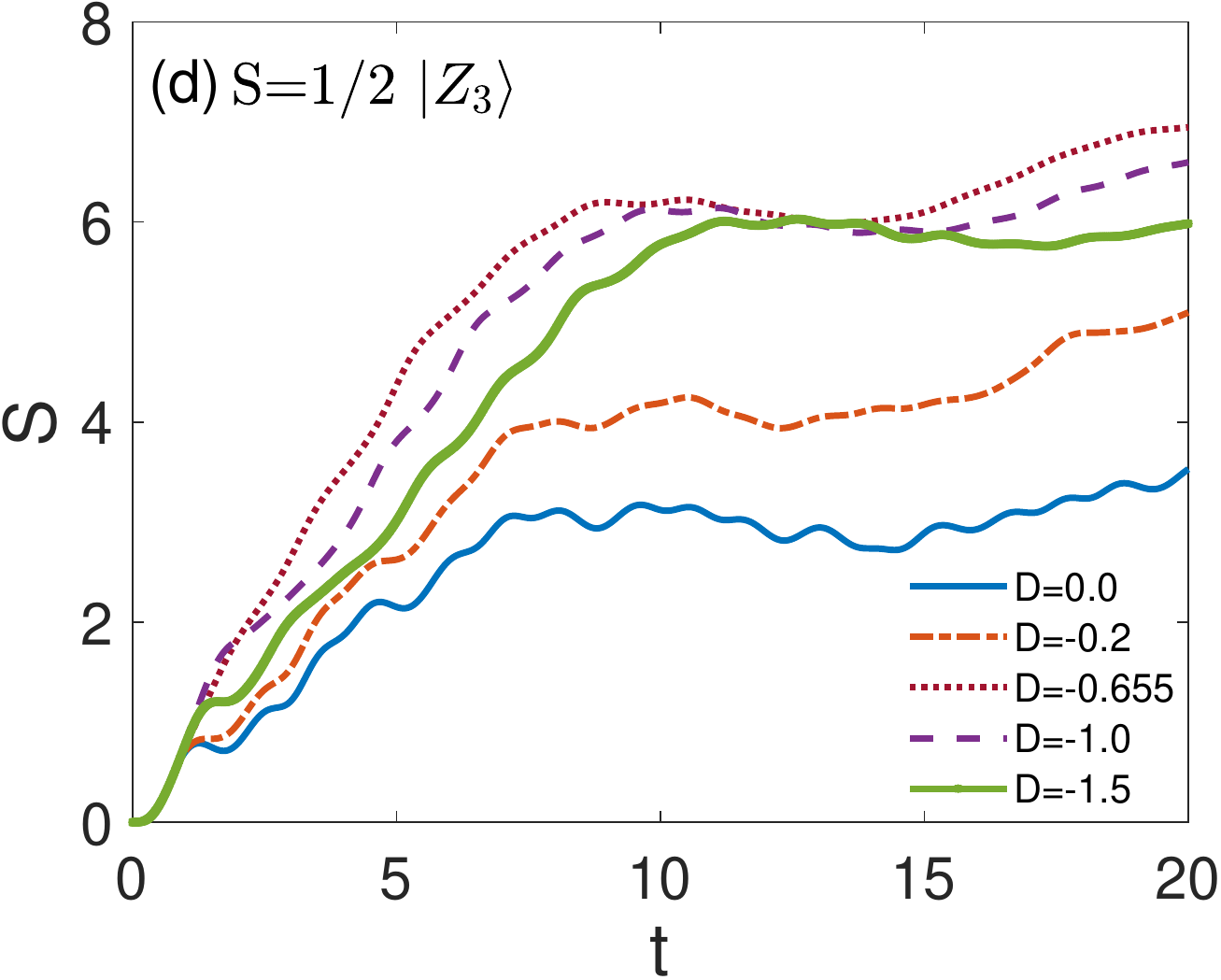}
\vskip .2cm
\caption{ Evolution of the bipartite entanglement entropy in the quantum
quench from initial states:
(a) $\vert \widetilde{\mathbb{Z}}_2 \rangle$,
(b) $\vert \widetilde{\mathbb{Z}}_3 \rangle$ for the spin-1 KD model,
(c) $\vert {\mathbb{Z}}_2 \rangle$ and
(d) $\vert {\mathbb{Z}}_3 \rangle$ states for the spin-1/2 detuned PXP model
with $N=24$.   }
\label{fig:SA}
\end{figure}

We next investigate the dynamics of bipartite entanglement entropies in both the KD model and the detuned PXP model.
We choose the region $A$ to be one half of the chain, and compare the
dynamics of half-chain entanglement entropy ${\cal S}$ in a quantum
quench from different initial states for both the spin-1 KD model and
the spin-1/2 detuned PXP model with periodic boundary conditions. In our numerical calculation for $S=1$ KD model [cf., Figs. \ref{fig:SA}(a-b)], we utilize the time-evolving MPS approach with MPOs~\cite{ITensor}, where the bond dimension is set as $\chi = 500$ and the time step is $dt=0.025$. The
bipartite entanglement of the evolved state starting from the initial
state $\vert\widetilde{\mathbb{Z}}_2\rangle$ is shown in Fig.
\ref{fig:SA}(a). When $D$ is small negative, ${\cal S}$ increases slowly
over time while exhibits coherent oscillations, featuring the
many-body revivals. As $D$ becomes more negative, the temporal growth
rate of the bipartite entanglement increases, and the
coherent oscillation becomes weak. When $D$ approaches $D_c$, the
entanglement ${\cal S}$ almost increases linearly with time until
saturation, and the coherent oscillations disappear, implying that
the system quickly thermalizes.

When $D$ is smaller than $D_c$, the
linear growth rate of entanglement decreases with a smaller saturated
value of entanglement. We find that the growth of entanglement entropy
starting from the initial state $\vert\widetilde{\mathbb{Z}}_3\rangle$
exhibits a similar trend, as shown in Fig. \ref{fig:SA}(b). For
comparison, we show the evolution of the bipartite von Neumann
entanglement entropy of the spin-1/2 detuned PXP model when the system
is initially prepared in the state $\vert {\mathbb{Z}}_2\rangle$
($\vert{\mathbb{Z}}_3 \rangle$) in Fig. \ref{fig:SA}(c)
[\ref{fig:SA}(d)]. One observes ${\cal S}$ is gradually growing for $D$
being small negative, while undergoes an extremely fast growth until saturation at
$D\ \approx D_c$. When $D < D_c$, the
growth of entanglement entropy slows down again.
Although the two Hamiltonians and their corresponding initial states are unitarily equivalent under the local transformation (\ref{statesmap}), the entanglement evolution displays noticeable differences. Notably, the coherent oscillations for $S=1/2$ become considerably weaker compared to those for $S=1$ when $0\ge D>D_c$. The bipartite entanglement entropy heavily depends on the choice of presentations and bipartition methods. This can be perceived by an analytical example presented in the Appendix A of Ref.~\cite{You2022prr}.

\section{Quantum phase transition of spin-1 Kitaev \\
chain with uniaxial single-ion anisotropy}
\label{sec:QPTKD}

In the previous section, we discovered a close relationship between QMBS
states and quantum criticality. Accordingly, we proceed to
investigate the quantum phase transition of the spin-1 KD model given by
Eq. (\ref{equ:KD}). We adopt the iTEBD algorithm with a bond dimension of $\chi=120$~\cite{mynote}. In our calculations, we set the imaginary time as $10^{-5.5}$ to ensure a truncation error smaller than $10^{-8}$. The advantage of using iTEBD is its capability to treat infinite-size systems directly, providing numerical evidence for the emergence of a symmetry-breaking phase. According to the core spirit of the
Landau-Ginzburg-Wilson paradigm, the quantum phase transition of a
many-body system can be described by a well-defined order parameter.
We calculate the two-point correlations between the $i$-th and $j$-th sites,
\begin{eqnarray}
\label{eq:C(i,j)}
C^{ab}(i,j)=\left\langle S^{a}_i \exp\left(i\theta\sum_{l=i+1}^{j-1}S_l^a\right) S^{b}_j \right\rangle, a,b=x,y,z, ~~
\end{eqnarray}
which can detect different symmetry-breaking phases. Equation (\ref{eq:C(i,j)}) reduces to two-point correlations for $\theta=0$, while it becomes the den Nijs-Rommelse
string order parameter for $\theta=\pi$~\cite{den.Nijs.Marcel1989,Tasaki.Hal1991}. Note that there is no phase accumulated for two nearest-neighboring sites and a general angle $\theta$ could capture the hidden topological orders
~\cite{Liu2015}.
The Hamiltonian in Eq. (\ref{equ:KD}) is invariant under a joint operation that combines
a $\pi/2$-rotation about the $z$-axis
and a single-site translation, which implies that on a finite-size system
\begin{eqnarray}
\label{jointsymmetry}
&&C^{xx}(1,2) = C^{yy}(2,3),\quad C^{yy}(1,2) = C^{xx}(2,3), \nonumber\\
&&C^{zz}(1,2) = C^{zz}(2,3).
\end{eqnarray}
The joint symmetry can be expressed in the rotated Hamiltonian
(\ref{equ:tilde HkD}) as $\tilde{C}^{xy}(1,2)=\tilde{C}^{xy}(2,3)$,
$\tilde{C}^{yx}(1,2)=\tilde{C}^{yx}(2,3)$, and
$\tilde{C}^{zz}(1,2)=\tilde{C}^{zz}(2,3)$.

\begin{figure}[t!]
\centering
\includegraphics[width=1.07\columnwidth]{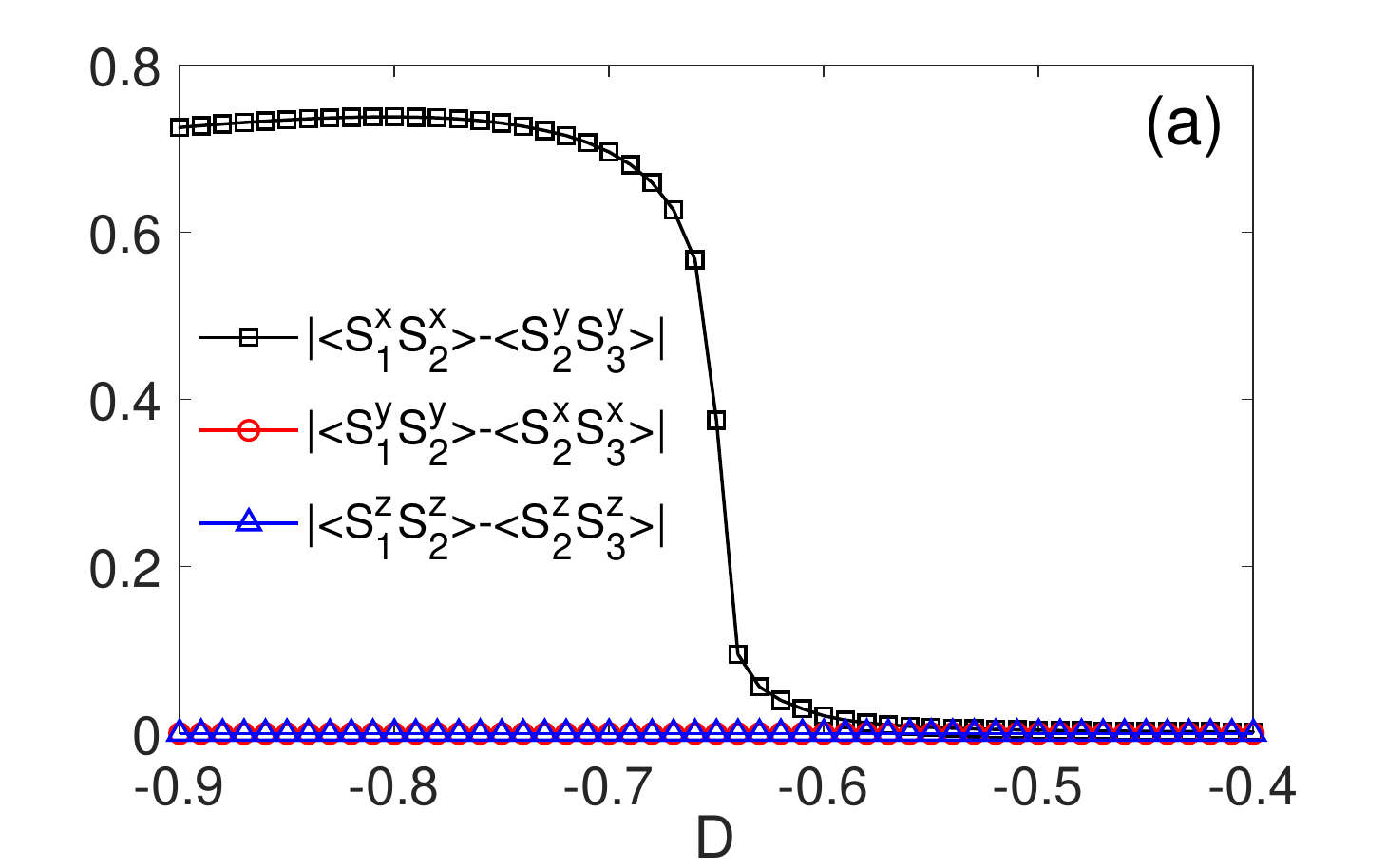}
\includegraphics[width=1.07\columnwidth]{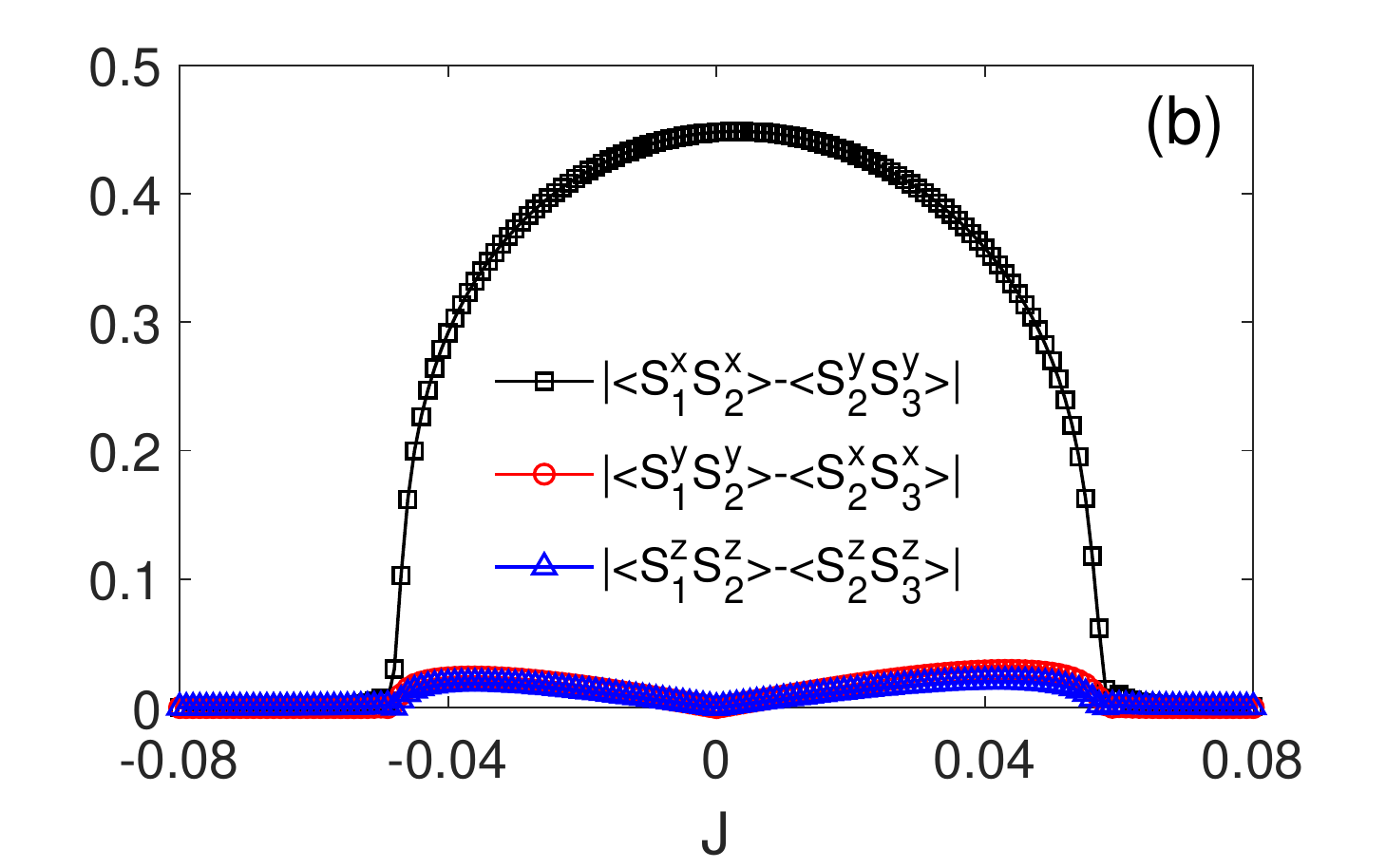}
\caption{ The three components of dimer order parameter $O_D$ as a function of $D$.
(a) From the dimer phase to the KSL phase at \mbox{$J=0$}.
(b)  From the FM$_z$ phase to the dimer phase to the AF$_z$ phase at
\mbox{$D=-2$.}
Here we use the iTEBD method and the bond dimension is set as $\chi=120$.  }
\label{fig:Dimer}
\end{figure}

In the zero-field limit, the ground state is a gapped Kitaev spin liquid
(KSL), which is stable against nonzero perturbations~\cite{You2020}.
Upon applying the uniaxial single-ion anisotropy $D$, the ground state
remains in the flux-free sector, i.e., $\vec{w}{=}\{1,1,\cdots,1\}$.
At a large positive $D$, the spins are confined to $\vert z \rangle$
(i.e., $\langle S_{j}^z \rangle=0$), while for a large negative $D$,
the ground states are restricted to $\vert x \rangle$ or
$\vert y\rangle$ ($\langle S_{j}^z\rangle=\pm 1$ ).
Surprisingly, unlike the KSL phase, a notable difference between
$C^{xx}(1,2)$ and $C^{yy}(2,3)$, or equivalently, $\tilde{C}^{xy}(1,2)$
and $\tilde{C}^{xy}(2,3)$, implies the spontaneous breaking of the
translational symmetry, in the way how the system hosts the dimer order.
The dimer phase is characterized by an alternation of nearest-neighbor
spin-spin correlations, which is characterized by the difference of
$\langle\boldsymbol{S}_i\cdot\boldsymbol{S}_j\rangle$ between the odd
bonds and even bonds. A~finite dimer order parameter is defined by
\begin{eqnarray}
\label{dimerorderparameter}
O_D &=& \vert \langle \boldsymbol{S}_{2j-1} \cdot \boldsymbol{S}_{2j} \rangle  -  \langle \boldsymbol{S}_{2j} \cdot \boldsymbol{S}_{2j+1} \rangle  \vert.
\end{eqnarray}
 To be more specific, we can also examine the $x$, $y$, and $z$ components of the dimer order parameter, such as
\begin{eqnarray}
\label{dimerorderparametercomponents}
O_D^x&=&\vert \langle {S}_{2j-1}^x {S}_{2j}^x \rangle
 -  \langle {S}_{2j}^y   {S}_{2j+1}^y \rangle  \vert, \nonumber\\
O_D^y&=&\vert \langle {S}_{2j-1}^y {S}_{2j}^y \rangle
 -  \langle {S}_{2j}^x   {S}_{2j+1}^x \rangle  \vert, \nonumber\\
O_D^z&=&\vert \langle {S}_{2j-1}^z {S}_{2j}^z \rangle
-  \langle {S}_{2j}^z  {S}_{2j+1}^z \rangle  \vert .
\end{eqnarray}
Note that the dimer order arises from the Kitaev interactions (\ref{equ:KD}), leading to the characterization of the $x$ and $y$ components as the differences between distinct types of Ising interactions on odd and even bonds.
Figure \ref{fig:Dimer}(a) illustrates that the $x$ component of the
dimer order parameter increases smoothly from zero to a finite value as
the parameter $D$ is decreased and crosses the critical value
$D_c = -0.655$,  indicating a second-order transition occurs at $D_c$.
The presence of nonvanishing dimer correlations for $D > D_c$ can be attributed to the limitations imposed by the finite bond dimension.
The dimer orders are associated with the spontaneous breaking of
translational symmetry of $O_D^x$ in an infinite system. The emergence
of the dimer ordering is distinct from the general mechanism for the
formation of dimerized phases, which is typically induced by inherent
bond alternation and the resulting breaking of translational symmetry.
The ground state is two-fold degenerate for $D<D_c$ in the thermodynamic
limit, which is in contrast to the gapped ground state for $D>D_c$.

\section{Effect of Heisenberg interactions}
\label{sec:Heisenberg}

\begin{figure}[t!]
\centering
\includegraphics[width=\columnwidth]{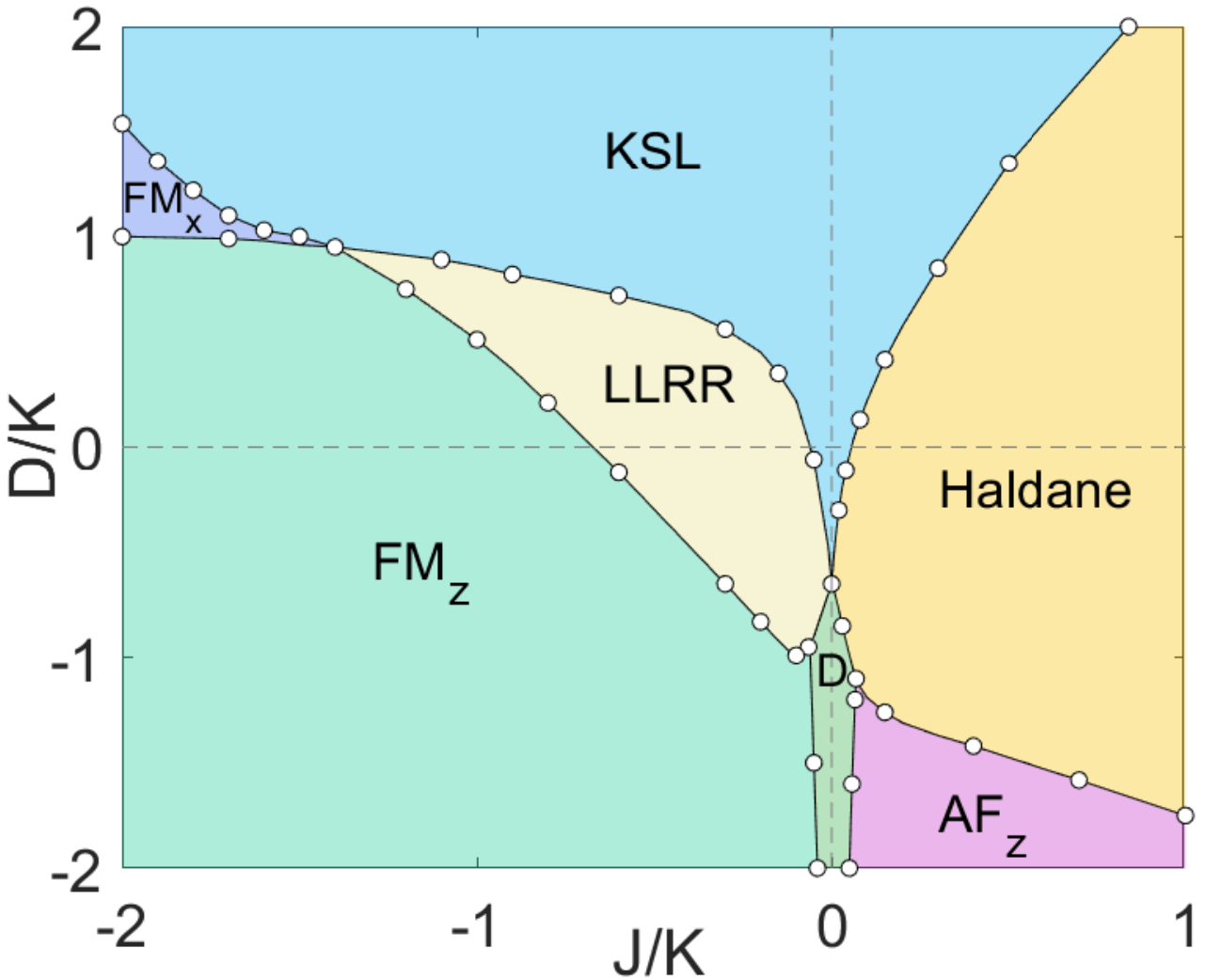}
\caption{Quantum phase diagram of the spin-1 Kitaev-Heisenberg model
with uniaxial single-ion anisotropy calculated by the iTEBD method with
bond dimension $\chi=120$.
The quantum phase transition from the dimer phase (D) to the KSL phase
occurs at \mbox{$D_c=-0.655$} for $J = 0$ (vertical dashed line).
At $D=0$ (horizontal dashed line), the KSL is stable in the range of
$\vert J\vert<0.08$.
 }
\label{fig:phase diagram}
\end{figure}

It has been recognized the scarred states display anomalous
stability in the Kitaev phase in the vicinity of $D=0$~\cite{You2022prr}.  Quantum spin liquids are widely believed to be crucially driven by the Kitaev interactions
in spin-orbit-coupled materials.
While Kitaev interactions are highly anisotropic, the isotropic Heisenberg interaction, ubiquitous in real materials, can also play an essential role in the emergence of exotic phenomena in many-body systems. The relevance of the Kitaev phase in a broader regime becomes paramount for understanding scar stability and its potential applications in solid-state systems.  To address this point, we investigate the evolution of the phase boundaries of the Kitaev spin liquid and the dimer phase by introducing Heisenberg interactions that disrupt the $\mathbb{Z}_2$ gauge fields, as given by
\begin{eqnarray}
\hat{H}_{\rm J} &=& J \sum_{j=1}^{N}\boldsymbol{S}_{j} \cdot \boldsymbol{S}_{j+1}. \label{HamJ}
\end{eqnarray}
When the parameters $\{D,J\}$ vary, the competitions of various
correlations trigger miscellaneous phase transitions. Figure
\ref{fig:phase diagram} depicts the phase diagram for the
Kitaev-Heisenberg chain with uniaxial single-ion anisotropy (KHD model).
The phase diagram is much richer than expected. Seven distinct phases
are identified, including the KSL phase, dimer phase ($D$), the spin
nematic phase with a left-left-right-right pattern (LLRR), Haldane phase,
$x$-component ferromagnetic (FM$_x$) phase, \mbox{$z$-component}
ferromagnetic (FM$_z$) phase and $z$-component antiferromagnetic
(AF$_z$) phase.

\begin{figure}[t!]
\centering
\vskip .2cm
\includegraphics[width=0.49\columnwidth]{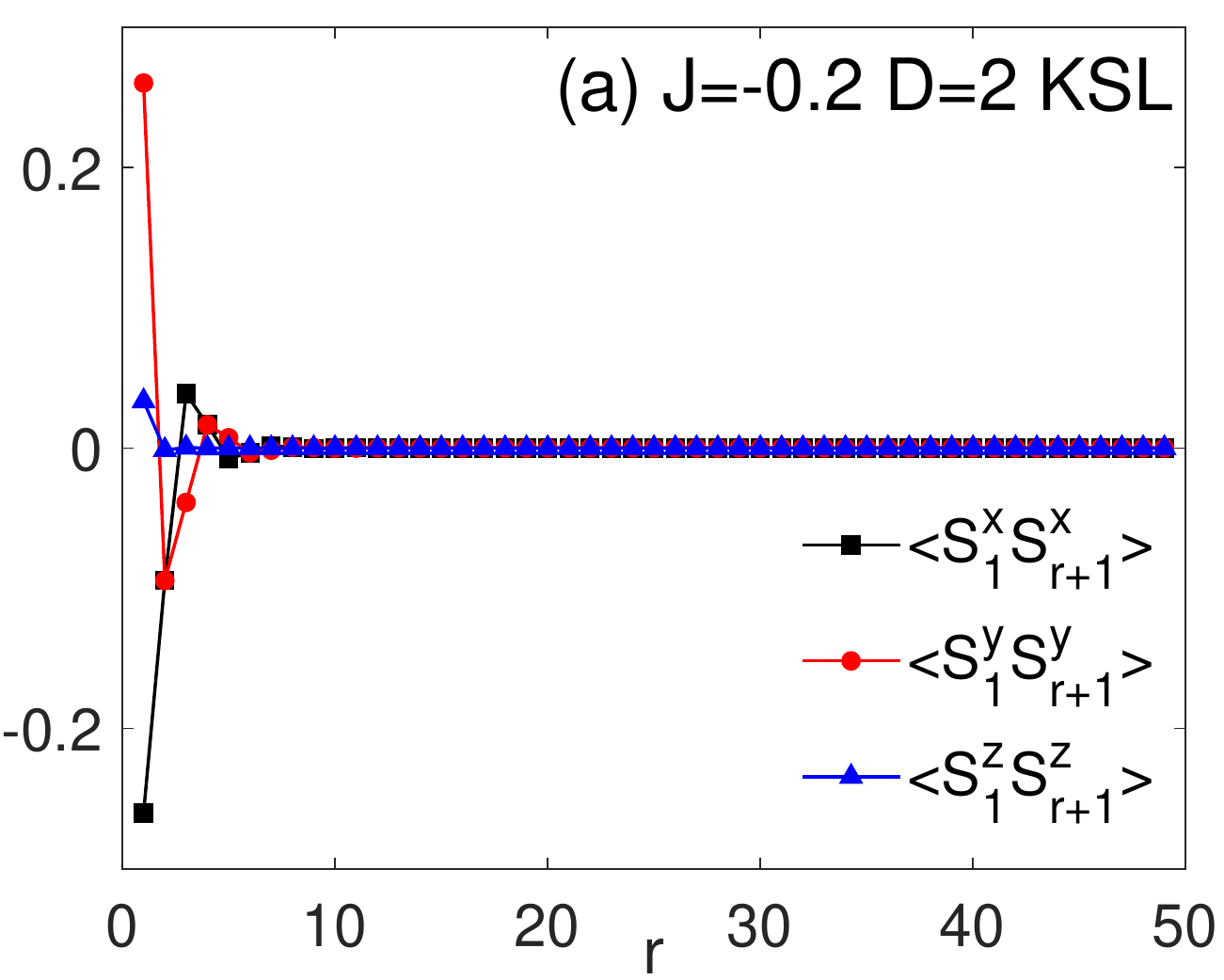}
\includegraphics[width=0.49\columnwidth]{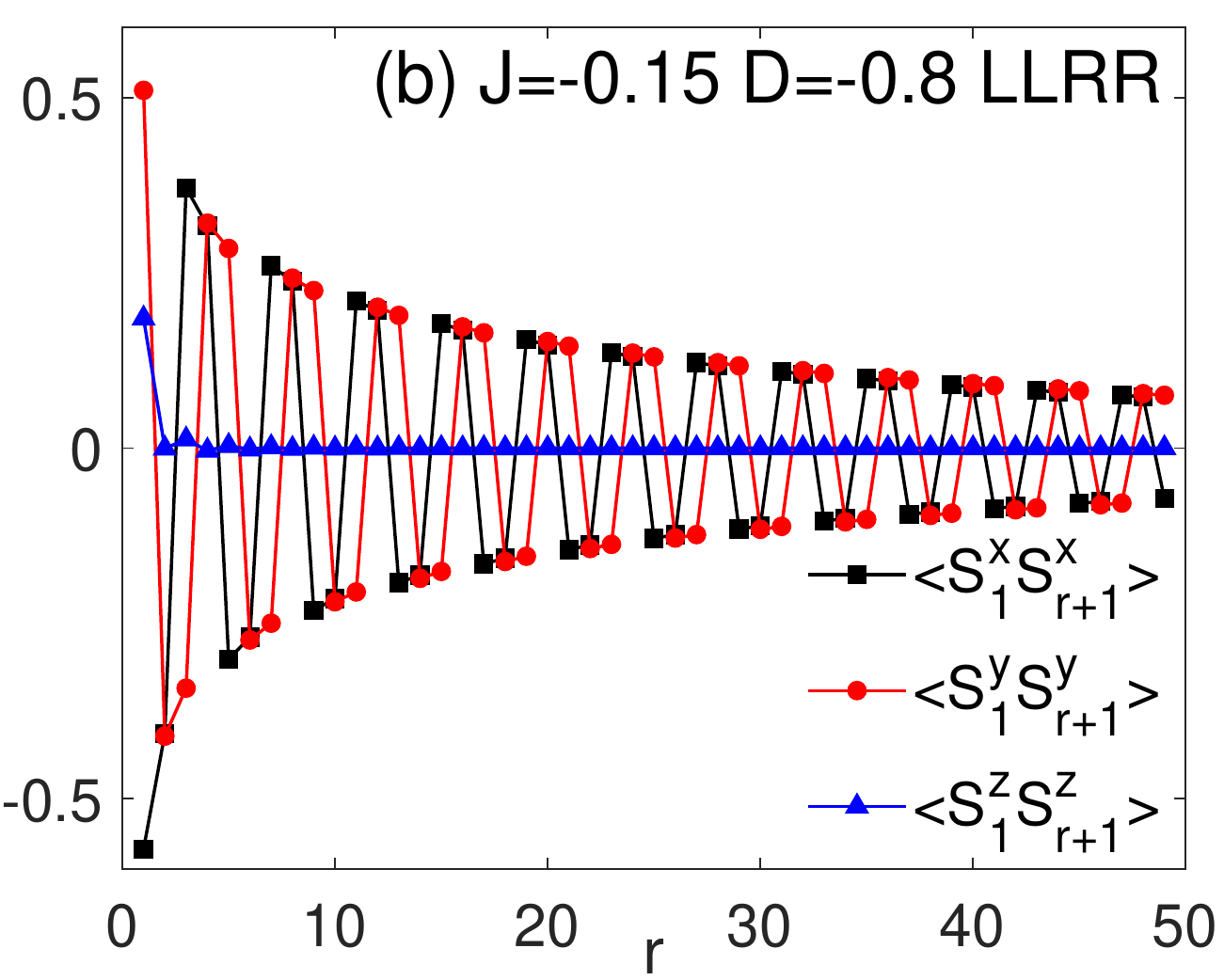}
\vskip .2cm
\includegraphics[width=0.49\columnwidth]{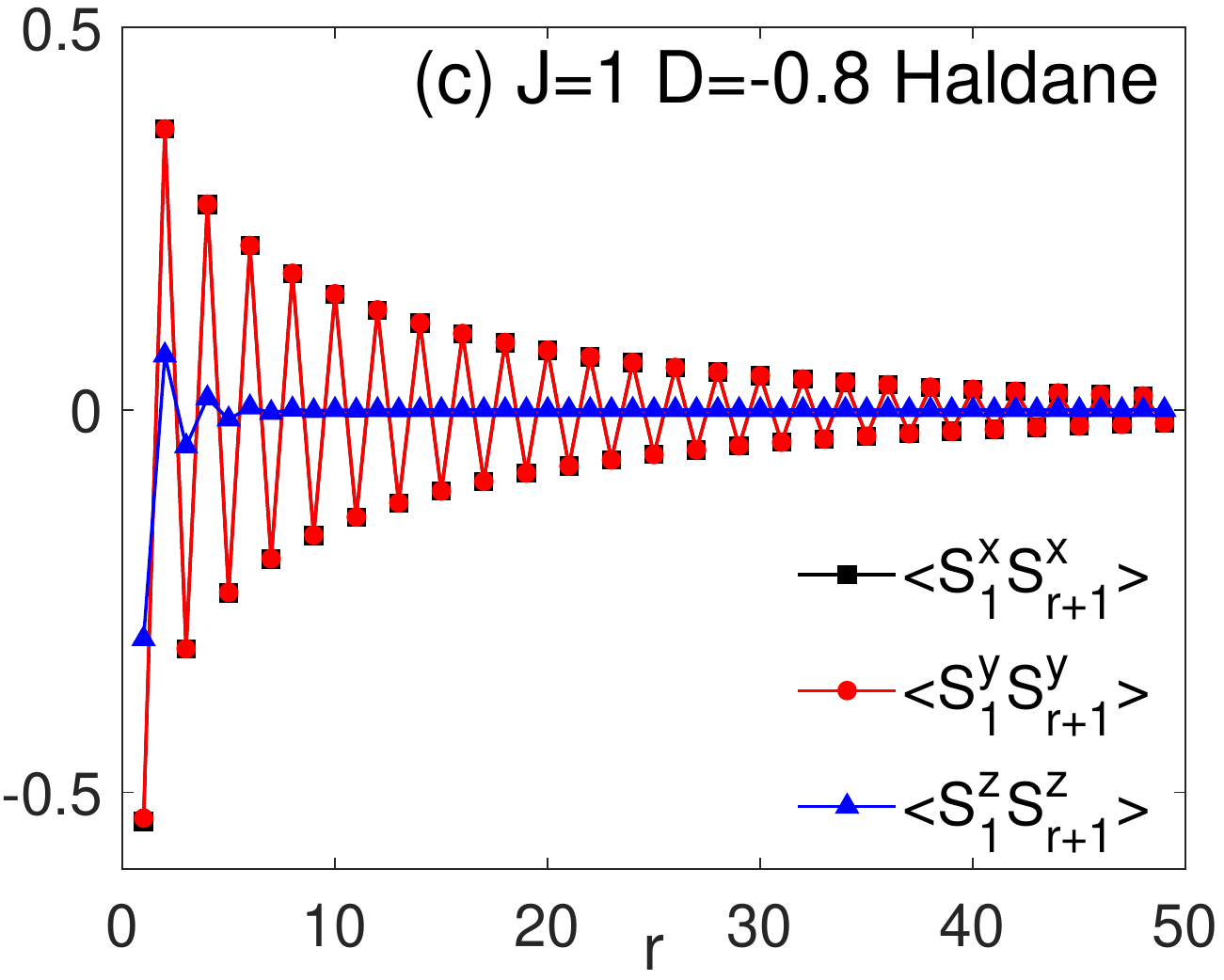}
\includegraphics[width=0.49\columnwidth]{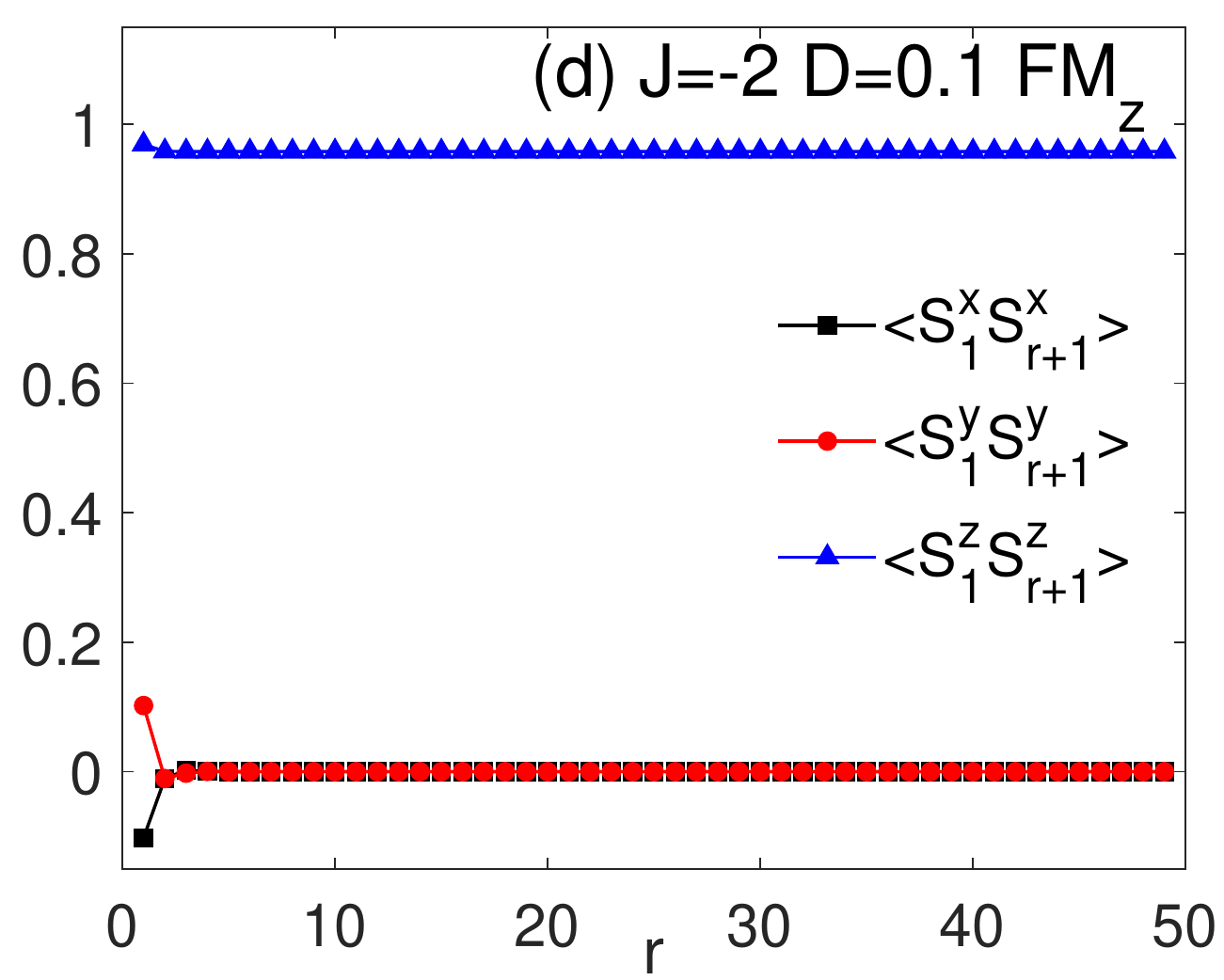}
\vskip .2cm
\includegraphics[width=0.49\columnwidth]{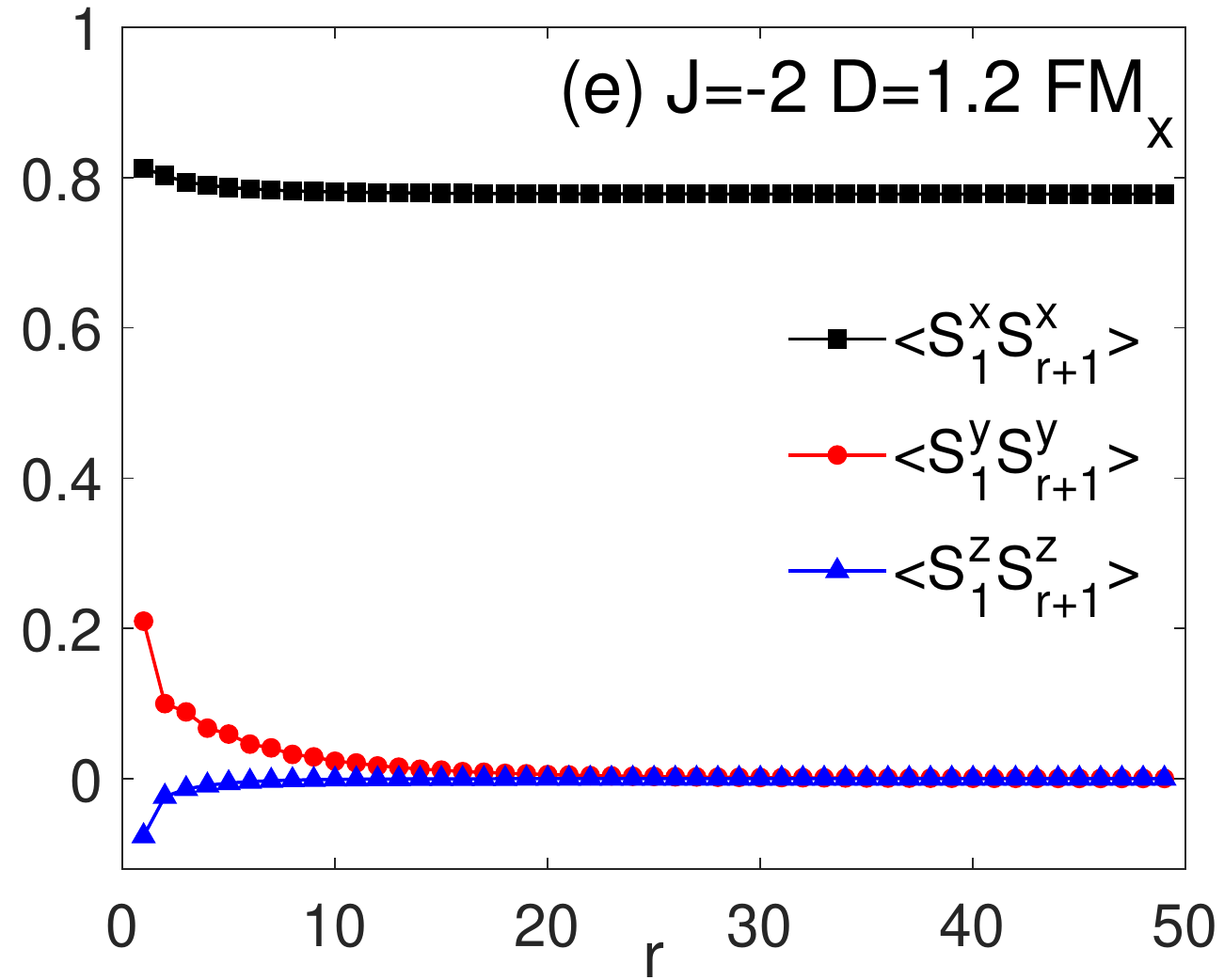}
\includegraphics[width=0.49\columnwidth]{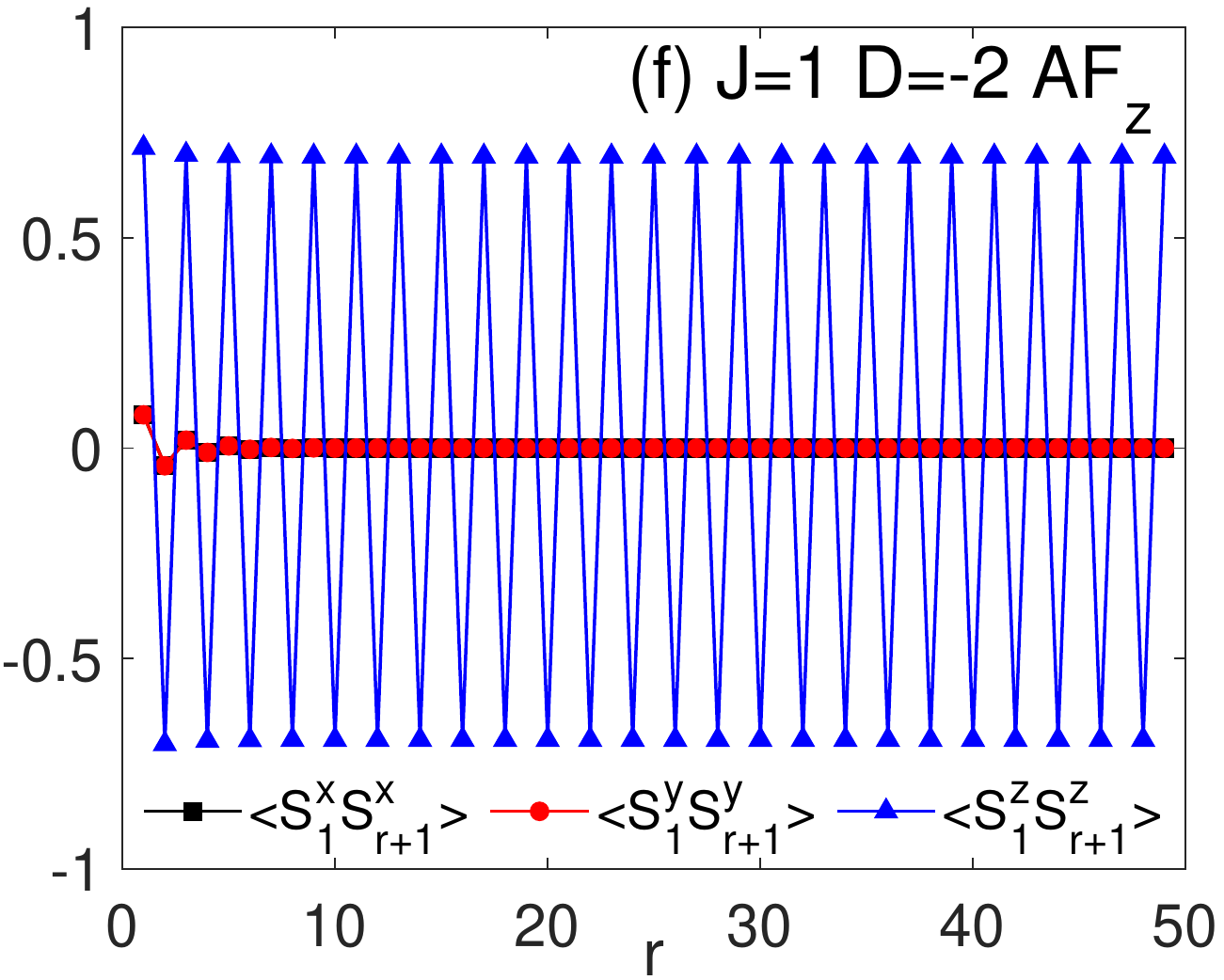}
\vskip .2cm
\caption{The correlation between site $1$ and site $j$ with $\theta=0$ for
increasing distance $r \equiv\vert j-1\vert$ for representative points in:
(a) KSL phase with $J=-0.2$, $D=2$; (b) LLRR phase with $J=-0.15$, $D=-0.8$;
(c) Haldane phase with $J=1$, $D=-0.8$;
(d) FM$_z$ phase $J=-2$, $D=0.1$;
(e) FM$_x$ phase $J=-2$, $D=1.2$; and
(f) AF$_z$ phase $J=1$, $D=-2$.
Here we use the iTEBD method and the bond dimension set as $\chi=120$.}
\label{fig:Twositecorrelation}
\end{figure}

The joint symmetry (\ref{jointsymmetry}) is preserved in the whole KSL
phase for the infinite system.
It has been
reported
that on the line of $D=0$ (the horizontal dashed line in Fig.
\ref{fig:phase diagram}), the ground state of the Kitaev-Heisenberg
model undergoes the FM$_z$ phase, the LLRR phase, the KSL and the
Haldane phase with increasing $J$.  The successive second-order
quantum phase transitions occur at $J_c=-0.6$, $-0.08$, and $0.08$, respectively~\cite{You2020}.
For $J=0$ and $D=0$, the pure Kitaev chain hosts only two nearest neighboring antiferromagnetic orders $\langle S_{2j-1}^x S_{2j}^x \rangle$ and $\langle S_{2j}^y S_{2j+1}^y \rangle$ while other correlations vanish, similar to the spin-1/2 Kitaev honeycomb model
\cite{Winter_2017}. Away from the Kitav limit, the two-spin correlation functions are found to decay exponentially and the short correlation length $\xi$  will extend to a few sites, as shown in Fig. \ref{fig:Twositecorrelation}(a).
Note that the ground-state properties of integer spin chains are in stark contrast to those of half-odd integer spin. In comparison, the ground state of spin-1/2 Kitaev chain is $2^{N/2-1}$-fold degenerate~\cite{W.L.You2008},
and the macroscopic degeneracy makes the ground state vulnerable. As such, an infinitesimal Heisenberg coupling is sufficient to
lift the ground-state degeneracy and
generate magnetic long-range order
~\cite{Trousselet_2010,Trousselet2012}. In contrast, the spin-1 chain
supports a gapped KSL ground state, which can sustain a finite Heisenberg coupling.
It is remarkable that the KSL phase becomes more robust against the Heisenberg interactions for large positive $D$ . One can further observe that the size of the KSL phase enlarges with increasing positive $D$ and becomes narrower for negative $D$. However,
it is found that the three components of the dimer order parameter $O_D$ are all mismatched in a fairly small region of the parameter space
except for $J=0$, as exhibited in Fig. \ref{fig:Dimer}(b).

\begin{figure}[t!]
\begin{center}
\includegraphics[width=1.05\columnwidth]{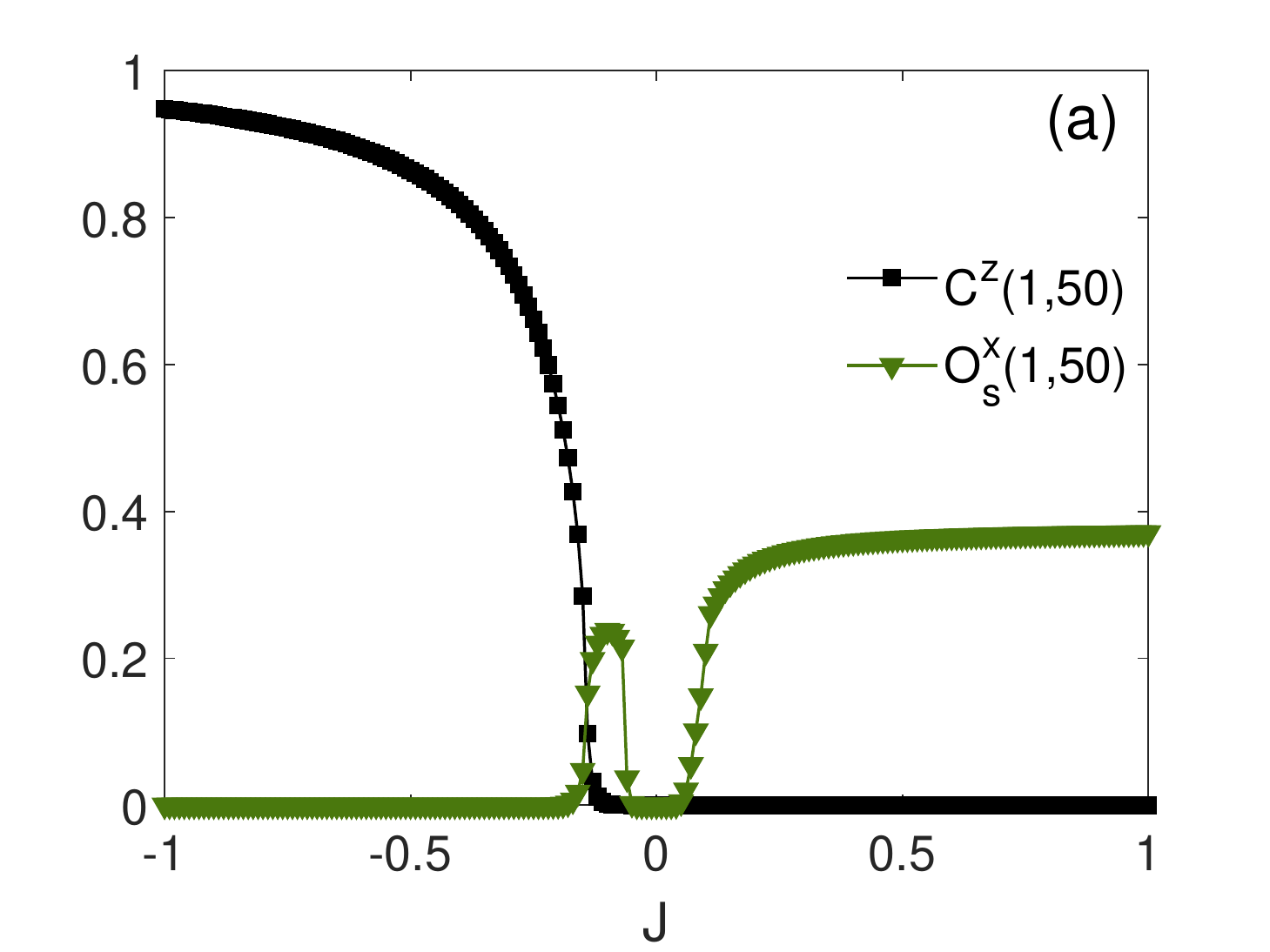}
\includegraphics[width=1.05\columnwidth]{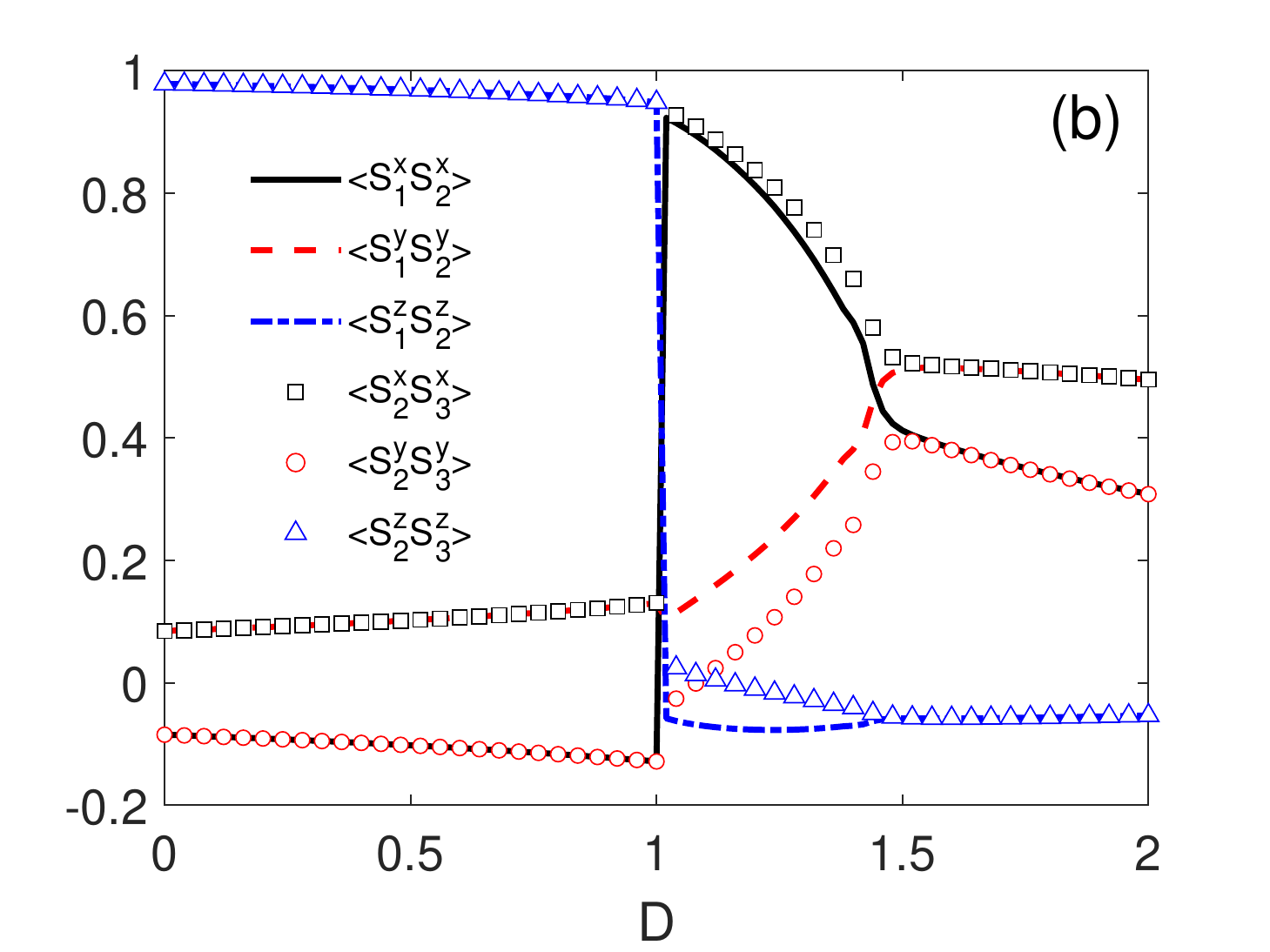}
\end{center}
\caption{(a) Two-point correlation $C^z(i,j)$ (\ref{eq:C(i,j)}) with
$\theta=0$ and the string order parameter $O_S^x(i,j)$ (\ref{equ:O_S})
between sites $1$ and $50$ for $D = -0.8$ and $J$ from $-1$ to $1$.
(b) Two-point spin-spin correlations $C^{\alpha}(i,j)$ (\ref{eq:C(i,j)})
with $\theta=0$ at $J = -2$. Here we use the iTEBD method and the bond
dimension is set as $\chi=120$.
 }
\label{fig:Dimer_S12_S23}
\end{figure}

A hallmark of the Haldane phase is the non-local string order parameter,
which was first introduced by den Nijs and Rommelse \cite{den.Nijs.Marcel1989} and later refined by Tasaki \cite{Tasaki.Hal1991}. Its limiting value
reveals the hidden  $\mathbb{Z}_2 \times \mathbb{Z}_2$ symmetry breaking
\begin{eqnarray}
O_S^{\alpha}(i,j) &\equiv&- {\rm lim}_{|i-j|\rightarrow\infty} C^{aa}(i,j). \label{equ:O_S}
\end{eqnarray}
This order parameter serves as a distinct feature of the Haldane phase.
Figure \ref{fig:Dimer_S12_S23}(a) illustrates two-site correlations
between sites $1$ and $50$. One observes that $C^z(1,50)$ is finite in
the FM$_z$ phase, while the string order parameter $O_S^x(1,50)$ is
nonvanishing in two regions, i.e., $-0.22 \lesssim J\lesssim-0.04$ and
$J\gtrsim 0.03$. To distinguish the two phases, we plot the spin-spin
correlations between site 1 and site $1+r$ for typical parameters in Fig. \ref{fig:Twositecorrelation}. One can observe in Fig.
\ref{fig:Twositecorrelation}(b) that  both $C^x(1,1+r)$ and $C^y(1,1+r)$
alternate between two successive positive and negative values as the
distance of two sites $r$ increases, indicating the onset of the spin
nematic ordering~\cite{Zvyagin2023}, while $C^x(1,1+r)$ and $C^y(1,1+r)$
decay exponentially with respect to $r$, manifesting the existence of
the Haldane phase as demonstrated in Fig. \ref{fig:Twositecorrelation}(c).

Furthermore, Fig. \ref{fig:Twositecorrelation}(d) depicts the
correlations for $J=-2$, $D=0.1$, in which the $z$-component
correlations $C^z(1,j)$ dominates with a value close to $1$, implying
the FM$_z$ ground state, while in Fig. \ref{fig:Twositecorrelation}(e)
the dominant correlations $C^x(1,j)$ characterizes the FM$_x$ phase for
$J=-2$, $D=1.2$. Upon increasing $D$ at $J=-2$, the transition from
FM$_z$ to FM$_x$ takes place at $D_c=1$, see Fig.
\ref{fig:Dimer_S12_S23}(b). Unlike the FM$_z$ phase, the joint symmetry (\ref{jointsymmetry}) is broken in the FM$_x$ phase. By further
increasing the value of $D$, the ground state evolves from the FM$_x$
state into the KSL phase, in which the joint symmetry is restored again.
In contrast, the two-site correlation functions in Fig.
\ref{fig:Twositecorrelation}(f) exhibit a distinct behavior.
Specifically, $C^z(1,j)$ shows a periodic oscillation between values
close to $-1$ and 1, while $C^x(1,j)$ and $C^y(1,j)$ nearly vanish.
These observations provide strong evidence that the system is in the
AF$_z$ phase.

\section{Summary and conclusions}
\label{Summary and conclusions}

To summarize, we have explored the physics arising from the cooperative
effect of uniaxial single-ion anisotropy (SIA) and Heisenberg
interactions in the spin-1 Kitaev chain. We studied quantum many-body
scar (QMBS) states and quantum phase transitions in spin-1 Kitaev chain
with SIA. We find that the local $\mathbb{Z}_2$ gauge fields, a hallmark
of Kitaev model, are still conserved in the spin-1 Kitaev chain with SIA
\mbox{(KD model)}. In this case, the Hilbert space is fragmented into
$2^N$ unequal subspaces characterized by $\vec{w}=\{w_1,w_n,\cdots,w_N\}$.
Among an exponential number of Krylov subspaces, it has been recognized
that in the uniform sector with local $\mathbb{Z}_2$ gauge fields, i.e.,
$\vec{w}=\{1,1,\cdots,1\}$, is the largest Krylov subspace, in which a
local transformation maps the spin-1 KD model onto the detuned PXP model
with spin 1/2 degrees of freedom, and the SIA acts as a static detuning.
The dual transformation suggests a solid-state-based realization of the
PXP model based on Mott insulator with strong spin-orbit and Hund’s
\mbox{couplings}.

Considering the ground states becomes twofold degenerate
$\vert\mathbb{Z}_2\rangle$ state in the limit of $D\to -\infty$ due to
the Hilbert space constraint, a continuous transition in the detuned PXP
model occurs at $D_c=-0.655$. This quantum phase transition corresponds
to the emergence of dimer phase induced by the spontaneous breaking of
translational symmetry in the flux-free sector which can be described by
the dimer order parameter (\ref{dimerorderparameter}). We find the most
prominent coherent oscillations of quantum fidelity in the quantum
quench from initial states $\vert\mathbb{Z}_2\rangle$ and
$\vert\mathbb{Z}_3\rangle$, a characteristic of the embedded
prototypical PXP model for $D=0$. We demonstrate that these fidelity
revivals are robust against small SIA perturbation. The non-thermalizing
dynamics can be also reflected by measuring the expectation values of
certain local observables, which will vanish for $D<D_c$.
Finally, we provide a complete
phase diagram for the spin-1 KD model by describing the interplay between Kitaev interactions, Heisenberg interactions and SIA. In particular, we underline the evolution of the Kitave phase in a broader regime,
therefore showing the relevance for the scar stability and possible solid-state applications.
Seven phases are identified by The numerical methods through the corresponding spin-spin
correlations, including the Kitaev spin liquid, dimer phase, LLRR phase,
Haldane phase, FM$_{x}$ phase, FM$_z$ phase and AF$_z$ phase. Our study
on the higher-spin Kitaev chain will likely help to identify candidate
materials for Kitaev spin liquid.

\begin{acknowledgments}
The authors appreciate very insightful discussions with Hosho Katsura, Gaoyong Sun and Zhi-Xiang Sun. We acknowledges Ming Xue for bringing Ref.~\cite{Yu2017} to our attention. This work is supported by the National Natural Science Foundation of
China (NSFC) under Grant No. 12174194, Postgraduate Research \&
Practice Innovation Program of Jiangsu Province, under Grant No. KYCX23\_0347,
Opening Fund of the Key Laboratory of Aerospace Information Materials
and Physics (Nanjing University of Aeronautics and Astronautics), MIIT,
Top-notch Academic Programs Project of Jiangsu Higher Education
Institutions (TAPP), and stable supports for basic institute research
under Grant No. 190101. \mbox{A.M.O. kindly} acknowledges Narodowe
Centrum Nauki (NCN, Poland) Project No. 2021/43/B/ST3/02166 and is
grateful for support via the Alexander von Humboldt \mbox{Foundation}
\mbox{Fellowship} \mbox{(Humboldt-Forschungspreis).}
\end{acknowledgments}

\appendix

\section{Mapping the Kitaev model with single-ion \\
anisotropy within flux-free sector to the detuned PXP model}
\label{APPENDIX A}

For convenience, we use the rotated Hamiltonian Eq. (\ref{equ:tilde HkD})
and set $K_j=K$. The local two-spin Hamiltonian is given by
\begin{eqnarray}
\tilde{H}_{j,j+1} = K  S_j^x S_{j+1}^y +D [ (S_j^z)^2+(S_{j+1}^z)^2].
\end{eqnarray}

For the 5 states satisfying $w_j=1$,  we have
\begin{eqnarray}
\tilde{H}_{j,j+1} \vert xy \rangle &=&  2 D \vert xy \rangle,\quad
\tilde{H}_{j,j+1} \vert xz \rangle =  D \vert xz \rangle,\nonumber \\
\tilde{H}_{j,j+1} \vert zy \rangle &=&  D \vert zy \rangle,\quad\;\;
\tilde{H}_{j,j+1} \vert zz \rangle = K \vert yx \rangle,\nonumber \\
\tilde{H}_{j,j+1} \vert yx \rangle &=& K \vert zz \rangle +2D \vert yx \rangle.
\end{eqnarray}
Accordingly, the Hamiltonian can be written in the matrix form as
\begin{eqnarray}
\tilde{H}_{j,j+1} = \left(
                 \begin{array}{ccccc}
                   2D & 0 & 0 & 0 &0\\
                   0 & D & 0 & 0 &0\\
                   0 & 0 & D & 0 &0\\
                   0 & 0 & 0 & 2D &K\\
                   0 & 0 & 0 & K &0\\
                 \end{array}
               \right),
\end{eqnarray}
which yields 5 energy eigenvalues $2D$,$D$,$D$,$D\pm\sqrt{D^2 +K^2}$.
Hence, within the lowest-state manifold residing in the \mbox{$w_j=1$}
sector that is spanned by  $\{\vert zz \rangle, \vert yx \rangle\}$,
Eq. (\ref{equ:tilde HkD}) can be written for an effective model of
spin-1/2 degrees of freedom, which can be simplified as
\begin{eqnarray}
\label{equ:H_KD_eff}
\hat{H}_{\rm KD,eff} = X_i +  2D  n_i.
\end{eqnarray}
It is noted that the Hilbert space constraint is imposed by the
projector onto the low-energy subspace spanned by configurations with
no adjacent excited states, which is written as
\begin{eqnarray}
\mathcal{P}=\prod_j (\mathbbm{1}-n_jn_{j+1}).
\end{eqnarray}

The so-called detuned PXP model (\ref{equ:dPXP}) can be derived via the
Schrieffer-Wolff transformation in the limit of strong interactions
(small $\epsilon$) of the following Hamiltonian
\begin{equation}
\label{eq:H_B1}
\hat{H}=\hat{H}_0 + \epsilon \hat{H}_1.
\end{equation}
The leading part of Eq. (\ref{eq:H_B1})
$\hat{H}_0 = \sum_{j=1}^N n_j n_{j+1}$
vanishes in this subspace, we
must consider the first non-trivial order that is given by
$H_{SW}=\epsilon \mathcal{P}H_1\mathcal{P}$.
If $\hat{H}_1$ describes the transverse field term,
\begin{eqnarray}
    \hat{H}_1= \sum_{j=1}^N X_j,
\end{eqnarray}
where $X_j$ is defined in Eq. (\ref{eq:H_PXP}), we have
\begin{eqnarray}
&&\mathcal{P}H_1\mathcal{P}=\prod_{i=1}^N (1-n_in_{i+1}) \sum_{j=1}^N  X_j \prod_{k=1}^N  (1-n_kn_{k+1})\nonumber\\
    &&=\sum_j\;(1-n_{j-1}n_j)(1-n_jn_{j+1})X_j \nonumber \\
    &&\quad\quad\times (1-n_{j-1}n_j)(1-n_jn_{j+1})\nonumber\\
    &&=\sum_j(X_j-n_{j-1}X_j-X_jn_{j+1}+n_{j-1}X_jn_{j+1})\nonumber\\
    &&=\sum_j(1-n_{j-1})X_j(1-n_{j+1})\nonumber\\
    &&=\sum_jP_{j-1}X_jP_{j+1}.
\end{eqnarray}

Then we consider the detuned term,
\begin{eqnarray}
\hat{H}_1= \sum_{j=1}^N n_j.
\end{eqnarray}
In this case, we have
\begin{eqnarray}
&& \mathcal{P}H_1\mathcal{P}=\prod_i (1-n_in_{i+1}) \sum_j n_j \prod_k (1-n_kn_{k+1})\nonumber\\
    &&=\sum_j(1\!-n_{j-1}n_j)(1\!-n_jn_{j+1})n_j (1\!-n_{j-1}n_j)(1\!-n_jn_{j+1})\nonumber\\
    &&=\sum_j(n_j-n_{j-1}n_j-n_jn_{j+1}+n_{j-1}n_jn_{j+1})\nonumber\\
    &&=\sum_j(1-n_{j-1})n_j(1-n_{j+1})\nonumber\\
    &&=\sum_jP_{j-1}n_jP_{j+1}.
\end{eqnarray}

Note that a similar form can be derived even when $\hat{H}_1$ is a non-Hermitian matrix, given by
\begin{eqnarray}
    \hat{H}_1= \sum_{j=1}^N i Y_j,
\end{eqnarray}
we also have
\begin{eqnarray}
&&\mathcal{P}H_1\mathcal{P}=\prod_i (1-n_in_{i+1}) \sum_j (iY_j) \prod_k (1-n_kn_{k+1})\nonumber\\
    &&=\sum_j(1-n_{j-1}n_j)(1-n_jn_{j+1})(iY_j) \nonumber\\
    &&\quad\quad\times  (1-n_{j-1}n_j)(1-n_jn_{j+1})\nonumber\\
    &&=\sum_j\left[(iY_j)-n_{j-1}(iY_j)-(iY_j)n_{j+1}+n_{j-1}(iY_j)n_{j+1}\right]\nonumber\\
    &&=\sum_j(1-n_{j-1})(iY_j)(1-n_{j+1})\nonumber\\
    &&=\sum_jP_{j-1}(iY_j)P_{j+1}.
\end{eqnarray}
However, if $\hat{H}_1= \sum_{j=1}^N Z_j$, the $PZP$ cannot be derived.
A crucial difference is that $X_j P_j =n_j X_j$, $Y_j P_j =n_j Y_j$,
$n_j^2=n_j$, while $Z_j P_j \neq n_j Z_j$.

\section{Dynamical evolution in a quench \\ from initial product states}
\label{APPENDIX B}

Here we use the similar strategy that is introduced for the dynamics of the generalized Hubbard models~\cite{Yu2017}. Considering a Hamiltonian $\hat{H}$ that can be separated into $\hat{H}_0$ and $\lambda \hat{V}$, where $\lambda$ denotes the perturbation strength, an initial state $\vert \psi(0) \rangle$ evolves into $\vert \psi(t) \rangle = e^{-i\hat{H} t} \vert \psi(0) \rangle$ at any time $t$.  If we can find an antiunitary operator $\hat{U}$ satisfies the following conditions: \hfill\break
(i) $\hat{U}$ anticommutes with $\hat{H}_0$ and commutes with $\hat{V}$,
i.e.,
\begin{eqnarray}
    \{\hat{U},\hat{H}_0\} = 0 ,\quad [\hat{U},\hat{V}] = 0.
\end{eqnarray}
(ii) The initial state $\vert\psi(0)\rangle$ only acquires a global
phase factor under $\hat{U}$, i.e.,
\begin{eqnarray}
    \hat{U}^{-1} \vert \psi(0)\rangle = e^{i\chi} \vert \psi(0) \rangle.
\end{eqnarray}
(iii) We consider a given Hermitian operator $\hat{O}$ that is even
or odd under symmetry operation by $\hat{U}$, i.e.,
\begin{eqnarray}
    \hat{U}^{-1}\hat{O} \hat{U} = \pm \hat{O},
\end{eqnarray}
then we can conclude
\begin{eqnarray}
\langle\hat{O}\rangle_{+\lambda} = \pm\langle\hat{O}\rangle_{-\lambda}.
\end{eqnarray}

Back to the KD model in Eq.(\ref{equ:tilde HkD}), which can be rewritten
as $\tilde{H} = \tilde{H}_{\rm K} + \hat{H}_{\rm D}$.
We then apply $\hat{U}=\exp(i \pi S_j^x)$, which will yield
$S_j^x\rightarrow S_j^x,S_j^y\rightarrow -S_j^y,S_j^z\rightarrow -S_j^z$.
Considering the condition (i), one finds
\begin{eqnarray}
\hat{U}^{-1}e^{-i(\widetilde{H}_{\rm K} + \hat{H}_{\rm D})t}\hat{U} = e^{-i(\widetilde{H}_{\rm K} - \hat{H}_{\rm D})t}.
\end{eqnarray}
To this end, in the quantum quench starting from the initial states
$\vert\widetilde{\mathbb{Z}}_k\rangle$, e.g.,
$\vert\widetilde{\mathbb{Z}}_2\rangle$,
we have
\begin{eqnarray}
\langle \hat{O} \rangle_{+D}\!&=& \langle \widetilde{\mathbb{Z}}_2 \vert e^{i(\widetilde{H}_{\rm K} + \hat{H}_{\rm D})t} \hat{O}
e^{-i(\widetilde{H}_{\rm K} + \hat{H}_{\rm D})t}
\vert \widetilde{\mathbb{Z}}_2 \rangle  \nonumber \\
   \!&=&\langle \widetilde{\mathbb{Z}}_2 \vert U e^{i(\widetilde{H}_{\rm K} - \hat{H}_{\rm D})t} (U^{-1} \hat{O} U) e^{-i(\widetilde{H}_{\rm K} - \hat{H}_{\rm D})t} U^{-1}   \vert \widetilde{\mathbb{Z}}_2 \rangle  \nonumber \\
   \!&=& \langle \hat{O} \rangle_{-D},\\
F(t)_{+D}\!&=&  \langle \widetilde{\mathbb{Z}}_2 \vert  e^{-i(\widetilde{H}_{\rm K} + \hat{H}_{\rm D})t}    \vert \widetilde{\mathbb{Z}}_2 \rangle  \nonumber \\
  \!&=&\langle \widetilde{\mathbb{Z}}_2 \vert U  e^{-i(\widetilde{H}_{\rm K} - \hat{H}_{\rm D})t}  U^{-1}  \vert \widetilde{\mathbb{Z}}_2 \rangle
     = F(t)_{-D},
\end{eqnarray}
where the following simple relations
\begin{eqnarray}
    \hat{U}^{-1}\vert \widetilde{\mathbb{Z}}_2 \rangle \! = \! \vert x(-y) \cdots x(-y) \rangle \! = \! (-1)^{N/2}\vert xy \cdots xy \rangle,\nonumber \\
    \langle \widetilde{\mathbb{Z}}_2 \vert \hat{U} \! = \! \langle x(-y) \cdots x(-y) \vert \! = \! (-1)^{N/2}\langle xy \cdots xy \vert, \nonumber
\end{eqnarray}
are used. Therefore, the values of $\langle \hat{O} \rangle$ and  $F(t)$ are symmetric between positive and negative  $D$ in the quantum quench starting from the $\vert \widetilde{\mathbb{Z}}_2 \rangle$ state. It is straightforward to generalize the theorem to other initial product states $\vert \widetilde{\mathbb{Z}}_k \rangle$ $(k\neq 2)$.


\begin{thebibliography}{138}%
\makeatletter
\providecommand \@ifxundefined [1]{%
 \@ifx{#1\undefined}
}%
\providecommand \@ifnum [1]{%
 \ifnum #1\expandafter \@firstoftwo
 \else \expandafter \@secondoftwo
 \fi
}%
\providecommand \@ifx [1]{%
 \ifx #1\expandafter \@firstoftwo
 \else \expandafter \@secondoftwo
 \fi
}%
\providecommand \natexlab [1]{#1}%
\providecommand \enquote  [1]{``#1''}%
\providecommand \bibnamefont  [1]{#1}%
\providecommand \bibfnamefont [1]{#1}%
\providecommand \citenamefont [1]{#1}%
\providecommand \href@noop [0]{\@secondoftwo}%
\providecommand \href [0]{\begingroup \@sanitize@url \@href}%
\providecommand \@href[1]{\@@startlink{#1}\@@href}%
\providecommand \@@href[1]{\endgroup#1\@@endlink}%
\providecommand \@sanitize@url [0]{\catcode `\\12\catcode `\$12\catcode
  `\&12\catcode `\#12\catcode `\^12\catcode `\_12\catcode `\%12\relax}%
\providecommand \@@startlink[1]{}%
\providecommand \@@endlink[0]{}%
\providecommand \url  [0]{\begingroup\@sanitize@url \@url }%
\providecommand \@url [1]{\endgroup\@href {#1}{\urlprefix }}%
\providecommand \urlprefix  [0]{URL }%
\providecommand \Eprint [0]{\href }%
\providecommand \doibase [0]{https://doi.org/}%
\providecommand \selectlanguage [0]{\@gobble}%
\providecommand \bibinfo  [0]{\@secondoftwo}%
\providecommand \bibfield  [0]{\@secondoftwo}%
\providecommand \translation [1]{[#1]}%
\providecommand \BibitemOpen [0]{}%
\providecommand \bibitemStop [0]{}%
\providecommand \bibitemNoStop [0]{.\EOS\space}%
\providecommand \EOS [0]{\spacefactor3000\relax}%
\providecommand \BibitemShut  [1]{\csname bibitem#1\endcsname}%
\let\auto@bib@innerbib\@empty
\bibitem [{\citenamefont {Polkovnikov}\ \emph {et~al.}(2011)\citenamefont
  {Polkovnikov}, \citenamefont {Sengupta}, \citenamefont {Silva},\ and\
  \citenamefont {Vengalattore}}]{Polkovnikov2011Nonequilibrium}%
  \BibitemOpen
  \bibfield  {author} {\bibinfo {author} {\bibfnamefont {A.}~\bibnamefont
  {Polkovnikov}}, \bibinfo {author} {\bibfnamefont {K.}~\bibnamefont
  {Sengupta}}, \bibinfo {author} {\bibfnamefont {A.}~\bibnamefont {Silva}},\
  and\ \bibinfo {author} {\bibfnamefont {M.}~\bibnamefont {Vengalattore}},\
  }\bibfield  {title} {\bibinfo {title} {Colloquium: Nonequilibrium dynamics of
  closed interacting quantum systems},\ }\href
  {https://doi.org/10.1103/RevModPhys.83.863} {\bibfield  {journal} {\bibinfo
  {journal} {Rev. Mod. Phys.}\ }\textbf {\bibinfo {volume} {83}},\ \bibinfo
  {pages} {863} (\bibinfo {year} {2011})}\BibitemShut {NoStop}%
\bibitem [{\citenamefont {Deutsch}(1991)}]{Deutsch1991}%
  \BibitemOpen
  \bibfield  {author} {\bibinfo {author} {\bibfnamefont {J.~M.}\ \bibnamefont
  {Deutsch}},\ }\bibfield  {title} {\bibinfo {title} {Quantum statistical
  mechanics in a closed system},\ }\href
  {https://doi.org/10.1103/PhysRevA.43.2046} {\bibfield  {journal} {\bibinfo
  {journal} {Phys. Rev. A}\ }\textbf {\bibinfo {volume} {43}},\ \bibinfo
  {pages} {2046} (\bibinfo {year} {1991})}\BibitemShut {NoStop}%
\bibitem [{\citenamefont {Srednicki}(1994)}]{Srednicki1994}%
  \BibitemOpen
  \bibfield  {author} {\bibinfo {author} {\bibfnamefont {M.}~\bibnamefont
  {Srednicki}},\ }\bibfield  {title} {\bibinfo {title} {Chaos and quantum
  thermalization},\ }\href {https://doi.org/10.1103/PhysRevE.50.888} {\bibfield
   {journal} {\bibinfo  {journal} {Phys. Rev. E}\ }\textbf {\bibinfo {volume}
  {50}},\ \bibinfo {pages} {888} (\bibinfo {year} {1994})}\BibitemShut
  {NoStop}%
\bibitem [{\citenamefont {Rigol}\ \emph {et~al.}(2008)\citenamefont {Rigol},
  \citenamefont {Dunjko},\ and\ \citenamefont {Olshanii}}]{Rigol2008}%
  \BibitemOpen
  \bibfield  {author} {\bibinfo {author} {\bibfnamefont {M.}~\bibnamefont
  {Rigol}}, \bibinfo {author} {\bibfnamefont {V.}~\bibnamefont {Dunjko}},\ and\
  \bibinfo {author} {\bibfnamefont {M.}~\bibnamefont {Olshanii}},\ }\bibfield
  {title} {\bibinfo {title} {Thermalization and its mechanism for generic
  isolated quantum systems},\ }\href {https://doi.org/10.1038/nature06838}
  {\bibfield  {journal} {\bibinfo  {journal} {Nature}\ }\textbf {\bibinfo
  {volume} {452}},\ \bibinfo {pages} {854} (\bibinfo {year}
  {2008})}\BibitemShut {NoStop}%
\bibitem [{\citenamefont {Rigol}\ and\ \citenamefont
  {Srednicki}(2012)}]{Rigol2012}%
  \BibitemOpen
  \bibfield  {author} {\bibinfo {author} {\bibfnamefont {M.}~\bibnamefont
  {Rigol}}\ and\ \bibinfo {author} {\bibfnamefont {M.}~\bibnamefont
  {Srednicki}},\ }\bibfield  {title} {\bibinfo {title} {Alternatives to
  eigenstate thermalization},\ }\href
  {https://doi.org/10.1103/PhysRevLett.108.110601} {\bibfield  {journal}
  {\bibinfo  {journal} {Phys. Rev. Lett.}\ }\textbf {\bibinfo {volume} {108}},\
  \bibinfo {pages} {110601} (\bibinfo {year} {2012})}\BibitemShut {NoStop}%
\bibitem [{\citenamefont {Kim}\ \emph {et~al.}(2014)\citenamefont {Kim},
  \citenamefont {Ikeda},\ and\ \citenamefont {Huse}}]{Kim2014}%
  \BibitemOpen
  \bibfield  {author} {\bibinfo {author} {\bibfnamefont {H.}~\bibnamefont
  {Kim}}, \bibinfo {author} {\bibfnamefont {T.~N.}\ \bibnamefont {Ikeda}},\
  and\ \bibinfo {author} {\bibfnamefont {D.~A.}\ \bibnamefont {Huse}},\
  }\bibfield  {title} {\bibinfo {title} {Testing whether all eigenstates obey
  the eigenstate thermalization hypothesis},\ }\href
  {https://doi.org/10.1103/PhysRevE.90.052105} {\bibfield  {journal} {\bibinfo
  {journal} {Phys. Rev. E}\ }\textbf {\bibinfo {volume} {90}},\ \bibinfo
  {pages} {052105} (\bibinfo {year} {2014})}\BibitemShut {NoStop}%
\bibitem [{\citenamefont {Deutsch}(2018)}]{Deutsch_2018}%
  \BibitemOpen
  \bibfield  {author} {\bibinfo {author} {\bibfnamefont {J.~M.}\ \bibnamefont
  {Deutsch}},\ }\bibfield  {title} {\bibinfo {title} {Eigenstate thermalization
  hypothesis},\ }\href {https://doi.org/10.1088/1361-6633/aac9f1} {\bibfield
  {journal} {\bibinfo  {journal} {Reports on Progress in Physics}\ }\textbf
  {\bibinfo {volume} {81}},\ \bibinfo {pages} {082001} (\bibinfo {year}
  {2018})}\BibitemShut {NoStop}%
\bibitem [{\citenamefont {Anderson}(1958)}]{Anderson1958}%
  \BibitemOpen
  \bibfield  {author} {\bibinfo {author} {\bibfnamefont {P.~W.}\ \bibnamefont
  {Anderson}},\ }\bibfield  {title} {\bibinfo {title} {Absence of diffusion in
  certain random lattices},\ }\href {https://doi.org/10.1103/PhysRev.109.1492}
  {\bibfield  {journal} {\bibinfo  {journal} {Phys. Rev.}\ }\textbf {\bibinfo
  {volume} {109}},\ \bibinfo {pages} {1492} (\bibinfo {year}
  {1958})}\BibitemShut {NoStop}%
\bibitem [{\citenamefont {Rigol}\ \emph {et~al.}(2007)\citenamefont {Rigol},
  \citenamefont {Dunjko}, \citenamefont {Yurovsky},\ and\ \citenamefont
  {Olshanii}}]{Rigol2007}%
  \BibitemOpen
  \bibfield  {author} {\bibinfo {author} {\bibfnamefont {M.}~\bibnamefont
  {Rigol}}, \bibinfo {author} {\bibfnamefont {V.}~\bibnamefont {Dunjko}},
  \bibinfo {author} {\bibfnamefont {V.}~\bibnamefont {Yurovsky}},\ and\
  \bibinfo {author} {\bibfnamefont {M.}~\bibnamefont {Olshanii}},\ }\bibfield
  {title} {\bibinfo {title} {{Relaxation in a Completely Integrable Many-Body
  Quantum System: An Ab Initio Study of the Dynamics of the Highly Excited
  States of 1D Lattice Hard-Core Bosons}},\ }\href
  {https://doi.org/10.1103/PhysRevLett.98.050405} {\bibfield  {journal}
  {\bibinfo  {journal} {Phys. Rev. Lett.}\ }\textbf {\bibinfo {volume} {98}},\
  \bibinfo {pages} {050405} (\bibinfo {year} {2007})}\BibitemShut {NoStop}%
\bibitem [{\citenamefont {Biroli}\ \emph {et~al.}(2010)\citenamefont {Biroli},
  \citenamefont {Kollath},\ and\ \citenamefont {L\"auchli}}]{Biroli2010}%
  \BibitemOpen
  \bibfield  {author} {\bibinfo {author} {\bibfnamefont {G.}~\bibnamefont
  {Biroli}}, \bibinfo {author} {\bibfnamefont {C.}~\bibnamefont {Kollath}},\
  and\ \bibinfo {author} {\bibfnamefont {A.~M.}\ \bibnamefont {L\"auchli}},\
  }\bibfield  {title} {\bibinfo {title} {Effect of rare fluctuations on the
  thermalization of isolated quantum systems},\ }\href
  {https://doi.org/10.1103/PhysRevLett.105.250401} {\bibfield  {journal}
  {\bibinfo  {journal} {Phys. Rev. Lett.}\ }\textbf {\bibinfo {volume} {105}},\
  \bibinfo {pages} {250401} (\bibinfo {year} {2010})}\BibitemShut {NoStop}%
\bibitem [{\citenamefont {Palzer}\ \emph {et~al.}(2009)\citenamefont {Palzer},
  \citenamefont {Zipkes}, \citenamefont {Sias},\ and\ \citenamefont
  {K\"ohl}}]{Stefan2009}%
  \BibitemOpen
  \bibfield  {author} {\bibinfo {author} {\bibfnamefont {S.}~\bibnamefont
  {Palzer}}, \bibinfo {author} {\bibfnamefont {C.}~\bibnamefont {Zipkes}},
  \bibinfo {author} {\bibfnamefont {C.}~\bibnamefont {Sias}},\ and\ \bibinfo
  {author} {\bibfnamefont {M.}~\bibnamefont {K\"ohl}},\ }\bibfield  {title}
  {\bibinfo {title} {{Quantum Transport through a Tonks-Girardeau Gas}},\
  }\href {https://doi.org/10.1103/PhysRevLett.103.150601} {\bibfield  {journal}
  {\bibinfo  {journal} {Phys. Rev. Lett.}\ }\textbf {\bibinfo {volume} {103}},\
  \bibinfo {pages} {150601} (\bibinfo {year} {2009})}\BibitemShut {NoStop}%
\bibitem [{\citenamefont {Gamayun}\ \emph {et~al.}(2014)\citenamefont
  {Gamayun}, \citenamefont {Lychkovskiy},\ and\ \citenamefont
  {Cheianov}}]{Gamayun2014}%
  \BibitemOpen
  \bibfield  {author} {\bibinfo {author} {\bibfnamefont {O.}~\bibnamefont
  {Gamayun}}, \bibinfo {author} {\bibfnamefont {O.}~\bibnamefont
  {Lychkovskiy}},\ and\ \bibinfo {author} {\bibfnamefont {V.}~\bibnamefont
  {Cheianov}},\ }\bibfield  {title} {\bibinfo {title} {{Kinetic theory for a
  mobile impurity in a degenerate Tonks-Girardeau gas}},\ }\href
  {https://doi.org/10.1103/PhysRevE.90.032132} {\bibfield  {journal} {\bibinfo
  {journal} {Phys. Rev. E}\ }\textbf {\bibinfo {volume} {90}},\ \bibinfo
  {pages} {032132} (\bibinfo {year} {2014})}\BibitemShut {NoStop}%
\bibitem [{\citenamefont {Barnett}\ and\ \citenamefont
  {Seth}(2015)}]{Lionel2015}%
  \BibitemOpen
  \bibfield  {author} {\bibinfo {author} {\bibfnamefont {L.}~\bibnamefont
  {Barnett}}\ and\ \bibinfo {author} {\bibfnamefont {A.~K.}\ \bibnamefont
  {Seth}},\ }\bibfield  {title} {\bibinfo {title} {Granger causality for
  state-space models},\ }\href {https://doi.org/10.1103/PhysRevE.91.040101}
  {\bibfield  {journal} {\bibinfo  {journal} {Phys. Rev. E}\ }\textbf {\bibinfo
  {volume} {91}},\ \bibinfo {pages} {040101} (\bibinfo {year}
  {2015})}\BibitemShut {NoStop}%
\bibitem [{\citenamefont {Vidmar}\ and\ \citenamefont
  {Rigol}(2016)}]{Vidmar_2016}%
  \BibitemOpen
  \bibfield  {author} {\bibinfo {author} {\bibfnamefont {L.}~\bibnamefont
  {Vidmar}}\ and\ \bibinfo {author} {\bibfnamefont {M.}~\bibnamefont {Rigol}},\
  }\bibfield  {title} {\bibinfo {title} {{Generalized Gibbs ensemble in
  integrable lattice models}},\ }\href
  {https://doi.org/10.1088/1742-5468/2016/06/064007} {\bibfield  {journal}
  {\bibinfo  {journal} {Journal of Statistical Mechanics: Theory and
  Experiment}\ }\textbf {\bibinfo {volume} {2016}},\ \bibinfo {pages} {064007}
  (\bibinfo {year} {2016})}\BibitemShut {NoStop}%
\bibitem [{\citenamefont {Gornyi}\ \emph {et~al.}(2005)\citenamefont {Gornyi},
  \citenamefont {Mirlin},\ and\ \citenamefont {Polyakov}}]{Gornyi2005}%
  \BibitemOpen
  \bibfield  {author} {\bibinfo {author} {\bibfnamefont {I.~V.}\ \bibnamefont
  {Gornyi}}, \bibinfo {author} {\bibfnamefont {A.~D.}\ \bibnamefont {Mirlin}},\
  and\ \bibinfo {author} {\bibfnamefont {D.~G.}\ \bibnamefont {Polyakov}},\
  }\bibfield  {title} {\bibinfo {title} {Interacting electrons in disordered
  wires: Anderson localization and low-{$T$} transport},\ }\href
  {https://doi.org/10.1103/PhysRevLett.95.206603} {\bibfield  {journal}
  {\bibinfo  {journal} {Phys. Rev. Lett.}\ }\textbf {\bibinfo {volume} {95}},\
  \bibinfo {pages} {206603} (\bibinfo {year} {2005})}\BibitemShut {NoStop}%
\bibitem [{\citenamefont {Basko}\ \emph {et~al.}(2006)\citenamefont {Basko},
  \citenamefont {Aleiner},\ and\ \citenamefont {Altshuler}}]{BASKO20061126}%
  \BibitemOpen
  \bibfield  {author} {\bibinfo {author} {\bibfnamefont {D.}~\bibnamefont
  {Basko}}, \bibinfo {author} {\bibfnamefont {I.}~\bibnamefont {Aleiner}},\
  and\ \bibinfo {author} {\bibfnamefont {B.}~\bibnamefont {Altshuler}},\
  }\bibfield  {title} {\bibinfo {title} {Metal–insulator transition in a
  weakly interacting many-electron system with localized single-particle
  states},\ }\href {https://doi.org/https://doi.org/10.1016/j.aop.2005.11.014}
  {\bibfield  {journal} {\bibinfo  {journal} {Annals of Physics}\ }\textbf
  {\bibinfo {volume} {321}},\ \bibinfo {pages} {1126} (\bibinfo {year}
  {2006})}\BibitemShut {NoStop}%
\bibitem [{\citenamefont {Pal}\ and\ \citenamefont {Huse}(2010)}]{Pal2010}%
  \BibitemOpen
  \bibfield  {author} {\bibinfo {author} {\bibfnamefont {A.}~\bibnamefont
  {Pal}}\ and\ \bibinfo {author} {\bibfnamefont {D.~A.}\ \bibnamefont {Huse}},\
  }\bibfield  {title} {\bibinfo {title} {Many-body localization phase
  transition},\ }\href {https://doi.org/10.1103/PhysRevB.82.174411} {\bibfield
  {journal} {\bibinfo  {journal} {Phys. Rev. B}\ }\textbf {\bibinfo {volume}
  {82}},\ \bibinfo {pages} {174411} (\bibinfo {year} {2010})}\BibitemShut
  {NoStop}%
\bibitem [{\citenamefont {Lazarides}\ \emph {et~al.}(2015)\citenamefont
  {Lazarides}, \citenamefont {Das},\ and\ \citenamefont
  {Moessner}}]{Lazarides2015}%
  \BibitemOpen
  \bibfield  {author} {\bibinfo {author} {\bibfnamefont {A.}~\bibnamefont
  {Lazarides}}, \bibinfo {author} {\bibfnamefont {A.}~\bibnamefont {Das}},\
  and\ \bibinfo {author} {\bibfnamefont {R.}~\bibnamefont {Moessner}},\
  }\bibfield  {title} {\bibinfo {title} {Fate of many-body localization under
  periodic driving},\ }\href {https://doi.org/10.1103/PhysRevLett.115.030402}
  {\bibfield  {journal} {\bibinfo  {journal} {Phys. Rev. Lett.}\ }\textbf
  {\bibinfo {volume} {115}},\ \bibinfo {pages} {030402} (\bibinfo {year}
  {2015})}\BibitemShut {NoStop}%
\bibitem [{\citenamefont {Altman}(2018)}]{Altman2018}%
  \BibitemOpen
  \bibfield  {author} {\bibinfo {author} {\bibfnamefont {E.}~\bibnamefont
  {Altman}},\ }\bibfield  {title} {\bibinfo {title} {Many-body localization and
  quantum thermalization},\ }\href {https://doi.org/10.1038/s41567-018-0305-7}
  {\bibfield  {journal} {\bibinfo  {journal} {Nature Physics}\ }\textbf
  {\bibinfo {volume} {14}},\ \bibinfo {pages} {979} (\bibinfo {year}
  {2018})}\BibitemShut {NoStop}%
\bibitem [{\citenamefont {Schreiber}\ \emph {et~al.}(2015)\citenamefont
  {Schreiber}, \citenamefont {Hodgman}, \citenamefont {Bordia}, \citenamefont
  {Lüschen}, \citenamefont {Fischer}, \citenamefont {Vosk}, \citenamefont
  {Altman}, \citenamefont {Schneider},\ and\ \citenamefont
  {Bloch}}]{Schreiber2015}%
  \BibitemOpen
  \bibfield  {author} {\bibinfo {author} {\bibfnamefont {M.}~\bibnamefont
  {Schreiber}}, \bibinfo {author} {\bibfnamefont {S.~S.}\ \bibnamefont
  {Hodgman}}, \bibinfo {author} {\bibfnamefont {P.}~\bibnamefont {Bordia}},
  \bibinfo {author} {\bibfnamefont {H.~P.}\ \bibnamefont {Lüschen}}, \bibinfo
  {author} {\bibfnamefont {M.~H.}\ \bibnamefont {Fischer}}, \bibinfo {author}
  {\bibfnamefont {R.}~\bibnamefont {Vosk}}, \bibinfo {author} {\bibfnamefont
  {E.}~\bibnamefont {Altman}}, \bibinfo {author} {\bibfnamefont
  {U.}~\bibnamefont {Schneider}},\ and\ \bibinfo {author} {\bibfnamefont
  {I.}~\bibnamefont {Bloch}},\ }\bibfield  {title} {\bibinfo {title}
  {Observation of many-body localization of interacting fermions in a
  quasirandom optical lattice},\ }\href
  {https://doi.org/10.1126/science.aaa7432} {\bibfield  {journal} {\bibinfo
  {journal} {Science}\ }\textbf {\bibinfo {volume} {349}},\ \bibinfo {pages}
  {842} (\bibinfo {year} {2015})}\BibitemShut {NoStop}%
\bibitem [{\citenamefont {Pakrouski}\ \emph {et~al.}(2020)\citenamefont
  {Pakrouski}, \citenamefont {Pallegar}, \citenamefont {Popov},\ and\
  \citenamefont {Klebanov}}]{Pakrouski2020}%
  \BibitemOpen
  \bibfield  {author} {\bibinfo {author} {\bibfnamefont {K.}~\bibnamefont
  {Pakrouski}}, \bibinfo {author} {\bibfnamefont {P.~N.}\ \bibnamefont
  {Pallegar}}, \bibinfo {author} {\bibfnamefont {F.~K.}\ \bibnamefont
  {Popov}},\ and\ \bibinfo {author} {\bibfnamefont {I.~R.}\ \bibnamefont
  {Klebanov}},\ }\bibfield  {title} {\bibinfo {title} {{Many-Body Scars as a
  Group Invariant Sector of Hilbert Space}},\ }\href
  {https://doi.org/10.1103/PhysRevLett.125.230602} {\bibfield  {journal}
  {\bibinfo  {journal} {Phys. Rev. Lett.}\ }\textbf {\bibinfo {volume} {125}},\
  \bibinfo {pages} {230602} (\bibinfo {year} {2020})}\BibitemShut {NoStop}%
\bibitem [{\citenamefont {Rigol}(2009)}]{Rigol2009}%
  \BibitemOpen
  \bibfield  {author} {\bibinfo {author} {\bibfnamefont {M.}~\bibnamefont
  {Rigol}},\ }\bibfield  {title} {\bibinfo {title} {Breakdown of thermalization
  in finite one-dimensional systems},\ }\href
  {https://doi.org/10.1103/PhysRevLett.103.100403} {\bibfield  {journal}
  {\bibinfo  {journal} {Phys. Rev. Lett.}\ }\textbf {\bibinfo {volume} {103}},\
  \bibinfo {pages} {100403} (\bibinfo {year} {2009})}\BibitemShut {NoStop}%
\bibitem [{\citenamefont {Abanin}\ \emph {et~al.}(2019)\citenamefont {Abanin},
  \citenamefont {Altman}, \citenamefont {Bloch},\ and\ \citenamefont
  {Serbyn}}]{Abanin2019}%
  \BibitemOpen
  \bibfield  {author} {\bibinfo {author} {\bibfnamefont {D.~A.}\ \bibnamefont
  {Abanin}}, \bibinfo {author} {\bibfnamefont {E.}~\bibnamefont {Altman}},
  \bibinfo {author} {\bibfnamefont {I.}~\bibnamefont {Bloch}},\ and\ \bibinfo
  {author} {\bibfnamefont {M.}~\bibnamefont {Serbyn}},\ }\bibfield  {title}
  {\bibinfo {title} {Colloquium: Many-body localization, thermalization, and
  entanglement},\ }\href {https://doi.org/10.1103/RevModPhys.91.021001}
  {\bibfield  {journal} {\bibinfo  {journal} {Rev. Mod. Phys.}\ }\textbf
  {\bibinfo {volume} {91}},\ \bibinfo {pages} {021001} (\bibinfo {year}
  {2019})}\BibitemShut {NoStop}%
\bibitem [{\citenamefont {Bernien}\ \emph {et~al.}(2017)\citenamefont
  {Bernien}, \citenamefont {Schwartz}, \citenamefont {Keesling}, \citenamefont
  {Levine}, \citenamefont {Omran}, \citenamefont {Pichler}, \citenamefont
  {Choi}, \citenamefont {Zibrov}, \citenamefont {Endres}, \citenamefont
  {Greiner}, \citenamefont {Vuleti{\'{c}}},\ and\ \citenamefont
  {Lukin}}]{Bernien2017}%
  \BibitemOpen
  \bibfield  {author} {\bibinfo {author} {\bibfnamefont {H.}~\bibnamefont
  {Bernien}}, \bibinfo {author} {\bibfnamefont {S.}~\bibnamefont {Schwartz}},
  \bibinfo {author} {\bibfnamefont {A.}~\bibnamefont {Keesling}}, \bibinfo
  {author} {\bibfnamefont {H.}~\bibnamefont {Levine}}, \bibinfo {author}
  {\bibfnamefont {A.}~\bibnamefont {Omran}}, \bibinfo {author} {\bibfnamefont
  {H.}~\bibnamefont {Pichler}}, \bibinfo {author} {\bibfnamefont
  {S.}~\bibnamefont {Choi}}, \bibinfo {author} {\bibfnamefont {A.~S.}\
  \bibnamefont {Zibrov}}, \bibinfo {author} {\bibfnamefont {M.}~\bibnamefont
  {Endres}}, \bibinfo {author} {\bibfnamefont {M.}~\bibnamefont {Greiner}},
  \bibinfo {author} {\bibfnamefont {V.}~\bibnamefont {Vuleti{\'{c}}}},\ and\
  \bibinfo {author} {\bibfnamefont {M.~D.}\ \bibnamefont {Lukin}},\ }\bibfield
  {title} {\bibinfo {title} {Probing many-body dynamics on a 51-atom quantum
  simulator},\ }\href {https://doi.org/10.1038/nature24622} {\bibfield
  {journal} {\bibinfo  {journal} {Nature}\ }\textbf {\bibinfo {volume} {551}},\
  \bibinfo {pages} {579} (\bibinfo {year} {2017})}\BibitemShut {NoStop}%
\bibitem [{\citenamefont {Choi}\ \emph {et~al.}(2019)\citenamefont {Choi},
  \citenamefont {Turner}, \citenamefont {Pichler}, \citenamefont {Ho},
  \citenamefont {Michailidis}, \citenamefont {Papi\ifmmode~\acute{c}\else
  \'{c}\fi{}}, \citenamefont {Serbyn}, \citenamefont {Lukin},\ and\
  \citenamefont {Abanin}}]{Choi2019}%
  \BibitemOpen
  \bibfield  {author} {\bibinfo {author} {\bibfnamefont {S.}~\bibnamefont
  {Choi}}, \bibinfo {author} {\bibfnamefont {C.~J.}\ \bibnamefont {Turner}},
  \bibinfo {author} {\bibfnamefont {H.}~\bibnamefont {Pichler}}, \bibinfo
  {author} {\bibfnamefont {W.~W.}\ \bibnamefont {Ho}}, \bibinfo {author}
  {\bibfnamefont {A.~A.}\ \bibnamefont {Michailidis}}, \bibinfo {author}
  {\bibfnamefont {Z.}~\bibnamefont {Papi\ifmmode~\acute{c}\else \'{c}\fi{}}},
  \bibinfo {author} {\bibfnamefont {M.}~\bibnamefont {Serbyn}}, \bibinfo
  {author} {\bibfnamefont {M.~D.}\ \bibnamefont {Lukin}},\ and\ \bibinfo
  {author} {\bibfnamefont {D.~A.}\ \bibnamefont {Abanin}},\ }\bibfield  {title}
  {\bibinfo {title} {{Emergent SU(2) Dynamics and Perfect Quantum Many-Body
  Scars}},\ }\href {https://doi.org/10.1103/PhysRevLett.122.220603} {\bibfield
  {journal} {\bibinfo  {journal} {Phys. Rev. Lett.}\ }\textbf {\bibinfo
  {volume} {122}},\ \bibinfo {pages} {220603} (\bibinfo {year}
  {2019})}\BibitemShut {NoStop}%
\bibitem [{\citenamefont {Iadecola}\ \emph {et~al.}(2019)\citenamefont
  {Iadecola}, \citenamefont {Schecter},\ and\ \citenamefont
  {Xu}}]{Iadecola2019}%
  \BibitemOpen
  \bibfield  {author} {\bibinfo {author} {\bibfnamefont {T.}~\bibnamefont
  {Iadecola}}, \bibinfo {author} {\bibfnamefont {M.}~\bibnamefont {Schecter}},\
  and\ \bibinfo {author} {\bibfnamefont {S.}~\bibnamefont {Xu}},\ }\bibfield
  {title} {\bibinfo {title} {Quantum many-body scars from magnon
  condensation},\ }\href {https://doi.org/10.1103/PhysRevB.100.184312}
  {\bibfield  {journal} {\bibinfo  {journal} {Phys. Rev. B}\ }\textbf {\bibinfo
  {volume} {100}},\ \bibinfo {pages} {184312} (\bibinfo {year}
  {2019})}\BibitemShut {NoStop}%
\bibitem [{\citenamefont {Lin}\ \emph {et~al.}(2020{\natexlab{a}})\citenamefont
  {Lin}, \citenamefont {Calvera},\ and\ \citenamefont {Hsieh}}]{Lin2020}%
  \BibitemOpen
  \bibfield  {author} {\bibinfo {author} {\bibfnamefont {C.-J.}\ \bibnamefont
  {Lin}}, \bibinfo {author} {\bibfnamefont {V.}~\bibnamefont {Calvera}},\ and\
  \bibinfo {author} {\bibfnamefont {T.~H.}\ \bibnamefont {Hsieh}},\ }\bibfield
  {title} {\bibinfo {title} {{Quantum many-body scar states in two-dimensional
  Rydberg atom arrays}},\ }\href {https://doi.org/10.1103/PhysRevB.101.220304}
  {\bibfield  {journal} {\bibinfo  {journal} {Phys. Rev. B}\ }\textbf {\bibinfo
  {volume} {101}},\ \bibinfo {pages} {220304} (\bibinfo {year}
  {2020}{\natexlab{a}})}\BibitemShut {NoStop}%
\bibitem [{\citenamefont {Iadecola}\ and\ \citenamefont
  {Schecter}(2020)}]{Iadecola2020}%
  \BibitemOpen
  \bibfield  {author} {\bibinfo {author} {\bibfnamefont {T.}~\bibnamefont
  {Iadecola}}\ and\ \bibinfo {author} {\bibfnamefont {M.}~\bibnamefont
  {Schecter}},\ }\bibfield  {title} {\bibinfo {title} {Quantum many-body scar
  states with emergent kinetic constraints and finite-entanglement revivals},\
  }\href {https://doi.org/10.1103/PhysRevB.101.024306} {\bibfield  {journal}
  {\bibinfo  {journal} {Phys. Rev. B}\ }\textbf {\bibinfo {volume} {101}},\
  \bibinfo {pages} {024306} (\bibinfo {year} {2020})}\BibitemShut {NoStop}%
\bibitem [{\citenamefont {Bull}\ \emph {et~al.}(2020)\citenamefont {Bull},
  \citenamefont {Desaules},\ and\ \citenamefont {Papi\ifmmode~\acute{c}\else
  \'{c}\fi{}}}]{Bull2020}%
  \BibitemOpen
  \bibfield  {author} {\bibinfo {author} {\bibfnamefont {K.}~\bibnamefont
  {Bull}}, \bibinfo {author} {\bibfnamefont {J.-Y.}\ \bibnamefont {Desaules}},\
  and\ \bibinfo {author} {\bibfnamefont {Z.}~\bibnamefont
  {Papi\ifmmode~\acute{c}\else \'{c}\fi{}}},\ }\bibfield  {title} {\bibinfo
  {title} {{Quantum scars as embeddings of weakly broken Lie algebra
  representations}},\ }\href {https://doi.org/10.1103/PhysRevB.101.165139}
  {\bibfield  {journal} {\bibinfo  {journal} {Phys. Rev. B}\ }\textbf {\bibinfo
  {volume} {101}},\ \bibinfo {pages} {165139} (\bibinfo {year}
  {2020})}\BibitemShut {NoStop}%
\bibitem [{\citenamefont {Turner}\ \emph {et~al.}(2021)\citenamefont {Turner},
  \citenamefont {Desaules}, \citenamefont {Bull},\ and\ \citenamefont
  {Papi\ifmmode~\acute{c}\else \'{c}\fi{}}}]{Turner2021}%
  \BibitemOpen
  \bibfield  {author} {\bibinfo {author} {\bibfnamefont {C.~J.}\ \bibnamefont
  {Turner}}, \bibinfo {author} {\bibfnamefont {J.-Y.}\ \bibnamefont
  {Desaules}}, \bibinfo {author} {\bibfnamefont {K.}~\bibnamefont {Bull}},\
  and\ \bibinfo {author} {\bibfnamefont {Z.}~\bibnamefont
  {Papi\ifmmode~\acute{c}\else \'{c}\fi{}}},\ }\bibfield  {title} {\bibinfo
  {title} {{Correspondence Principle for Many-Body Scars in Ultracold Rydberg
  Atoms}},\ }\href {https://doi.org/10.1103/PhysRevX.11.021021} {\bibfield
  {journal} {\bibinfo  {journal} {Phys. Rev. X}\ }\textbf {\bibinfo {volume}
  {11}},\ \bibinfo {pages} {021021} (\bibinfo {year} {2021})}\BibitemShut
  {NoStop}%
\bibitem [{\citenamefont {Ljubotina}\ \emph {et~al.}(2022)\citenamefont
  {Ljubotina}, \citenamefont {Roos}, \citenamefont {Abanin},\ and\
  \citenamefont {Serbyn}}]{Ljubotina2022}%
  \BibitemOpen
  \bibfield  {author} {\bibinfo {author} {\bibfnamefont {M.}~\bibnamefont
  {Ljubotina}}, \bibinfo {author} {\bibfnamefont {B.}~\bibnamefont {Roos}},
  \bibinfo {author} {\bibfnamefont {D.~A.}\ \bibnamefont {Abanin}},\ and\
  \bibinfo {author} {\bibfnamefont {M.}~\bibnamefont {Serbyn}},\ }\bibfield
  {title} {\bibinfo {title} {Optimal steering of matrix product states and
  quantum many-body scars},\ }\href
  {https://doi.org/10.1103/PRXQuantum.3.030343} {\bibfield  {journal} {\bibinfo
   {journal} {PRX Quantum}\ }\textbf {\bibinfo {volume} {3}},\ \bibinfo {pages}
  {030343} (\bibinfo {year} {2022})}\BibitemShut {NoStop}%
\bibitem [{\citenamefont {Windt}\ and\ \citenamefont
  {Pichler}(2022)}]{Windt2022}%
  \BibitemOpen
  \bibfield  {author} {\bibinfo {author} {\bibfnamefont {B.}~\bibnamefont
  {Windt}}\ and\ \bibinfo {author} {\bibfnamefont {H.}~\bibnamefont
  {Pichler}},\ }\bibfield  {title} {\bibinfo {title} {Squeezing quantum
  many-body scars},\ }\href {https://doi.org/10.1103/PhysRevLett.128.090606}
  {\bibfield  {journal} {\bibinfo  {journal} {Phys. Rev. Lett.}\ }\textbf
  {\bibinfo {volume} {128}},\ \bibinfo {pages} {090606} (\bibinfo {year}
  {2022})}\BibitemShut {NoStop}%
\bibitem [{\citenamefont {Ren}\ \emph {et~al.}(2022)\citenamefont {Ren},
  \citenamefont {Liang},\ and\ \citenamefont {Fang}}]{Ren2022}%
  \BibitemOpen
  \bibfield  {author} {\bibinfo {author} {\bibfnamefont {J.}~\bibnamefont
  {Ren}}, \bibinfo {author} {\bibfnamefont {C.}~\bibnamefont {Liang}},\ and\
  \bibinfo {author} {\bibfnamefont {C.}~\bibnamefont {Fang}},\ }\bibfield
  {title} {\bibinfo {title} {Deformed symmetry structures and quantum many-body
  scar subspaces},\ }\href {https://doi.org/10.1103/PhysRevResearch.4.013155}
  {\bibfield  {journal} {\bibinfo  {journal} {Phys. Rev. Res.}\ }\textbf
  {\bibinfo {volume} {4}},\ \bibinfo {pages} {013155} (\bibinfo {year}
  {2022})}\BibitemShut {NoStop}%
\bibitem [{\citenamefont {Dooley}\ \emph {et~al.}(2023)\citenamefont {Dooley},
  \citenamefont {Pappalardi},\ and\ \citenamefont {Goold}}]{Dooley2023}%
  \BibitemOpen
  \bibfield  {author} {\bibinfo {author} {\bibfnamefont {S.}~\bibnamefont
  {Dooley}}, \bibinfo {author} {\bibfnamefont {S.}~\bibnamefont {Pappalardi}},\
  and\ \bibinfo {author} {\bibfnamefont {J.}~\bibnamefont {Goold}},\ }\bibfield
   {title} {\bibinfo {title} {Entanglement enhanced metrology with quantum
  many-body scars},\ }\href {https://doi.org/10.1103/PhysRevB.107.035123}
  {\bibfield  {journal} {\bibinfo  {journal} {Phys. Rev. B}\ }\textbf {\bibinfo
  {volume} {107}},\ \bibinfo {pages} {035123} (\bibinfo {year}
  {2023})}\BibitemShut {NoStop}%
\bibitem [{\citenamefont {Zhang}\ \emph {et~al.}(2023)\citenamefont {Zhang},
  \citenamefont {Dong}, \citenamefont {Gao}, \citenamefont {Zhao},
  \citenamefont {Hao}, \citenamefont {Desaules}, \citenamefont {Guo},
  \citenamefont {Chen}, \citenamefont {Deng}, \citenamefont {Liu},
  \citenamefont {Ren}, \citenamefont {Yao}, \citenamefont {Zhang},
  \citenamefont {Xu}, \citenamefont {Wang}, \citenamefont {Jin}, \citenamefont
  {Zhu}, \citenamefont {Zhang}, \citenamefont {Li}, \citenamefont {Song},
  \citenamefont {Wang}, \citenamefont {Liu}, \citenamefont {Papi{\'{c}}},
  \citenamefont {Ying}, \citenamefont {Wang},\ and\ \citenamefont
  {Lai}}]{Zhang2023}%
  \BibitemOpen
  \bibfield  {author} {\bibinfo {author} {\bibfnamefont {P.}~\bibnamefont
  {Zhang}}, \bibinfo {author} {\bibfnamefont {H.}~\bibnamefont {Dong}},
  \bibinfo {author} {\bibfnamefont {Y.}~\bibnamefont {Gao}}, \bibinfo {author}
  {\bibfnamefont {L.}~\bibnamefont {Zhao}}, \bibinfo {author} {\bibfnamefont
  {J.}~\bibnamefont {Hao}}, \bibinfo {author} {\bibfnamefont {J.-Y.}\
  \bibnamefont {Desaules}}, \bibinfo {author} {\bibfnamefont {Q.}~\bibnamefont
  {Guo}}, \bibinfo {author} {\bibfnamefont {J.}~\bibnamefont {Chen}}, \bibinfo
  {author} {\bibfnamefont {J.}~\bibnamefont {Deng}}, \bibinfo {author}
  {\bibfnamefont {B.}~\bibnamefont {Liu}}, \bibinfo {author} {\bibfnamefont
  {W.}~\bibnamefont {Ren}}, \bibinfo {author} {\bibfnamefont {Y.}~\bibnamefont
  {Yao}}, \bibinfo {author} {\bibfnamefont {X.}~\bibnamefont {Zhang}}, \bibinfo
  {author} {\bibfnamefont {S.}~\bibnamefont {Xu}}, \bibinfo {author}
  {\bibfnamefont {K.}~\bibnamefont {Wang}}, \bibinfo {author} {\bibfnamefont
  {F.}~\bibnamefont {Jin}}, \bibinfo {author} {\bibfnamefont {X.}~\bibnamefont
  {Zhu}}, \bibinfo {author} {\bibfnamefont {B.}~\bibnamefont {Zhang}}, \bibinfo
  {author} {\bibfnamefont {H.}~\bibnamefont {Li}}, \bibinfo {author}
  {\bibfnamefont {C.}~\bibnamefont {Song}}, \bibinfo {author} {\bibfnamefont
  {Z.}~\bibnamefont {Wang}}, \bibinfo {author} {\bibfnamefont {F.}~\bibnamefont
  {Liu}}, \bibinfo {author} {\bibfnamefont {Z.}~\bibnamefont {Papi{\'{c}}}},
  \bibinfo {author} {\bibfnamefont {L.}~\bibnamefont {Ying}}, \bibinfo {author}
  {\bibfnamefont {H.}~\bibnamefont {Wang}},\ and\ \bibinfo {author}
  {\bibfnamefont {Y.-C.}\ \bibnamefont {Lai}},\ }\bibfield  {title} {\bibinfo
  {title} {{Many-body Hilbert space scarring on a superconducting processor}},\
  }\href {https://doi.org/10.1038/s41567-022-01784-9} {\bibfield  {journal}
  {\bibinfo  {journal} {Nature Physics}\ }\textbf {\bibinfo {volume} {19}},\
  \bibinfo {pages} {120} (\bibinfo {year} {2023})}\BibitemShut {NoStop}%
\bibitem [{\citenamefont {Geraedts}\ \emph {et~al.}(2016)\citenamefont
  {Geraedts}, \citenamefont {Nandkishore},\ and\ \citenamefont
  {Regnault}}]{Geraedts2016}%
  \BibitemOpen
  \bibfield  {author} {\bibinfo {author} {\bibfnamefont {S.~D.}\ \bibnamefont
  {Geraedts}}, \bibinfo {author} {\bibfnamefont {R.}~\bibnamefont
  {Nandkishore}},\ and\ \bibinfo {author} {\bibfnamefont {N.}~\bibnamefont
  {Regnault}},\ }\bibfield  {title} {\bibinfo {title} {Many-body localization
  and thermalization: Insights from the entanglement spectrum},\ }\href
  {https://doi.org/10.1103/PhysRevB.93.174202} {\bibfield  {journal} {\bibinfo
  {journal} {Phys. Rev. B}\ }\textbf {\bibinfo {volume} {93}},\ \bibinfo
  {pages} {174202} (\bibinfo {year} {2016})}\BibitemShut {NoStop}%
\bibitem [{\citenamefont {Ho}\ \emph {et~al.}(2019)\citenamefont {Ho},
  \citenamefont {Choi}, \citenamefont {Pichler},\ and\ \citenamefont
  {Lukin}}]{M.D.Lukin2019}%
  \BibitemOpen
  \bibfield  {author} {\bibinfo {author} {\bibfnamefont {W.~W.}\ \bibnamefont
  {Ho}}, \bibinfo {author} {\bibfnamefont {S.}~\bibnamefont {Choi}}, \bibinfo
  {author} {\bibfnamefont {H.}~\bibnamefont {Pichler}},\ and\ \bibinfo {author}
  {\bibfnamefont {M.~D.}\ \bibnamefont {Lukin}},\ }\bibfield  {title} {\bibinfo
  {title} {Periodic orbits, entanglement, and quantum many-body scars in
  constrained models: Matrix product state approach},\ }\href
  {https://doi.org/10.1103/PhysRevLett.122.040603} {\bibfield  {journal}
  {\bibinfo  {journal} {Phys. Rev. Lett.}\ }\textbf {\bibinfo {volume} {122}},\
  \bibinfo {pages} {040603} (\bibinfo {year} {2019})}\BibitemShut {NoStop}%
\bibitem [{\citenamefont {Moudgalya}\ \emph {et~al.}(2022)\citenamefont
  {Moudgalya}, \citenamefont {Bernevig},\ and\ \citenamefont
  {Regnault}}]{Moudgalya_2022}%
  \BibitemOpen
  \bibfield  {author} {\bibinfo {author} {\bibfnamefont {S.}~\bibnamefont
  {Moudgalya}}, \bibinfo {author} {\bibfnamefont {B.~A.}\ \bibnamefont
  {Bernevig}},\ and\ \bibinfo {author} {\bibfnamefont {N.}~\bibnamefont
  {Regnault}},\ }\bibfield  {title} {\bibinfo {title} {{Quantum many-body scars
  and Hilbert space fragmentation: a review of exact results}},\ }\href
  {https://doi.org/10.1088/1361-6633/ac73a0} {\bibfield  {journal} {\bibinfo
  {journal} {Reports on Progress in Physics}\ }\textbf {\bibinfo {volume}
  {85}},\ \bibinfo {pages} {086501} (\bibinfo {year} {2022})}\BibitemShut
  {NoStop}%
\bibitem [{\citenamefont {Gong}\ and\ \citenamefont
  {Duan}(2013)}]{ZheXuanGong2013}%
  \BibitemOpen
  \bibfield  {author} {\bibinfo {author} {\bibfnamefont {Z.-X.}\ \bibnamefont
  {Gong}}\ and\ \bibinfo {author} {\bibfnamefont {L.-M.}\ \bibnamefont
  {Duan}},\ }\bibfield  {title} {\bibinfo {title} {Prethermalization and
  dynamic phase transition in an isolated trapped ion spin chain},\ }\href
  {https://doi.org/10.1088/1367-2630/15/11/113051} {\bibfield  {journal}
  {\bibinfo  {journal} {New Journal of Physics}\ }\textbf {\bibinfo {volume}
  {15}},\ \bibinfo {pages} {113051} (\bibinfo {year} {2013})}\BibitemShut
  {NoStop}%
\bibitem [{\citenamefont {Neyenhuis}\ \emph {et~al.}(2017)\citenamefont
  {Neyenhuis}, \citenamefont {Zhang}, \citenamefont {Hess}, \citenamefont
  {Smith}, \citenamefont {Lee}, \citenamefont {Richerme}, \citenamefont {Gong},
  \citenamefont {Gorshkov},\ and\ \citenamefont {Monroe}}]{Neyenhuis2017}%
  \BibitemOpen
  \bibfield  {author} {\bibinfo {author} {\bibfnamefont {B.}~\bibnamefont
  {Neyenhuis}}, \bibinfo {author} {\bibfnamefont {J.}~\bibnamefont {Zhang}},
  \bibinfo {author} {\bibfnamefont {P.~W.}\ \bibnamefont {Hess}}, \bibinfo
  {author} {\bibfnamefont {J.}~\bibnamefont {Smith}}, \bibinfo {author}
  {\bibfnamefont {A.~C.}\ \bibnamefont {Lee}}, \bibinfo {author} {\bibfnamefont
  {P.}~\bibnamefont {Richerme}}, \bibinfo {author} {\bibfnamefont {Z.-X.}\
  \bibnamefont {Gong}}, \bibinfo {author} {\bibfnamefont {A.~V.}\ \bibnamefont
  {Gorshkov}},\ and\ \bibinfo {author} {\bibfnamefont {C.}~\bibnamefont
  {Monroe}},\ }\bibfield  {title} {\bibinfo {title} {Observation of
  prethermalization in long-range interacting spin chains},\ }\href
  {https://doi.org/10.1126/sciadv.1700672} {\bibfield  {journal} {\bibinfo
  {journal} {Science Advances}\ }\textbf {\bibinfo {volume} {3}},\ \bibinfo
  {pages} {e1700672} (\bibinfo {year} {2017})}\BibitemShut {NoStop}%
\bibitem [{\citenamefont {Tang}\ \emph {et~al.}(2018)\citenamefont {Tang},
  \citenamefont {Kao}, \citenamefont {Li}, \citenamefont {Seo}, \citenamefont
  {Mallayya}, \citenamefont {Rigol}, \citenamefont {Gopalakrishnan},\ and\
  \citenamefont {Lev}}]{Tang2018}%
  \BibitemOpen
  \bibfield  {author} {\bibinfo {author} {\bibfnamefont {Y.}~\bibnamefont
  {Tang}}, \bibinfo {author} {\bibfnamefont {W.}~\bibnamefont {Kao}}, \bibinfo
  {author} {\bibfnamefont {K.-Y.}\ \bibnamefont {Li}}, \bibinfo {author}
  {\bibfnamefont {S.}~\bibnamefont {Seo}}, \bibinfo {author} {\bibfnamefont
  {K.}~\bibnamefont {Mallayya}}, \bibinfo {author} {\bibfnamefont
  {M.}~\bibnamefont {Rigol}}, \bibinfo {author} {\bibfnamefont
  {S.}~\bibnamefont {Gopalakrishnan}},\ and\ \bibinfo {author} {\bibfnamefont
  {B.~L.}\ \bibnamefont {Lev}},\ }\bibfield  {title} {\bibinfo {title}
  {{Thermalization near Integrability in a Dipolar Quantum Newton's Cradle}},\
  }\href {https://doi.org/10.1103/PhysRevX.8.021030} {\bibfield  {journal}
  {\bibinfo  {journal} {Phys. Rev. X}\ }\textbf {\bibinfo {volume} {8}},\
  \bibinfo {pages} {021030} (\bibinfo {year} {2018})}\BibitemShut {NoStop}%
\bibitem [{\citenamefont {Kao}\ \emph {et~al.}(2021)\citenamefont {Kao},
  \citenamefont {Li}, \citenamefont {Lin}, \citenamefont {Gopalakrishnan},\
  and\ \citenamefont {Lev}}]{W.Kao2021}%
  \BibitemOpen
  \bibfield  {author} {\bibinfo {author} {\bibfnamefont {W.}~\bibnamefont
  {Kao}}, \bibinfo {author} {\bibfnamefont {K.-Y.}\ \bibnamefont {Li}},
  \bibinfo {author} {\bibfnamefont {K.-Y.}\ \bibnamefont {Lin}}, \bibinfo
  {author} {\bibfnamefont {S.}~\bibnamefont {Gopalakrishnan}},\ and\ \bibinfo
  {author} {\bibfnamefont {B.~L.}\ \bibnamefont {Lev}},\ }\bibfield  {title}
  {\bibinfo {title} {{Topological pumping of a 1D dipolar gas into strongly
  correlated prethermal states}},\ }\href
  {https://doi.org/10.1126/science.abb4928} {\bibfield  {journal} {\bibinfo
  {journal} {Science}\ }\textbf {\bibinfo {volume} {371}},\ \bibinfo {pages}
  {296} (\bibinfo {year} {2021})}\BibitemShut {NoStop}%
\bibitem [{\citenamefont {Kinoshita}\ \emph {et~al.}(2006)\citenamefont
  {Kinoshita}, \citenamefont {Wenger},\ and\ \citenamefont
  {Weiss}}]{Kinoshita2006}%
  \BibitemOpen
  \bibfield  {author} {\bibinfo {author} {\bibfnamefont {T.}~\bibnamefont
  {Kinoshita}}, \bibinfo {author} {\bibfnamefont {T.}~\bibnamefont {Wenger}},\
  and\ \bibinfo {author} {\bibfnamefont {D.~S.}\ \bibnamefont {Weiss}},\
  }\bibfield  {title} {\bibinfo {title} {{A quantum Newton's cradle}},\ }\href
  {https://doi.org/10.1038/nature04693} {\bibfield  {journal} {\bibinfo
  {journal} {Nature}\ }\textbf {\bibinfo {volume} {440}},\ \bibinfo {pages}
  {900} (\bibinfo {year} {2006})}\BibitemShut {NoStop}%
\bibitem [{\citenamefont {Xu}\ \emph {et~al.}(2018{\natexlab{a}})\citenamefont
  {Xu}, \citenamefont {Chen}, \citenamefont {Zeng}, \citenamefont {Zhang},
  \citenamefont {Song}, \citenamefont {Liu}, \citenamefont {Guo}, \citenamefont
  {Zhang}, \citenamefont {Xu}, \citenamefont {Deng}, \citenamefont {Huang},
  \citenamefont {Wang}, \citenamefont {Zhu}, \citenamefont {Zheng},\ and\
  \citenamefont {Fan}}]{XuKai2018}%
  \BibitemOpen
  \bibfield  {author} {\bibinfo {author} {\bibfnamefont {K.}~\bibnamefont
  {Xu}}, \bibinfo {author} {\bibfnamefont {J.-J.}\ \bibnamefont {Chen}},
  \bibinfo {author} {\bibfnamefont {Y.}~\bibnamefont {Zeng}}, \bibinfo {author}
  {\bibfnamefont {Y.-R.}\ \bibnamefont {Zhang}}, \bibinfo {author}
  {\bibfnamefont {C.}~\bibnamefont {Song}}, \bibinfo {author} {\bibfnamefont
  {W.}~\bibnamefont {Liu}}, \bibinfo {author} {\bibfnamefont {Q.}~\bibnamefont
  {Guo}}, \bibinfo {author} {\bibfnamefont {P.}~\bibnamefont {Zhang}}, \bibinfo
  {author} {\bibfnamefont {D.}~\bibnamefont {Xu}}, \bibinfo {author}
  {\bibfnamefont {H.}~\bibnamefont {Deng}}, \bibinfo {author} {\bibfnamefont
  {K.}~\bibnamefont {Huang}}, \bibinfo {author} {\bibfnamefont
  {H.}~\bibnamefont {Wang}}, \bibinfo {author} {\bibfnamefont {X.}~\bibnamefont
  {Zhu}}, \bibinfo {author} {\bibfnamefont {D.}~\bibnamefont {Zheng}},\ and\
  \bibinfo {author} {\bibfnamefont {H.}~\bibnamefont {Fan}},\ }\bibfield
  {title} {\bibinfo {title} {Emulating many-body localization with a
  superconducting quantum processor},\ }\href
  {https://doi.org/10.1103/PhysRevLett.120.050507} {\bibfield  {journal}
  {\bibinfo  {journal} {Phys. Rev. Lett.}\ }\textbf {\bibinfo {volume} {120}},\
  \bibinfo {pages} {050507} (\bibinfo {year} {2018}{\natexlab{a}})}\BibitemShut
  {NoStop}%
\bibitem [{\citenamefont {Guo}\ \emph {et~al.}(2021)\citenamefont {Guo},
  \citenamefont {Cheng}, \citenamefont {Sun}, \citenamefont {Song},
  \citenamefont {Li}, \citenamefont {Wang}, \citenamefont {Ren}, \citenamefont
  {Dong}, \citenamefont {Zheng}, \citenamefont {Zhang}, \citenamefont
  {Mondaini}, \citenamefont {Fan},\ and\ \citenamefont {Wang}}]{Guo2021}%
  \BibitemOpen
  \bibfield  {author} {\bibinfo {author} {\bibfnamefont {Q.}~\bibnamefont
  {Guo}}, \bibinfo {author} {\bibfnamefont {C.}~\bibnamefont {Cheng}}, \bibinfo
  {author} {\bibfnamefont {Z.-H.}\ \bibnamefont {Sun}}, \bibinfo {author}
  {\bibfnamefont {Z.}~\bibnamefont {Song}}, \bibinfo {author} {\bibfnamefont
  {H.}~\bibnamefont {Li}}, \bibinfo {author} {\bibfnamefont {Z.}~\bibnamefont
  {Wang}}, \bibinfo {author} {\bibfnamefont {W.}~\bibnamefont {Ren}}, \bibinfo
  {author} {\bibfnamefont {H.}~\bibnamefont {Dong}}, \bibinfo {author}
  {\bibfnamefont {D.}~\bibnamefont {Zheng}}, \bibinfo {author} {\bibfnamefont
  {Y.-R.}\ \bibnamefont {Zhang}}, \bibinfo {author} {\bibfnamefont
  {R.}~\bibnamefont {Mondaini}}, \bibinfo {author} {\bibfnamefont
  {H.}~\bibnamefont {Fan}},\ and\ \bibinfo {author} {\bibfnamefont
  {H.}~\bibnamefont {Wang}},\ }\bibfield  {title} {\bibinfo {title}
  {Observation of energy-resolved many-body localization},\ }\href
  {https://doi.org/10.1038/s41567-020-1035-1} {\bibfield  {journal} {\bibinfo
  {journal} {Nature Physics}\ }\textbf {\bibinfo {volume} {17}},\ \bibinfo
  {pages} {234} (\bibinfo {year} {2021})}\BibitemShut {NoStop}%
\bibitem [{\citenamefont {Ren}\ \emph {et~al.}(2021)\citenamefont {Ren},
  \citenamefont {Liang},\ and\ \citenamefont {Fang}}]{RenJie2021}%
  \BibitemOpen
  \bibfield  {author} {\bibinfo {author} {\bibfnamefont {J.}~\bibnamefont
  {Ren}}, \bibinfo {author} {\bibfnamefont {C.}~\bibnamefont {Liang}},\ and\
  \bibinfo {author} {\bibfnamefont {C.}~\bibnamefont {Fang}},\ }\bibfield
  {title} {\bibinfo {title} {Quasisymmetry groups and many-body scar
  dynamics},\ }\href {https://doi.org/10.1103/PhysRevLett.126.120604}
  {\bibfield  {journal} {\bibinfo  {journal} {Phys. Rev. Lett.}\ }\textbf
  {\bibinfo {volume} {126}},\ \bibinfo {pages} {120604} (\bibinfo {year}
  {2021})}\BibitemShut {NoStop}%
\bibitem [{\citenamefont {Affleck}\ \emph {et~al.}(1987)\citenamefont
  {Affleck}, \citenamefont {Kennedy}, \citenamefont {Lieb},\ and\ \citenamefont
  {Tasaki}}]{Affleck1987}%
  \BibitemOpen
  \bibfield  {author} {\bibinfo {author} {\bibfnamefont {I.}~\bibnamefont
  {Affleck}}, \bibinfo {author} {\bibfnamefont {T.}~\bibnamefont {Kennedy}},
  \bibinfo {author} {\bibfnamefont {E.~H.}\ \bibnamefont {Lieb}},\ and\
  \bibinfo {author} {\bibfnamefont {H.}~\bibnamefont {Tasaki}},\ }\bibfield
  {title} {\bibinfo {title} {Rigorous results on valence-bond ground states in
  antiferromagnets},\ }\href {https://doi.org/10.1103/PhysRevLett.59.799}
  {\bibfield  {journal} {\bibinfo  {journal} {Phys. Rev. Lett.}\ }\textbf
  {\bibinfo {volume} {59}},\ \bibinfo {pages} {799} (\bibinfo {year}
  {1987})}\BibitemShut {NoStop}%
\bibitem [{\citenamefont {Moudgalya}\ \emph {et~al.}(2018)\citenamefont
  {Moudgalya}, \citenamefont {Regnault},\ and\ \citenamefont
  {Bernevig}}]{Moudgalya2018}%
  \BibitemOpen
  \bibfield  {author} {\bibinfo {author} {\bibfnamefont {S.}~\bibnamefont
  {Moudgalya}}, \bibinfo {author} {\bibfnamefont {N.}~\bibnamefont
  {Regnault}},\ and\ \bibinfo {author} {\bibfnamefont {B.~A.}\ \bibnamefont
  {Bernevig}},\ }\bibfield  {title} {\bibinfo {title} {Entanglement of exact
  excited states of {A}ffleck-{K}ennedy-{L}ieb-{T}asaki models: Exact results,
  many-body scars, and violation of the strong eigenstate thermalization
  hypothesis},\ }\href {https://doi.org/10.1103/PhysRevB.98.235156} {\bibfield
  {journal} {\bibinfo  {journal} {Phys. Rev. B}\ }\textbf {\bibinfo {volume}
  {98}},\ \bibinfo {pages} {235156} (\bibinfo {year} {2018})}\BibitemShut
  {NoStop}%
\bibitem [{\citenamefont {Schecter}\ and\ \citenamefont
  {Iadecola}(2019)}]{Schecter2019}%
  \BibitemOpen
  \bibfield  {author} {\bibinfo {author} {\bibfnamefont {M.}~\bibnamefont
  {Schecter}}\ and\ \bibinfo {author} {\bibfnamefont {T.}~\bibnamefont
  {Iadecola}},\ }\bibfield  {title} {\bibinfo {title} {Weak ergodicity breaking
  and quantum many-body scars in spin-1 ${XY}$ magnets},\ }\href
  {https://doi.org/10.1103/PhysRevLett.123.147201} {\bibfield  {journal}
  {\bibinfo  {journal} {Phys. Rev. Lett.}\ }\textbf {\bibinfo {volume} {123}},\
  \bibinfo {pages} {147201} (\bibinfo {year} {2019})}\BibitemShut {NoStop}%
\bibitem [{\citenamefont {Desaules}\ \emph {et~al.}(2021)\citenamefont
  {Desaules}, \citenamefont {Hudomal}, \citenamefont {Turner},\ and\
  \citenamefont {Papi\ifmmode~\acute{c}\else \'{c}\fi{}}}]{Desaules2021}%
  \BibitemOpen
  \bibfield  {author} {\bibinfo {author} {\bibfnamefont {J.-Y.}\ \bibnamefont
  {Desaules}}, \bibinfo {author} {\bibfnamefont {A.}~\bibnamefont {Hudomal}},
  \bibinfo {author} {\bibfnamefont {C.~J.}\ \bibnamefont {Turner}},\ and\
  \bibinfo {author} {\bibfnamefont {Z.}~\bibnamefont
  {Papi\ifmmode~\acute{c}\else \'{c}\fi{}}},\ }\bibfield  {title} {\bibinfo
  {title} {{Proposal for Realizing Quantum Scars in the Tilted 1D Fermi-Hubbard
  Model}},\ }\href {https://doi.org/10.1103/PhysRevLett.126.210601} {\bibfield
  {journal} {\bibinfo  {journal} {Phys. Rev. Lett.}\ }\textbf {\bibinfo
  {volume} {126}},\ \bibinfo {pages} {210601} (\bibinfo {year}
  {2021})}\BibitemShut {NoStop}%
\bibitem [{\citenamefont {Jaksch}\ \emph {et~al.}(2000)\citenamefont {Jaksch},
  \citenamefont {Cirac}, \citenamefont {Zoller}, \citenamefont {Rolston},
  \citenamefont {C\^ot\'e},\ and\ \citenamefont {Lukin}}]{Jaksch2000}%
  \BibitemOpen
  \bibfield  {author} {\bibinfo {author} {\bibfnamefont {D.}~\bibnamefont
  {Jaksch}}, \bibinfo {author} {\bibfnamefont {J.~I.}\ \bibnamefont {Cirac}},
  \bibinfo {author} {\bibfnamefont {P.}~\bibnamefont {Zoller}}, \bibinfo
  {author} {\bibfnamefont {S.~L.}\ \bibnamefont {Rolston}}, \bibinfo {author}
  {\bibfnamefont {R.}~\bibnamefont {C\^ot\'e}},\ and\ \bibinfo {author}
  {\bibfnamefont {M.~D.}\ \bibnamefont {Lukin}},\ }\bibfield  {title} {\bibinfo
  {title} {{Fast Quantum Gates for Neutral Atoms}},\ }\href
  {https://doi.org/10.1103/PhysRevLett.85.2208} {\bibfield  {journal} {\bibinfo
   {journal} {Phys. Rev. Lett.}\ }\textbf {\bibinfo {volume} {85}},\ \bibinfo
  {pages} {2208} (\bibinfo {year} {2000})}\BibitemShut {NoStop}%
\bibitem [{\citenamefont {Turner}\ \emph
  {et~al.}(2018{\natexlab{a}})\citenamefont {Turner}, \citenamefont
  {Michailidis}, \citenamefont {Abanin}, \citenamefont {Serbyn},\ and\
  \citenamefont {Papi{\'{c}}}}]{Turner2018}%
  \BibitemOpen
  \bibfield  {author} {\bibinfo {author} {\bibfnamefont {C.~J.}\ \bibnamefont
  {Turner}}, \bibinfo {author} {\bibfnamefont {A.~A.}\ \bibnamefont
  {Michailidis}}, \bibinfo {author} {\bibfnamefont {D.~A.}\ \bibnamefont
  {Abanin}}, \bibinfo {author} {\bibfnamefont {M.}~\bibnamefont {Serbyn}},\
  and\ \bibinfo {author} {\bibfnamefont {Z.}~\bibnamefont {Papi{\'{c}}}},\
  }\bibfield  {title} {\bibinfo {title} {Weak ergodicity breaking from quantum
  many-body scars},\ }\href {https://doi.org/10.1038/s41567-018-0137-5}
  {\bibfield  {journal} {\bibinfo  {journal} {Nature Physics}\ }\textbf
  {\bibinfo {volume} {14}},\ \bibinfo {pages} {745} (\bibinfo {year}
  {2018}{\natexlab{a}})}\BibitemShut {NoStop}%
\bibitem [{\citenamefont {Khemani}\ \emph {et~al.}(2019)\citenamefont
  {Khemani}, \citenamefont {Laumann},\ and\ \citenamefont
  {Chandran}}]{Khemani2019}%
  \BibitemOpen
  \bibfield  {author} {\bibinfo {author} {\bibfnamefont {V.}~\bibnamefont
  {Khemani}}, \bibinfo {author} {\bibfnamefont {C.~R.}\ \bibnamefont
  {Laumann}},\ and\ \bibinfo {author} {\bibfnamefont {A.}~\bibnamefont
  {Chandran}},\ }\bibfield  {title} {\bibinfo {title} {Signatures of
  integrability in the dynamics of {R}ydberg-blockaded chains},\ }\href
  {https://doi.org/10.1103/PhysRevB.99.161101} {\bibfield  {journal} {\bibinfo
  {journal} {Phys. Rev. B}\ }\textbf {\bibinfo {volume} {99}},\ \bibinfo
  {pages} {161101} (\bibinfo {year} {2019})}\BibitemShut {NoStop}%
\bibitem [{\citenamefont {Lin}\ and\ \citenamefont
  {Motrunich}(2019)}]{Lin2019ExactQuantum}%
  \BibitemOpen
  \bibfield  {author} {\bibinfo {author} {\bibfnamefont {C.-J.}\ \bibnamefont
  {Lin}}\ and\ \bibinfo {author} {\bibfnamefont {O.~I.}\ \bibnamefont
  {Motrunich}},\ }\bibfield  {title} {\bibinfo {title} {Exact quantum many-body
  scar states in the {R}ydberg-blockaded atom chain},\ }\href
  {https://doi.org/10.1103/PhysRevLett.122.173401} {\bibfield  {journal}
  {\bibinfo  {journal} {Phys. Rev. Lett.}\ }\textbf {\bibinfo {volume} {122}},\
  \bibinfo {pages} {173401} (\bibinfo {year} {2019})}\BibitemShut {NoStop}%
\bibitem [{\citenamefont {Mark}\ \emph {et~al.}(2020)\citenamefont {Mark},
  \citenamefont {Lin},\ and\ \citenamefont {Motrunich}}]{Mark2020}%
  \BibitemOpen
  \bibfield  {author} {\bibinfo {author} {\bibfnamefont {D.~K.}\ \bibnamefont
  {Mark}}, \bibinfo {author} {\bibfnamefont {C.-J.}\ \bibnamefont {Lin}},\ and\
  \bibinfo {author} {\bibfnamefont {O.~I.}\ \bibnamefont {Motrunich}},\
  }\bibfield  {title} {\bibinfo {title} {{Exact eigenstates in the Lesanovsky
  model, proximity to integrability and the PXP model, and approximate scar
  states}},\ }\href {https://doi.org/10.1103/PhysRevB.101.094308} {\bibfield
  {journal} {\bibinfo  {journal} {Phys. Rev. B}\ }\textbf {\bibinfo {volume}
  {101}},\ \bibinfo {pages} {094308} (\bibinfo {year} {2020})}\BibitemShut
  {NoStop}%
\bibitem [{\citenamefont {Mukherjee}\ \emph
  {et~al.}(2020{\natexlab{a}})\citenamefont {Mukherjee}, \citenamefont {Nandy},
  \citenamefont {Sen}, \citenamefont {Sen},\ and\ \citenamefont
  {Sengupta}}]{Mukherjee2020}%
  \BibitemOpen
  \bibfield  {author} {\bibinfo {author} {\bibfnamefont {B.}~\bibnamefont
  {Mukherjee}}, \bibinfo {author} {\bibfnamefont {S.}~\bibnamefont {Nandy}},
  \bibinfo {author} {\bibfnamefont {A.}~\bibnamefont {Sen}}, \bibinfo {author}
  {\bibfnamefont {D.}~\bibnamefont {Sen}},\ and\ \bibinfo {author}
  {\bibfnamefont {K.}~\bibnamefont {Sengupta}},\ }\bibfield  {title} {\bibinfo
  {title} {Collapse and revival of quantum many-body scars via floquet
  engineering},\ }\href {https://doi.org/10.1103/PhysRevB.101.245107}
  {\bibfield  {journal} {\bibinfo  {journal} {Phys. Rev. B}\ }\textbf {\bibinfo
  {volume} {101}},\ \bibinfo {pages} {245107} (\bibinfo {year}
  {2020}{\natexlab{a}})}\BibitemShut {NoStop}%
\bibitem [{\citenamefont {Serbyn}\ \emph {et~al.}(2021)\citenamefont {Serbyn},
  \citenamefont {Abanin},\ and\ \citenamefont {Papi{\'{c}}}}]{Serbyn2021}%
  \BibitemOpen
  \bibfield  {author} {\bibinfo {author} {\bibfnamefont {M.}~\bibnamefont
  {Serbyn}}, \bibinfo {author} {\bibfnamefont {D.~A.}\ \bibnamefont {Abanin}},\
  and\ \bibinfo {author} {\bibfnamefont {Z.}~\bibnamefont {Papi{\'{c}}}},\
  }\bibfield  {title} {\bibinfo {title} {Quantum many-body scars and weak
  breaking of ergodicity},\ }\href {https://doi.org/10.1038/s41567-021-01230-2}
  {\bibfield  {journal} {\bibinfo  {journal} {Nature Physics}\ }\textbf
  {\bibinfo {volume} {17}},\ \bibinfo {pages} {675} (\bibinfo {year}
  {2021})}\BibitemShut {NoStop}%
\bibitem [{\citenamefont {Roux}(2010)}]{Roux2010}%
  \BibitemOpen
  \bibfield  {author} {\bibinfo {author} {\bibfnamefont {G.}~\bibnamefont
  {Roux}},\ }\bibfield  {title} {\bibinfo {title} {{Finite-size effects in
  global quantum quenches: Examples from free bosons in an harmonic trap and
  the one-dimensional Bose-Hubbard model}},\ }\href
  {https://doi.org/10.1103/PhysRevA.81.053604} {\bibfield  {journal} {\bibinfo
  {journal} {Phys. Rev. A}\ }\textbf {\bibinfo {volume} {81}},\ \bibinfo
  {pages} {053604} (\bibinfo {year} {2010})}\BibitemShut {NoStop}%
\bibitem [{\citenamefont {Sierant}\ and\ \citenamefont
  {Zakrzewski}(2018)}]{Sierant_2018}%
  \BibitemOpen
  \bibfield  {author} {\bibinfo {author} {\bibfnamefont {P.}~\bibnamefont
  {Sierant}}\ and\ \bibinfo {author} {\bibfnamefont {J.}~\bibnamefont
  {Zakrzewski}},\ }\bibfield  {title} {\bibinfo {title} {Many-body localization
  of bosons in optical lattices},\ }\href
  {https://doi.org/10.1088/1367-2630/aabb17} {\bibfield  {journal} {\bibinfo
  {journal} {New Journal of Physics}\ }\textbf {\bibinfo {volume} {20}},\
  \bibinfo {pages} {043032} (\bibinfo {year} {2018})}\BibitemShut {NoStop}%
\bibitem [{\citenamefont {Zhao}\ \emph {et~al.}(2020)\citenamefont {Zhao},
  \citenamefont {Vovrosh}, \citenamefont {Mintert},\ and\ \citenamefont
  {Knolle}}]{ZhaoHongzheng2020}%
  \BibitemOpen
  \bibfield  {author} {\bibinfo {author} {\bibfnamefont {H.}~\bibnamefont
  {Zhao}}, \bibinfo {author} {\bibfnamefont {J.}~\bibnamefont {Vovrosh}},
  \bibinfo {author} {\bibfnamefont {F.}~\bibnamefont {Mintert}},\ and\ \bibinfo
  {author} {\bibfnamefont {J.}~\bibnamefont {Knolle}},\ }\bibfield  {title}
  {\bibinfo {title} {Quantum many-body scars in optical lattices},\ }\href
  {https://doi.org/10.1103/PhysRevLett.124.160604} {\bibfield  {journal}
  {\bibinfo  {journal} {Phys. Rev. Lett.}\ }\textbf {\bibinfo {volume} {124}},\
  \bibinfo {pages} {160604} (\bibinfo {year} {2020})}\BibitemShut {NoStop}%
\bibitem [{\citenamefont {Mukherjee}\ \emph
  {et~al.}(2020{\natexlab{b}})\citenamefont {Mukherjee}, \citenamefont {Sen},
  \citenamefont {Sen},\ and\ \citenamefont {Sengupta}}]{Mukherjee2020_2}%
  \BibitemOpen
  \bibfield  {author} {\bibinfo {author} {\bibfnamefont {B.}~\bibnamefont
  {Mukherjee}}, \bibinfo {author} {\bibfnamefont {A.}~\bibnamefont {Sen}},
  \bibinfo {author} {\bibfnamefont {D.}~\bibnamefont {Sen}},\ and\ \bibinfo
  {author} {\bibfnamefont {K.}~\bibnamefont {Sengupta}},\ }\bibfield  {title}
  {\bibinfo {title} {Restoring coherence via aperiodic drives in a many-body
  quantum system},\ }\href {https://doi.org/10.1103/PhysRevB.102.014301}
  {\bibfield  {journal} {\bibinfo  {journal} {Phys. Rev. B}\ }\textbf {\bibinfo
  {volume} {102}},\ \bibinfo {pages} {014301} (\bibinfo {year}
  {2020}{\natexlab{b}})}\BibitemShut {NoStop}%
\bibitem [{\citenamefont {Halimeh}\ \emph {et~al.}(2022)\citenamefont
  {Halimeh}, \citenamefont {McCulloch}, \citenamefont {Yang},\ and\
  \citenamefont {Hauke}}]{Halimeh2020}%
  \BibitemOpen
  \bibfield  {author} {\bibinfo {author} {\bibfnamefont {J.~C.}\ \bibnamefont
  {Halimeh}}, \bibinfo {author} {\bibfnamefont {I.~P.}\ \bibnamefont
  {McCulloch}}, \bibinfo {author} {\bibfnamefont {B.}~\bibnamefont {Yang}},\
  and\ \bibinfo {author} {\bibfnamefont {P.}~\bibnamefont {Hauke}},\ }\bibfield
   {title} {\bibinfo {title} {Tuning the topological
  $\ensuremath{\theta}$-angle in cold-atom quantum simulators of gauge
  theories},\ }\href {https://doi.org/10.1103/PRXQuantum.3.040316} {\bibfield
  {journal} {\bibinfo  {journal} {PRX Quantum}\ }\textbf {\bibinfo {volume}
  {3}},\ \bibinfo {pages} {040316} (\bibinfo {year} {2022})}\BibitemShut
  {NoStop}%
\bibitem [{\citenamefont {Fendley}\ \emph {et~al.}(2004)\citenamefont
  {Fendley}, \citenamefont {Sengupta},\ and\ \citenamefont
  {Sachdev}}]{Fendley2004}%
  \BibitemOpen
  \bibfield  {author} {\bibinfo {author} {\bibfnamefont {P.}~\bibnamefont
  {Fendley}}, \bibinfo {author} {\bibfnamefont {K.}~\bibnamefont {Sengupta}},\
  and\ \bibinfo {author} {\bibfnamefont {S.}~\bibnamefont {Sachdev}},\
  }\bibfield  {title} {\bibinfo {title} {Competing density-wave orders in a
  one-dimensional hard-boson model},\ }\href
  {https://doi.org/10.1103/PhysRevB.69.075106} {\bibfield  {journal} {\bibinfo
  {journal} {Phys. Rev. B}\ }\textbf {\bibinfo {volume} {69}},\ \bibinfo
  {pages} {075106} (\bibinfo {year} {2004})}\BibitemShut {NoStop}%
\bibitem [{\citenamefont {Trebst}\ \emph {et~al.}(2008)\citenamefont {Trebst},
  \citenamefont {Troyer}, \citenamefont {Wang},\ and\ \citenamefont
  {Ludwig}}]{Trebst2008}%
  \BibitemOpen
  \bibfield  {author} {\bibinfo {author} {\bibfnamefont {S.}~\bibnamefont
  {Trebst}}, \bibinfo {author} {\bibfnamefont {M.}~\bibnamefont {Troyer}},
  \bibinfo {author} {\bibfnamefont {Z.}~\bibnamefont {Wang}},\ and\ \bibinfo
  {author} {\bibfnamefont {A.~W.~W.}\ \bibnamefont {Ludwig}},\ }\bibfield
  {title} {\bibinfo {title} {{A Short Introduction to Fibonacci Anyon
  Models}},\ }\href {https://doi.org/10.1143/PTPS.176.384} {\bibfield
  {journal} {\bibinfo  {journal} {Progress of Theoretical Physics Supplement}\
  }\textbf {\bibinfo {volume} {176}},\ \bibinfo {pages} {384} (\bibinfo {year}
  {2008})}\BibitemShut {NoStop}%
\bibitem [{\citenamefont {Lesanovsky}\ and\ \citenamefont
  {Katsura}(2012)}]{Lesanovsky2012}%
  \BibitemOpen
  \bibfield  {author} {\bibinfo {author} {\bibfnamefont {I.}~\bibnamefont
  {Lesanovsky}}\ and\ \bibinfo {author} {\bibfnamefont {H.}~\bibnamefont
  {Katsura}},\ }\bibfield  {title} {\bibinfo {title} {{Interacting Fibonacci
  anyons in a Rydberg gas}},\ }\href
  {https://doi.org/10.1103/PhysRevA.86.041601} {\bibfield  {journal} {\bibinfo
  {journal} {Phys. Rev. A}\ }\textbf {\bibinfo {volume} {86}},\ \bibinfo
  {pages} {041601} (\bibinfo {year} {2012})}\BibitemShut {NoStop}%
\bibitem [{\citenamefont {Chandran}\ \emph {et~al.}(2020)\citenamefont
  {Chandran}, \citenamefont {Burnell},\ and\ \citenamefont
  {Sondhi}}]{Chandran2020}%
  \BibitemOpen
  \bibfield  {author} {\bibinfo {author} {\bibfnamefont {A.}~\bibnamefont
  {Chandran}}, \bibinfo {author} {\bibfnamefont {F.~J.}\ \bibnamefont
  {Burnell}},\ and\ \bibinfo {author} {\bibfnamefont {S.~L.}\ \bibnamefont
  {Sondhi}},\ }\bibfield  {title} {\bibinfo {title} {{Absence of Fibonacci
  anyons in Rydberg chains}},\ }\href
  {https://doi.org/10.1103/PhysRevB.101.075104} {\bibfield  {journal} {\bibinfo
   {journal} {Phys. Rev. B}\ }\textbf {\bibinfo {volume} {101}},\ \bibinfo
  {pages} {075104} (\bibinfo {year} {2020})}\BibitemShut {NoStop}%
\bibitem [{\citenamefont {Moessner}\ and\ \citenamefont
  {Sondhi}(2001)}]{Moessner2001}%
  \BibitemOpen
  \bibfield  {author} {\bibinfo {author} {\bibfnamefont {R.}~\bibnamefont
  {Moessner}}\ and\ \bibinfo {author} {\bibfnamefont {S.~L.}\ \bibnamefont
  {Sondhi}},\ }\bibfield  {title} {\bibinfo {title} {Ising models of quantum
  frustration},\ }\href {https://doi.org/10.1103/PhysRevB.63.224401} {\bibfield
   {journal} {\bibinfo  {journal} {Phys. Rev. B}\ }\textbf {\bibinfo {volume}
  {63}},\ \bibinfo {pages} {224401} (\bibinfo {year} {2001})}\BibitemShut
  {NoStop}%
\bibitem [{\citenamefont {Laumann}\ \emph {et~al.}(2012)\citenamefont
  {Laumann}, \citenamefont {Moessner}, \citenamefont {Scardicchio},\ and\
  \citenamefont {Sondhi}}]{Laumann2012}%
  \BibitemOpen
  \bibfield  {author} {\bibinfo {author} {\bibfnamefont {C.~R.}\ \bibnamefont
  {Laumann}}, \bibinfo {author} {\bibfnamefont {R.}~\bibnamefont {Moessner}},
  \bibinfo {author} {\bibfnamefont {A.}~\bibnamefont {Scardicchio}},\ and\
  \bibinfo {author} {\bibfnamefont {S.~L.}\ \bibnamefont {Sondhi}},\ }\bibfield
   {title} {\bibinfo {title} {Quantum adiabatic algorithm and scaling of gaps
  at first-order quantum phase transitions},\ }\href
  {https://doi.org/10.1103/PhysRevLett.109.030502} {\bibfield  {journal}
  {\bibinfo  {journal} {Phys. Rev. Lett.}\ }\textbf {\bibinfo {volume} {109}},\
  \bibinfo {pages} {030502} (\bibinfo {year} {2012})}\BibitemShut {NoStop}%
\bibitem [{\citenamefont {Surace}\ \emph {et~al.}(2020)\citenamefont {Surace},
  \citenamefont {Mazza}, \citenamefont {Giudici}, \citenamefont {Lerose},
  \citenamefont {Gambassi},\ and\ \citenamefont {Dalmonte}}]{Surace2020}%
  \BibitemOpen
  \bibfield  {author} {\bibinfo {author} {\bibfnamefont {F.~M.}\ \bibnamefont
  {Surace}}, \bibinfo {author} {\bibfnamefont {P.~P.}\ \bibnamefont {Mazza}},
  \bibinfo {author} {\bibfnamefont {G.}~\bibnamefont {Giudici}}, \bibinfo
  {author} {\bibfnamefont {A.}~\bibnamefont {Lerose}}, \bibinfo {author}
  {\bibfnamefont {A.}~\bibnamefont {Gambassi}},\ and\ \bibinfo {author}
  {\bibfnamefont {M.}~\bibnamefont {Dalmonte}},\ }\bibfield  {title} {\bibinfo
  {title} {{Lattice Gauge Theories and String Dynamics in Rydberg Atom Quantum
  Simulators}},\ }\href {https://doi.org/10.1103/PhysRevX.10.021041} {\bibfield
   {journal} {\bibinfo  {journal} {Phys. Rev. X}\ }\textbf {\bibinfo {volume}
  {10}},\ \bibinfo {pages} {021041} (\bibinfo {year} {2020})}\BibitemShut
  {NoStop}%
\bibitem [{\citenamefont {Chen}\ and\ \citenamefont
  {Iadecola}(2021)}]{Chen2021}%
  \BibitemOpen
  \bibfield  {author} {\bibinfo {author} {\bibfnamefont {I.-C.}\ \bibnamefont
  {Chen}}\ and\ \bibinfo {author} {\bibfnamefont {T.}~\bibnamefont
  {Iadecola}},\ }\bibfield  {title} {\bibinfo {title} {{Emergent symmetries and
  slow quantum dynamics in a Rydberg-atom chain with confinement}},\ }\href
  {https://doi.org/10.1103/PhysRevB.103.214304} {\bibfield  {journal} {\bibinfo
   {journal} {Phys. Rev. B}\ }\textbf {\bibinfo {volume} {103}},\ \bibinfo
  {pages} {214304} (\bibinfo {year} {2021})}\BibitemShut {NoStop}%
\bibitem [{\citenamefont {Desaules}\ \emph
  {et~al.}(2023{\natexlab{a}})\citenamefont {Desaules}, \citenamefont
  {Banerjee}, \citenamefont {Hudomal}, \citenamefont
  {Papi\ifmmode~\acute{c}\else \'{c}\fi{}}, \citenamefont {Sen},\ and\
  \citenamefont {Halimeh}}]{Desaules12023}%
  \BibitemOpen
  \bibfield  {author} {\bibinfo {author} {\bibfnamefont {J.-Y.}\ \bibnamefont
  {Desaules}}, \bibinfo {author} {\bibfnamefont {D.}~\bibnamefont {Banerjee}},
  \bibinfo {author} {\bibfnamefont {A.}~\bibnamefont {Hudomal}}, \bibinfo
  {author} {\bibfnamefont {Z.}~\bibnamefont {Papi\ifmmode~\acute{c}\else
  \'{c}\fi{}}}, \bibinfo {author} {\bibfnamefont {A.}~\bibnamefont {Sen}},\
  and\ \bibinfo {author} {\bibfnamefont {J.~C.}\ \bibnamefont {Halimeh}},\
  }\bibfield  {title} {\bibinfo {title} {Weak ergodicity breaking in the
  schwinger model},\ }\href {https://doi.org/10.1103/PhysRevB.107.L201105}
  {\bibfield  {journal} {\bibinfo  {journal} {Phys. Rev. B}\ }\textbf {\bibinfo
  {volume} {107}},\ \bibinfo {pages} {L201105} (\bibinfo {year}
  {2023}{\natexlab{a}})}\BibitemShut {NoStop}%
\bibitem [{\citenamefont {Desaules}\ \emph
  {et~al.}(2023{\natexlab{b}})\citenamefont {Desaules}, \citenamefont
  {Hudomal}, \citenamefont {Banerjee}, \citenamefont {Sen}, \citenamefont
  {Papi\ifmmode~\acute{c}\else \'{c}\fi{}},\ and\ \citenamefont
  {Halimeh}}]{Desaules22023}%
  \BibitemOpen
  \bibfield  {author} {\bibinfo {author} {\bibfnamefont {J.-Y.}\ \bibnamefont
  {Desaules}}, \bibinfo {author} {\bibfnamefont {A.}~\bibnamefont {Hudomal}},
  \bibinfo {author} {\bibfnamefont {D.}~\bibnamefont {Banerjee}}, \bibinfo
  {author} {\bibfnamefont {A.}~\bibnamefont {Sen}}, \bibinfo {author}
  {\bibfnamefont {Z.}~\bibnamefont {Papi\ifmmode~\acute{c}\else \'{c}\fi{}}},\
  and\ \bibinfo {author} {\bibfnamefont {J.~C.}\ \bibnamefont {Halimeh}},\
  }\bibfield  {title} {\bibinfo {title} {Prominent quantum many-body scars in a
  truncated schwinger model},\ }\href
  {https://doi.org/10.1103/PhysRevB.107.205112} {\bibfield  {journal} {\bibinfo
   {journal} {Phys. Rev. B}\ }\textbf {\bibinfo {volume} {107}},\ \bibinfo
  {pages} {205112} (\bibinfo {year} {2023}{\natexlab{b}})}\BibitemShut
  {NoStop}%
\bibitem [{\citenamefont {Sala}\ \emph {et~al.}(2020)\citenamefont {Sala},
  \citenamefont {Rakovszky}, \citenamefont {Verresen}, \citenamefont {Knap},\
  and\ \citenamefont {Pollmann}}]{Sala2020}%
  \BibitemOpen
  \bibfield  {author} {\bibinfo {author} {\bibfnamefont {P.}~\bibnamefont
  {Sala}}, \bibinfo {author} {\bibfnamefont {T.}~\bibnamefont {Rakovszky}},
  \bibinfo {author} {\bibfnamefont {R.}~\bibnamefont {Verresen}}, \bibinfo
  {author} {\bibfnamefont {M.}~\bibnamefont {Knap}},\ and\ \bibinfo {author}
  {\bibfnamefont {F.}~\bibnamefont {Pollmann}},\ }\bibfield  {title} {\bibinfo
  {title} {{Ergodicity Breaking Arising from Hilbert Space Fragmentation in
  Dipole-Conserving Hamiltonians}},\ }\href
  {https://doi.org/10.1103/PhysRevX.10.011047} {\bibfield  {journal} {\bibinfo
  {journal} {Phys. Rev. X}\ }\textbf {\bibinfo {volume} {10}},\ \bibinfo
  {pages} {011047} (\bibinfo {year} {2020})}\BibitemShut {NoStop}%
\bibitem [{\citenamefont {Moudgalya}\ \emph
  {et~al.}(2020{\natexlab{a}})\citenamefont {Moudgalya}, \citenamefont
  {Regnault},\ and\ \citenamefont {Bernevig}}]{Moudgalya2020}%
  \BibitemOpen
  \bibfield  {author} {\bibinfo {author} {\bibfnamefont {S.}~\bibnamefont
  {Moudgalya}}, \bibinfo {author} {\bibfnamefont {N.}~\bibnamefont
  {Regnault}},\ and\ \bibinfo {author} {\bibfnamefont {B.~A.}\ \bibnamefont
  {Bernevig}},\ }\bibfield  {title} {\bibinfo {title}
  {$\ensuremath{\eta}$-pairing in hubbard models: From spectrum generating
  algebras to quantum many-body scars},\ }\href
  {https://doi.org/10.1103/PhysRevB.102.085140} {\bibfield  {journal} {\bibinfo
   {journal} {Phys. Rev. B}\ }\textbf {\bibinfo {volume} {102}},\ \bibinfo
  {pages} {085140} (\bibinfo {year} {2020}{\natexlab{a}})}\BibitemShut
  {NoStop}%
\bibitem [{\citenamefont {Mizuta}\ \emph {et~al.}(2020)\citenamefont {Mizuta},
  \citenamefont {Takasan},\ and\ \citenamefont {Kawakami}}]{Mizuta2020}%
  \BibitemOpen
  \bibfield  {author} {\bibinfo {author} {\bibfnamefont {K.}~\bibnamefont
  {Mizuta}}, \bibinfo {author} {\bibfnamefont {K.}~\bibnamefont {Takasan}},\
  and\ \bibinfo {author} {\bibfnamefont {N.}~\bibnamefont {Kawakami}},\
  }\bibfield  {title} {\bibinfo {title} {{Exact Floquet quantum many-body scars
  under Rydberg blockade}},\ }\href
  {https://doi.org/10.1103/PhysRevResearch.2.033284} {\bibfield  {journal}
  {\bibinfo  {journal} {Phys. Rev. Res.}\ }\textbf {\bibinfo {volume} {2}},\
  \bibinfo {pages} {033284} (\bibinfo {year} {2020})}\BibitemShut {NoStop}%
\bibitem [{\citenamefont {Mukherjee}\ \emph {et~al.}(2021)\citenamefont
  {Mukherjee}, \citenamefont {Cai},\ and\ \citenamefont {Liu}}]{Mukherjee2021}%
  \BibitemOpen
  \bibfield  {author} {\bibinfo {author} {\bibfnamefont {B.}~\bibnamefont
  {Mukherjee}}, \bibinfo {author} {\bibfnamefont {Z.}~\bibnamefont {Cai}},\
  and\ \bibinfo {author} {\bibfnamefont {W.~V.}\ \bibnamefont {Liu}},\
  }\bibfield  {title} {\bibinfo {title} {{Constraint-induced breaking and
  restoration of ergodicity in spin-1 PXP models}},\ }\href
  {https://doi.org/10.1103/PhysRevResearch.3.033201} {\bibfield  {journal}
  {\bibinfo  {journal} {Phys. Rev. Res.}\ }\textbf {\bibinfo {volume} {3}},\
  \bibinfo {pages} {033201} (\bibinfo {year} {2021})}\BibitemShut {NoStop}%
\bibitem [{\citenamefont {Michailidis}\ \emph {et~al.}(2020)\citenamefont
  {Michailidis}, \citenamefont {Turner}, \citenamefont
  {Papi\ifmmode~\acute{c}\else \'{c}\fi{}}, \citenamefont {Abanin},\ and\
  \citenamefont {Serbyn}}]{Michailidis2020}%
  \BibitemOpen
  \bibfield  {author} {\bibinfo {author} {\bibfnamefont {A.~A.}\ \bibnamefont
  {Michailidis}}, \bibinfo {author} {\bibfnamefont {C.~J.}\ \bibnamefont
  {Turner}}, \bibinfo {author} {\bibfnamefont {Z.}~\bibnamefont
  {Papi\ifmmode~\acute{c}\else \'{c}\fi{}}}, \bibinfo {author} {\bibfnamefont
  {D.~A.}\ \bibnamefont {Abanin}},\ and\ \bibinfo {author} {\bibfnamefont
  {M.}~\bibnamefont {Serbyn}},\ }\bibfield  {title} {\bibinfo {title}
  {Stabilizing two-dimensional quantum scars by deformation and
  synchronization},\ }\href {https://doi.org/10.1103/PhysRevResearch.2.022065}
  {\bibfield  {journal} {\bibinfo  {journal} {Phys. Rev. Res.}\ }\textbf
  {\bibinfo {volume} {2}},\ \bibinfo {pages} {022065} (\bibinfo {year}
  {2020})}\BibitemShut {NoStop}%
\bibitem [{\citenamefont {Turner}\ \emph
  {et~al.}(2018{\natexlab{b}})\citenamefont {Turner}, \citenamefont
  {Michailidis}, \citenamefont {Abanin}, \citenamefont {Serbyn},\ and\
  \citenamefont {Papi\ifmmode~\acute{c}\else \'{c}\fi{}}}]{Turner2018PhysRevB}%
  \BibitemOpen
  \bibfield  {author} {\bibinfo {author} {\bibfnamefont {C.~J.}\ \bibnamefont
  {Turner}}, \bibinfo {author} {\bibfnamefont {A.~A.}\ \bibnamefont
  {Michailidis}}, \bibinfo {author} {\bibfnamefont {D.~A.}\ \bibnamefont
  {Abanin}}, \bibinfo {author} {\bibfnamefont {M.}~\bibnamefont {Serbyn}},\
  and\ \bibinfo {author} {\bibfnamefont {Z.}~\bibnamefont
  {Papi\ifmmode~\acute{c}\else \'{c}\fi{}}},\ }\bibfield  {title} {\bibinfo
  {title} {{Quantum scarred eigenstates in a Rydberg atom chain: Entanglement,
  breakdown of thermalization, and stability to perturbations}},\ }\href
  {https://doi.org/10.1103/PhysRevB.98.155134} {\bibfield  {journal} {\bibinfo
  {journal} {Phys. Rev. B}\ }\textbf {\bibinfo {volume} {98}},\ \bibinfo
  {pages} {155134} (\bibinfo {year} {2018}{\natexlab{b}})}\BibitemShut
  {NoStop}%
\bibitem [{\citenamefont {Su}\ \emph {et~al.}(2023)\citenamefont {Su},
  \citenamefont {Sun}, \citenamefont {Hudomal}, \citenamefont {Desaules},
  \citenamefont {Zhou}, \citenamefont {Yang}, \citenamefont {Halimeh},
  \citenamefont {Yuan}, \citenamefont {Papi\ifmmode~\acute{c}\else
  \'{c}\fi{}},\ and\ \citenamefont {Pan}}]{JianWeiPan2023observation}%
  \BibitemOpen
  \bibfield  {author} {\bibinfo {author} {\bibfnamefont {G.-X.}\ \bibnamefont
  {Su}}, \bibinfo {author} {\bibfnamefont {H.}~\bibnamefont {Sun}}, \bibinfo
  {author} {\bibfnamefont {A.}~\bibnamefont {Hudomal}}, \bibinfo {author}
  {\bibfnamefont {J.-Y.}\ \bibnamefont {Desaules}}, \bibinfo {author}
  {\bibfnamefont {Z.-Y.}\ \bibnamefont {Zhou}}, \bibinfo {author}
  {\bibfnamefont {B.}~\bibnamefont {Yang}}, \bibinfo {author} {\bibfnamefont
  {J.~C.}\ \bibnamefont {Halimeh}}, \bibinfo {author} {\bibfnamefont {Z.-S.}\
  \bibnamefont {Yuan}}, \bibinfo {author} {\bibfnamefont {Z.}~\bibnamefont
  {Papi\ifmmode~\acute{c}\else \'{c}\fi{}}},\ and\ \bibinfo {author}
  {\bibfnamefont {J.-W.}\ \bibnamefont {Pan}},\ }\bibfield  {title} {\bibinfo
  {title} {{Observation of many-body scarring in a Bose-Hubbard quantum
  simulator}},\ }\href {https://doi.org/10.1103/PhysRevResearch.5.023010}
  {\bibfield  {journal} {\bibinfo  {journal} {Phys. Rev. Res.}\ }\textbf
  {\bibinfo {volume} {5}},\ \bibinfo {pages} {023010} (\bibinfo {year}
  {2023})}\BibitemShut {NoStop}%
\bibitem [{\citenamefont {Yao}\ \emph {et~al.}(2022)\citenamefont {Yao},
  \citenamefont {Pan}, \citenamefont {Liu},\ and\ \citenamefont
  {Zhai}}]{ZhaiHui2022}%
  \BibitemOpen
  \bibfield  {author} {\bibinfo {author} {\bibfnamefont {Z.}~\bibnamefont
  {Yao}}, \bibinfo {author} {\bibfnamefont {L.}~\bibnamefont {Pan}}, \bibinfo
  {author} {\bibfnamefont {S.}~\bibnamefont {Liu}},\ and\ \bibinfo {author}
  {\bibfnamefont {H.}~\bibnamefont {Zhai}},\ }\bibfield  {title} {\bibinfo
  {title} {Quantum many-body scars and quantum criticality},\ }\href
  {https://doi.org/10.1103/PhysRevB.105.125123} {\bibfield  {journal} {\bibinfo
   {journal} {Phys. Rev. B}\ }\textbf {\bibinfo {volume} {105}},\ \bibinfo
  {pages} {125123} (\bibinfo {year} {2022})}\BibitemShut {NoStop}%
\bibitem [{\citenamefont {Peng}\ and\ \citenamefont
  {Cui}(2022)}]{CuiXiaoling2022}%
  \BibitemOpen
  \bibfield  {author} {\bibinfo {author} {\bibfnamefont {C.}~\bibnamefont
  {Peng}}\ and\ \bibinfo {author} {\bibfnamefont {X.}~\bibnamefont {Cui}},\
  }\bibfield  {title} {\bibinfo {title} {{Bridging quantum many-body scars and
  quantum integrability in Ising chains with transverse and longitudinal
  fields}},\ }\href {https://doi.org/10.1103/PhysRevB.106.214311} {\bibfield
  {journal} {\bibinfo  {journal} {Phys. Rev. B}\ }\textbf {\bibinfo {volume}
  {106}},\ \bibinfo {pages} {214311} (\bibinfo {year} {2022})}\BibitemShut
  {NoStop}%
\bibitem [{\citenamefont {You}\ \emph {et~al.}(2020)\citenamefont {You},
  \citenamefont {Sun}, \citenamefont {Ren}, \citenamefont {Yu},\ and\
  \citenamefont {Ole\ifmmode~\acute{s}\else \'{s}\fi{}}}]{You2020}%
  \BibitemOpen
  \bibfield  {author} {\bibinfo {author} {\bibfnamefont {W.-L.}\ \bibnamefont
  {You}}, \bibinfo {author} {\bibfnamefont {G.}~\bibnamefont {Sun}}, \bibinfo
  {author} {\bibfnamefont {J.}~\bibnamefont {Ren}}, \bibinfo {author}
  {\bibfnamefont {W.~C.}\ \bibnamefont {Yu}},\ and\ \bibinfo {author}
  {\bibfnamefont {A.~M.}\ \bibnamefont {Ole\ifmmode~\acute{s}\else
  \'{s}\fi{}}},\ }\bibfield  {title} {\bibinfo {title} {Quantum phase
  transitions in the spin-1 {K}itaev-{H}eisenberg chain},\ }\href
  {https://doi.org/10.1103/PhysRevB.102.144437} {\bibfield  {journal} {\bibinfo
   {journal} {Phys. Rev. B}\ }\textbf {\bibinfo {volume} {102}},\ \bibinfo
  {pages} {144437} (\bibinfo {year} {2020})}\BibitemShut {NoStop}%
\bibitem [{\citenamefont {You}\ \emph {et~al.}(2022)\citenamefont {You},
  \citenamefont {Zhao}, \citenamefont {Ren}, \citenamefont {Sun}, \citenamefont
  {Li},\ and\ \citenamefont {Ole\ifmmode~\acute{s}\else
  \'{s}\fi{}}}]{You2022prr}%
  \BibitemOpen
  \bibfield  {author} {\bibinfo {author} {\bibfnamefont {W.-L.}\ \bibnamefont
  {You}}, \bibinfo {author} {\bibfnamefont {Z.}~\bibnamefont {Zhao}}, \bibinfo
  {author} {\bibfnamefont {J.}~\bibnamefont {Ren}}, \bibinfo {author}
  {\bibfnamefont {G.}~\bibnamefont {Sun}}, \bibinfo {author} {\bibfnamefont
  {L.}~\bibnamefont {Li}},\ and\ \bibinfo {author} {\bibfnamefont {A.~M.}\
  \bibnamefont {Ole\ifmmode~\acute{s}\else \'{s}\fi{}}},\ }\bibfield  {title}
  {\bibinfo {title} {Quantum many-body scars in spin-1 {K}itaev chains},\
  }\href {https://doi.org/10.1103/PhysRevResearch.4.013103} {\bibfield
  {journal} {\bibinfo  {journal} {Phys. Rev. Res.}\ }\textbf {\bibinfo {volume}
  {4}},\ \bibinfo {pages} {013103} (\bibinfo {year} {2022})}\BibitemShut
  {NoStop}%
\bibitem [{\citenamefont {Kitaev}(2006)}]{KITAEV20062}%
  \BibitemOpen
  \bibfield  {author} {\bibinfo {author} {\bibfnamefont {A.}~\bibnamefont
  {Kitaev}},\ }\bibfield  {title} {\bibinfo {title} {Anyons in an exactly
  solved model and beyond},\ }\href
  {https://doi.org/https://doi.org/10.1016/j.aop.2005.10.005} {\bibfield
  {journal} {\bibinfo  {journal} {Annals of Physics}\ }\textbf {\bibinfo
  {volume} {321}},\ \bibinfo {pages} {2} (\bibinfo {year} {2006})}\BibitemShut
  {NoStop}%
\bibitem [{\citenamefont {Jackeli}\ and\ \citenamefont
  {Khaliullin}(2009)}]{Jackeli2009}%
  \BibitemOpen
  \bibfield  {author} {\bibinfo {author} {\bibfnamefont {G.}~\bibnamefont
  {Jackeli}}\ and\ \bibinfo {author} {\bibfnamefont {G.}~\bibnamefont
  {Khaliullin}},\ }\bibfield  {title} {\bibinfo {title} {Mott insulators in the
  strong spin-orbit coupling limit: From {H}eisenberg to a quantum compass and
  {K}itaev models},\ }\href {https://doi.org/10.1103/PhysRevLett.102.017205}
  {\bibfield  {journal} {\bibinfo  {journal} {Phys. Rev. Lett.}\ }\textbf
  {\bibinfo {volume} {102}},\ \bibinfo {pages} {017205} (\bibinfo {year}
  {2009})}\BibitemShut {NoStop}%
\bibitem [{\citenamefont {Liu}\ \emph {et~al.}(2020)\citenamefont {Liu},
  \citenamefont {Chaloupka},\ and\ \citenamefont {Khaliullin}}]{Liu2020}%
  \BibitemOpen
  \bibfield  {author} {\bibinfo {author} {\bibfnamefont {H.}~\bibnamefont
  {Liu}}, \bibinfo {author} {\bibfnamefont {J.~c.~v.}\ \bibnamefont
  {Chaloupka}},\ and\ \bibinfo {author} {\bibfnamefont {G.}~\bibnamefont
  {Khaliullin}},\ }\bibfield  {title} {\bibinfo {title} {{Kitaev Spin Liquid in
  $3d$ Transition Metal Compounds}},\ }\href
  {https://doi.org/10.1103/PhysRevLett.125.047201} {\bibfield  {journal}
  {\bibinfo  {journal} {Phys. Rev. Lett.}\ }\textbf {\bibinfo {volume} {125}},\
  \bibinfo {pages} {047201} (\bibinfo {year} {2020})}\BibitemShut {NoStop}%
\bibitem [{\citenamefont {Li}\ \emph {et~al.}(2015)\citenamefont {Li},
  \citenamefont {Chen}, \citenamefont {Tong}, \citenamefont {Pi}, \citenamefont
  {Liu}, \citenamefont {Yang}, \citenamefont {Wang},\ and\ \citenamefont
  {Zhang}}]{LiYuesheng2015}%
  \BibitemOpen
  \bibfield  {author} {\bibinfo {author} {\bibfnamefont {Y.}~\bibnamefont
  {Li}}, \bibinfo {author} {\bibfnamefont {G.}~\bibnamefont {Chen}}, \bibinfo
  {author} {\bibfnamefont {W.}~\bibnamefont {Tong}}, \bibinfo {author}
  {\bibfnamefont {L.}~\bibnamefont {Pi}}, \bibinfo {author} {\bibfnamefont
  {J.}~\bibnamefont {Liu}}, \bibinfo {author} {\bibfnamefont {Z.}~\bibnamefont
  {Yang}}, \bibinfo {author} {\bibfnamefont {X.}~\bibnamefont {Wang}},\ and\
  \bibinfo {author} {\bibfnamefont {Q.}~\bibnamefont {Zhang}},\ }\bibfield
  {title} {\bibinfo {title} {Rare-earth triangular lattice spin liquid: A
  single-crystal study of $\mathrm{Yb}\mathrm{Mg}\mathrm{Ga}\mathrm{O}_{4}$},\
  }\href {https://doi.org/10.1103/PhysRevLett.115.167203} {\bibfield  {journal}
  {\bibinfo  {journal} {Phys. Rev. Lett.}\ }\textbf {\bibinfo {volume} {115}},\
  \bibinfo {pages} {167203} (\bibinfo {year} {2015})}\BibitemShut {NoStop}%
\bibitem [{\citenamefont {Ruan}\ \emph {et~al.}(2021)\citenamefont {Ruan},
  \citenamefont {Chen}, \citenamefont {Tang}, \citenamefont {Hwang},
  \citenamefont {Tsai}, \citenamefont {Lee}, \citenamefont {Wu}, \citenamefont
  {Ryu}, \citenamefont {Kahn}, \citenamefont {Liou}, \citenamefont {Jia},
  \citenamefont {Aikawa}, \citenamefont {Hwang}, \citenamefont {Wang},
  \citenamefont {Choi}, \citenamefont {Louie}, \citenamefont {Lee},
  \citenamefont {Shen}, \citenamefont {Mo},\ and\ \citenamefont
  {Crommie}}]{Ruan2021}%
  \BibitemOpen
  \bibfield  {author} {\bibinfo {author} {\bibfnamefont {W.}~\bibnamefont
  {Ruan}}, \bibinfo {author} {\bibfnamefont {Y.}~\bibnamefont {Chen}}, \bibinfo
  {author} {\bibfnamefont {S.}~\bibnamefont {Tang}}, \bibinfo {author}
  {\bibfnamefont {J.}~\bibnamefont {Hwang}}, \bibinfo {author} {\bibfnamefont
  {H.-Z.}\ \bibnamefont {Tsai}}, \bibinfo {author} {\bibfnamefont {R.~L.}\
  \bibnamefont {Lee}}, \bibinfo {author} {\bibfnamefont {M.}~\bibnamefont
  {Wu}}, \bibinfo {author} {\bibfnamefont {H.}~\bibnamefont {Ryu}}, \bibinfo
  {author} {\bibfnamefont {S.}~\bibnamefont {Kahn}}, \bibinfo {author}
  {\bibfnamefont {F.}~\bibnamefont {Liou}}, \bibinfo {author} {\bibfnamefont
  {C.}~\bibnamefont {Jia}}, \bibinfo {author} {\bibfnamefont {A.}~\bibnamefont
  {Aikawa}}, \bibinfo {author} {\bibfnamefont {C.}~\bibnamefont {Hwang}},
  \bibinfo {author} {\bibfnamefont {F.}~\bibnamefont {Wang}}, \bibinfo {author}
  {\bibfnamefont {Y.}~\bibnamefont {Choi}}, \bibinfo {author} {\bibfnamefont
  {S.~G.}\ \bibnamefont {Louie}}, \bibinfo {author} {\bibfnamefont {P.~A.}\
  \bibnamefont {Lee}}, \bibinfo {author} {\bibfnamefont {Z.-X.}\ \bibnamefont
  {Shen}}, \bibinfo {author} {\bibfnamefont {S.-K.}\ \bibnamefont {Mo}},\ and\
  \bibinfo {author} {\bibfnamefont {M.~F.}\ \bibnamefont {Crommie}},\
  }\bibfield  {title} {\bibinfo {title} {Evidence for quantum spin liquid
  behaviour in single-layer $\mathrm{1T}$-$\mathrm{TaSe}_2$ from scanning
  tunnelling microscopy},\ }\href {https://doi.org/10.1038/s41567-021-01321-0}
  {\bibfield  {journal} {\bibinfo  {journal} {Nature Physics}\ }\textbf
  {\bibinfo {volume} {17}},\ \bibinfo {pages} {1154} (\bibinfo {year}
  {2021})}\BibitemShut {NoStop}%
\bibitem [{\citenamefont {Wu}\ \emph {et~al.}(2022)\citenamefont {Wu},
  \citenamefont {Li}, \citenamefont {Zhang}, \citenamefont {Liu}, \citenamefont
  {Gao}, \citenamefont {Feng}, \citenamefont {Deng}, \citenamefont {Ren},
  \citenamefont {Wang}, \citenamefont {Chen}, \citenamefont {Embs},
  \citenamefont {Zhu}, \citenamefont {Huang}, \citenamefont {Xiang},
  \citenamefont {Chen}, \citenamefont {Wu}, \citenamefont {Choi}, \citenamefont
  {Qu}, \citenamefont {Li}, \citenamefont {Wang}, \citenamefont {Zhou},
  \citenamefont {Su}, \citenamefont {Wang}, \citenamefont {Chen}, \citenamefont
  {Zhang},\ and\ \citenamefont {Ma}}]{Wu2022}%
  \BibitemOpen
  \bibfield  {author} {\bibinfo {author} {\bibfnamefont {J.}~\bibnamefont
  {Wu}}, \bibinfo {author} {\bibfnamefont {J.}~\bibnamefont {Li}}, \bibinfo
  {author} {\bibfnamefont {Z.}~\bibnamefont {Zhang}}, \bibinfo {author}
  {\bibfnamefont {C.}~\bibnamefont {Liu}}, \bibinfo {author} {\bibfnamefont
  {Y.~H.}\ \bibnamefont {Gao}}, \bibinfo {author} {\bibfnamefont
  {E.}~\bibnamefont {Feng}}, \bibinfo {author} {\bibfnamefont {G.}~\bibnamefont
  {Deng}}, \bibinfo {author} {\bibfnamefont {Q.}~\bibnamefont {Ren}}, \bibinfo
  {author} {\bibfnamefont {Z.}~\bibnamefont {Wang}}, \bibinfo {author}
  {\bibfnamefont {R.}~\bibnamefont {Chen}}, \bibinfo {author} {\bibfnamefont
  {J.}~\bibnamefont {Embs}}, \bibinfo {author} {\bibfnamefont {F.}~\bibnamefont
  {Zhu}}, \bibinfo {author} {\bibfnamefont {Q.}~\bibnamefont {Huang}}, \bibinfo
  {author} {\bibfnamefont {Z.}~\bibnamefont {Xiang}}, \bibinfo {author}
  {\bibfnamefont {L.}~\bibnamefont {Chen}}, \bibinfo {author} {\bibfnamefont
  {Y.}~\bibnamefont {Wu}}, \bibinfo {author} {\bibfnamefont {E.~S.}\
  \bibnamefont {Choi}}, \bibinfo {author} {\bibfnamefont {Z.}~\bibnamefont
  {Qu}}, \bibinfo {author} {\bibfnamefont {L.}~\bibnamefont {Li}}, \bibinfo
  {author} {\bibfnamefont {J.}~\bibnamefont {Wang}}, \bibinfo {author}
  {\bibfnamefont {H.}~\bibnamefont {Zhou}}, \bibinfo {author} {\bibfnamefont
  {Y.}~\bibnamefont {Su}}, \bibinfo {author} {\bibfnamefont {X.}~\bibnamefont
  {Wang}}, \bibinfo {author} {\bibfnamefont {G.}~\bibnamefont {Chen}}, \bibinfo
  {author} {\bibfnamefont {Q.}~\bibnamefont {Zhang}},\ and\ \bibinfo {author}
  {\bibfnamefont {J.}~\bibnamefont {Ma}},\ }\bibfield  {title} {\bibinfo
  {title} {Magnetic field effects on the quantum spin liquid behaviors of
  $\mathrm{NaYbS_2}$},\ }\href {https://doi.org/10.1007/s44214-022-00011-z}
  {\bibfield  {journal} {\bibinfo  {journal} {Quantum Frontiers}\ }\textbf
  {\bibinfo {volume} {1}},\ \bibinfo {pages} {13} (\bibinfo {year}
  {2022})}\BibitemShut {NoStop}%
\bibitem [{\citenamefont {Khuntia}\ \emph {et~al.}(2020)\citenamefont
  {Khuntia}, \citenamefont {Velazquez}, \citenamefont {Barth{\'e}lemy},
  \citenamefont {Bert}, \citenamefont {Kermarrec}, \citenamefont {Legros},
  \citenamefont {Bernu}, \citenamefont {Messio}, \citenamefont {Zorko},\ and\
  \citenamefont {Mendels}}]{Khuntia2020}%
  \BibitemOpen
  \bibfield  {author} {\bibinfo {author} {\bibfnamefont {P.}~\bibnamefont
  {Khuntia}}, \bibinfo {author} {\bibfnamefont {M.}~\bibnamefont {Velazquez}},
  \bibinfo {author} {\bibfnamefont {Q.}~\bibnamefont {Barth{\'e}lemy}},
  \bibinfo {author} {\bibfnamefont {F.}~\bibnamefont {Bert}}, \bibinfo {author}
  {\bibfnamefont {E.}~\bibnamefont {Kermarrec}}, \bibinfo {author}
  {\bibfnamefont {A.}~\bibnamefont {Legros}}, \bibinfo {author} {\bibfnamefont
  {B.}~\bibnamefont {Bernu}}, \bibinfo {author} {\bibfnamefont
  {L.}~\bibnamefont {Messio}}, \bibinfo {author} {\bibfnamefont
  {A.}~\bibnamefont {Zorko}},\ and\ \bibinfo {author} {\bibfnamefont
  {P.}~\bibnamefont {Mendels}},\ }\bibfield  {title} {\bibinfo {title} {Gapless
  ground state in the archetypal quantum kagome antiferromagnet
  $\mathrm{ZnCu}_3\mathrm{(OH)}_6\mathrm{Cl}_2$},\ }\href
  {https://doi.org/10.1038/s41567-020-0792-1} {\bibfield  {journal} {\bibinfo
  {journal} {Nature Physics}\ }\textbf {\bibinfo {volume} {16}},\ \bibinfo
  {pages} {469} (\bibinfo {year} {2020})}\BibitemShut {NoStop}%
\bibitem [{\citenamefont {Shockley}\ \emph {et~al.}(2015)\citenamefont
  {Shockley}, \citenamefont {Bert}, \citenamefont {Orain}, \citenamefont
  {Okamoto},\ and\ \citenamefont {Mendels}}]{Shockley2015}%
  \BibitemOpen
  \bibfield  {author} {\bibinfo {author} {\bibfnamefont {A.~C.}\ \bibnamefont
  {Shockley}}, \bibinfo {author} {\bibfnamefont {F.}~\bibnamefont {Bert}},
  \bibinfo {author} {\bibfnamefont {J.-C.}\ \bibnamefont {Orain}}, \bibinfo
  {author} {\bibfnamefont {Y.}~\bibnamefont {Okamoto}},\ and\ \bibinfo {author}
  {\bibfnamefont {P.}~\bibnamefont {Mendels}},\ }\bibfield  {title} {\bibinfo
  {title} {{Frozen State and Spin Liquid Physics in
  $\mathrm{Na}_4\mathrm{Ir}_3\mathrm{O}_8$: An NMR Study}},\ }\href
  {https://doi.org/10.1103/PhysRevLett.115.047201} {\bibfield  {journal}
  {\bibinfo  {journal} {Phys. Rev. Lett.}\ }\textbf {\bibinfo {volume} {115}},\
  \bibinfo {pages} {047201} (\bibinfo {year} {2015})}\BibitemShut {NoStop}%
\bibitem [{\citenamefont {Banerjee}\ \emph {et~al.}(2017)\citenamefont
  {Banerjee}, \citenamefont {Yan}, \citenamefont {Knolle}, \citenamefont
  {Bridges}, \citenamefont {Stone}, \citenamefont {Lumsden}, \citenamefont
  {Mandrus}, \citenamefont {Tennant}, \citenamefont {Moessner},\ and\
  \citenamefont {Nagler}}]{A.Banerjee2017}%
  \BibitemOpen
  \bibfield  {author} {\bibinfo {author} {\bibfnamefont {A.}~\bibnamefont
  {Banerjee}}, \bibinfo {author} {\bibfnamefont {J.}~\bibnamefont {Yan}},
  \bibinfo {author} {\bibfnamefont {J.}~\bibnamefont {Knolle}}, \bibinfo
  {author} {\bibfnamefont {C.~A.}\ \bibnamefont {Bridges}}, \bibinfo {author}
  {\bibfnamefont {M.~B.}\ \bibnamefont {Stone}}, \bibinfo {author}
  {\bibfnamefont {M.~D.}\ \bibnamefont {Lumsden}}, \bibinfo {author}
  {\bibfnamefont {D.~G.}\ \bibnamefont {Mandrus}}, \bibinfo {author}
  {\bibfnamefont {D.~A.}\ \bibnamefont {Tennant}}, \bibinfo {author}
  {\bibfnamefont {R.}~\bibnamefont {Moessner}},\ and\ \bibinfo {author}
  {\bibfnamefont {S.~E.}\ \bibnamefont {Nagler}},\ }\bibfield  {title}
  {\bibinfo {title} {Neutron scattering in the proximate quantum spin liquid
  $\mathrm{\alpha}$-$\mathrm{Ru}\mathrm{Cl}_3$},\ }\href
  {https://doi.org/10.1126/science.aah6015} {\bibfield  {journal} {\bibinfo
  {journal} {Science}\ }\textbf {\bibinfo {volume} {356}},\ \bibinfo {pages}
  {1055} (\bibinfo {year} {2017})}\BibitemShut {NoStop}%
\bibitem [{\citenamefont {Yadav}\ \emph {et~al.}(2018)\citenamefont {Yadav},
  \citenamefont {Ray}, \citenamefont {Eldeeb}, \citenamefont {Nishimoto},
  \citenamefont {Hozoi},\ and\ \citenamefont {van~den Brink}}]{Y.Ravi2018}%
  \BibitemOpen
  \bibfield  {author} {\bibinfo {author} {\bibfnamefont {R.}~\bibnamefont
  {Yadav}}, \bibinfo {author} {\bibfnamefont {R.}~\bibnamefont {Ray}}, \bibinfo
  {author} {\bibfnamefont {M.~S.}\ \bibnamefont {Eldeeb}}, \bibinfo {author}
  {\bibfnamefont {S.}~\bibnamefont {Nishimoto}}, \bibinfo {author}
  {\bibfnamefont {L.}~\bibnamefont {Hozoi}},\ and\ \bibinfo {author}
  {\bibfnamefont {J.}~\bibnamefont {van~den Brink}},\ }\bibfield  {title}
  {\bibinfo {title} {Strong effect of hydrogen order on magnetic {K}itaev
  interactions in $\mathrm{H}_{3}\mathrm{LiIr}_{2}\mathrm{O}_{6}$},\ }\href
  {https://doi.org/10.1103/PhysRevLett.121.197203} {\bibfield  {journal}
  {\bibinfo  {journal} {Phys. Rev. Lett.}\ }\textbf {\bibinfo {volume} {121}},\
  \bibinfo {pages} {197203} (\bibinfo {year} {2018})}\BibitemShut {NoStop}%
\bibitem [{\citenamefont {Pal}\ \emph {et~al.}(2021)\citenamefont {Pal},
  \citenamefont {Seth}, \citenamefont {Sakrikar}, \citenamefont {Ali},
  \citenamefont {Bhattacharjee}, \citenamefont {Muthu}, \citenamefont {Singh},\
  and\ \citenamefont {Sood}}]{Srishti2021}%
  \BibitemOpen
  \bibfield  {author} {\bibinfo {author} {\bibfnamefont {S.}~\bibnamefont
  {Pal}}, \bibinfo {author} {\bibfnamefont {A.}~\bibnamefont {Seth}}, \bibinfo
  {author} {\bibfnamefont {P.}~\bibnamefont {Sakrikar}}, \bibinfo {author}
  {\bibfnamefont {A.}~\bibnamefont {Ali}}, \bibinfo {author} {\bibfnamefont
  {S.}~\bibnamefont {Bhattacharjee}}, \bibinfo {author} {\bibfnamefont
  {D.~V.~S.}\ \bibnamefont {Muthu}}, \bibinfo {author} {\bibfnamefont
  {Y.}~\bibnamefont {Singh}},\ and\ \bibinfo {author} {\bibfnamefont {A.~K.}\
  \bibnamefont {Sood}},\ }\bibfield  {title} {\bibinfo {title} {Probing
  signatures of fractionalization in the candidate quantum spin liquid
  $\mathrm{Cu}_{2}\mathrm{IrO}_{3}$ via anomalous raman scattering},\ }\href
  {https://doi.org/10.1103/PhysRevB.104.184420} {\bibfield  {journal} {\bibinfo
   {journal} {Phys. Rev. B}\ }\textbf {\bibinfo {volume} {104}},\ \bibinfo
  {pages} {184420} (\bibinfo {year} {2021})}\BibitemShut {NoStop}%
\bibitem [{\citenamefont {Imai}\ \emph {et~al.}(2022)\citenamefont {Imai},
  \citenamefont {Nawa}, \citenamefont {Shimizu}, \citenamefont {Yamada},
  \citenamefont {Fujihara}, \citenamefont {Aoyama}, \citenamefont {Takahashi},
  \citenamefont {Okuyama}, \citenamefont {Ohashi}, \citenamefont {Hagihala},
  \citenamefont {Torii}, \citenamefont {Morikawa}, \citenamefont {Terauchi},
  \citenamefont {Kawamata}, \citenamefont {Kato}, \citenamefont {Gotou},
  \citenamefont {Itoh}, \citenamefont {Sato},\ and\ \citenamefont
  {Ohgushi}}]{Yoshinori2022}%
  \BibitemOpen
  \bibfield  {author} {\bibinfo {author} {\bibfnamefont {Y.}~\bibnamefont
  {Imai}}, \bibinfo {author} {\bibfnamefont {K.}~\bibnamefont {Nawa}}, \bibinfo
  {author} {\bibfnamefont {Y.}~\bibnamefont {Shimizu}}, \bibinfo {author}
  {\bibfnamefont {W.}~\bibnamefont {Yamada}}, \bibinfo {author} {\bibfnamefont
  {H.}~\bibnamefont {Fujihara}}, \bibinfo {author} {\bibfnamefont
  {T.}~\bibnamefont {Aoyama}}, \bibinfo {author} {\bibfnamefont
  {R.}~\bibnamefont {Takahashi}}, \bibinfo {author} {\bibfnamefont
  {D.}~\bibnamefont {Okuyama}}, \bibinfo {author} {\bibfnamefont
  {T.}~\bibnamefont {Ohashi}}, \bibinfo {author} {\bibfnamefont
  {M.}~\bibnamefont {Hagihala}}, \bibinfo {author} {\bibfnamefont
  {S.}~\bibnamefont {Torii}}, \bibinfo {author} {\bibfnamefont
  {D.}~\bibnamefont {Morikawa}}, \bibinfo {author} {\bibfnamefont
  {M.}~\bibnamefont {Terauchi}}, \bibinfo {author} {\bibfnamefont
  {T.}~\bibnamefont {Kawamata}}, \bibinfo {author} {\bibfnamefont
  {M.}~\bibnamefont {Kato}}, \bibinfo {author} {\bibfnamefont {H.}~\bibnamefont
  {Gotou}}, \bibinfo {author} {\bibfnamefont {M.}~\bibnamefont {Itoh}},
  \bibinfo {author} {\bibfnamefont {T.~J.}\ \bibnamefont {Sato}},\ and\
  \bibinfo {author} {\bibfnamefont {K.}~\bibnamefont {Ohgushi}},\ }\bibfield
  {title} {\bibinfo {title} {{Zigzag magnetic order in the Kitaev spin-liquid
  candidate material $\mathrm{RuBr}_{3}$ with a honeycomb lattice}},\ }\href
  {https://doi.org/10.1103/PhysRevB.105.L041112} {\bibfield  {journal}
  {\bibinfo  {journal} {Phys. Rev. B}\ }\textbf {\bibinfo {volume} {105}},\
  \bibinfo {pages} {L041112} (\bibinfo {year} {2022})}\BibitemShut {NoStop}%
\bibitem [{\citenamefont {Halloran}\ \emph {et~al.}(2023)\citenamefont
  {Halloran}, \citenamefont {Desrochers}, \citenamefont {Zhang}, \citenamefont
  {Chen}, \citenamefont {Chern}, \citenamefont {Xu}, \citenamefont {Winn},
  \citenamefont {Graves-Brook}, \citenamefont {Stone}, \citenamefont
  {Kolesnikov}, \citenamefont {Qiu}, \citenamefont {Zhong}, \citenamefont
  {Cava}, \citenamefont {Kim},\ and\ \citenamefont {Broholm}}]{Halloran2023}%
  \BibitemOpen
  \bibfield  {author} {\bibinfo {author} {\bibfnamefont {T.}~\bibnamefont
  {Halloran}}, \bibinfo {author} {\bibfnamefont {F.}~\bibnamefont
  {Desrochers}}, \bibinfo {author} {\bibfnamefont {E.~Z.}\ \bibnamefont
  {Zhang}}, \bibinfo {author} {\bibfnamefont {T.}~\bibnamefont {Chen}},
  \bibinfo {author} {\bibfnamefont {L.~E.}\ \bibnamefont {Chern}}, \bibinfo
  {author} {\bibfnamefont {Z.}~\bibnamefont {Xu}}, \bibinfo {author}
  {\bibfnamefont {B.}~\bibnamefont {Winn}}, \bibinfo {author} {\bibfnamefont
  {M.}~\bibnamefont {Graves-Brook}}, \bibinfo {author} {\bibfnamefont {M.~B.}\
  \bibnamefont {Stone}}, \bibinfo {author} {\bibfnamefont {A.~I.}\ \bibnamefont
  {Kolesnikov}}, \bibinfo {author} {\bibfnamefont {Y.}~\bibnamefont {Qiu}},
  \bibinfo {author} {\bibfnamefont {R.}~\bibnamefont {Zhong}}, \bibinfo
  {author} {\bibfnamefont {R.}~\bibnamefont {Cava}}, \bibinfo {author}
  {\bibfnamefont {Y.~B.}\ \bibnamefont {Kim}},\ and\ \bibinfo {author}
  {\bibfnamefont {C.}~\bibnamefont {Broholm}},\ }\bibfield  {title} {\bibinfo
  {title} {{Geometrical frustration versus Kitaev interactions in
  $\mathrm{BaCo_2(AsO_4)_2}$}},\ }\href
  {https://doi.org/10.1073/pnas.2215509119} {\bibfield  {journal} {\bibinfo
  {journal} {Proceedings of the National Academy of Sciences}\ }\textbf
  {\bibinfo {volume} {120}},\ \bibinfo {pages} {e2215509119} (\bibinfo {year}
  {2023})}\BibitemShut {NoStop}%
\bibitem [{\citenamefont {Gao}\ \emph {et~al.}(2019)\citenamefont {Gao},
  \citenamefont {Chen}, \citenamefont {Tam}, \citenamefont {Huang},
  \citenamefont {Sasmal}, \citenamefont {Adroja}, \citenamefont {Ye},
  \citenamefont {Cao}, \citenamefont {Sala}, \citenamefont {Stone},
  \citenamefont {Baines}, \citenamefont {Verezhak}, \citenamefont {Hu},
  \citenamefont {Chung}, \citenamefont {Xu}, \citenamefont {Cheong},
  \citenamefont {Nallaiyan}, \citenamefont {Spagna}, \citenamefont {Maple},
  \citenamefont {Nevidomskyy}, \citenamefont {Morosan}, \citenamefont {Chen},\
  and\ \citenamefont {Dai}}]{Gao2019}%
  \BibitemOpen
  \bibfield  {author} {\bibinfo {author} {\bibfnamefont {B.}~\bibnamefont
  {Gao}}, \bibinfo {author} {\bibfnamefont {T.}~\bibnamefont {Chen}}, \bibinfo
  {author} {\bibfnamefont {D.~W.}\ \bibnamefont {Tam}}, \bibinfo {author}
  {\bibfnamefont {C.-L.}\ \bibnamefont {Huang}}, \bibinfo {author}
  {\bibfnamefont {K.}~\bibnamefont {Sasmal}}, \bibinfo {author} {\bibfnamefont
  {D.~T.}\ \bibnamefont {Adroja}}, \bibinfo {author} {\bibfnamefont
  {F.}~\bibnamefont {Ye}}, \bibinfo {author} {\bibfnamefont {H.}~\bibnamefont
  {Cao}}, \bibinfo {author} {\bibfnamefont {G.}~\bibnamefont {Sala}}, \bibinfo
  {author} {\bibfnamefont {M.~B.}\ \bibnamefont {Stone}}, \bibinfo {author}
  {\bibfnamefont {C.}~\bibnamefont {Baines}}, \bibinfo {author} {\bibfnamefont
  {J.~A.~T.}\ \bibnamefont {Verezhak}}, \bibinfo {author} {\bibfnamefont
  {H.}~\bibnamefont {Hu}}, \bibinfo {author} {\bibfnamefont {J.-H.}\
  \bibnamefont {Chung}}, \bibinfo {author} {\bibfnamefont {X.}~\bibnamefont
  {Xu}}, \bibinfo {author} {\bibfnamefont {S.-W.}\ \bibnamefont {Cheong}},
  \bibinfo {author} {\bibfnamefont {M.}~\bibnamefont {Nallaiyan}}, \bibinfo
  {author} {\bibfnamefont {S.}~\bibnamefont {Spagna}}, \bibinfo {author}
  {\bibfnamefont {M.~B.}\ \bibnamefont {Maple}}, \bibinfo {author}
  {\bibfnamefont {A.~H.}\ \bibnamefont {Nevidomskyy}}, \bibinfo {author}
  {\bibfnamefont {E.}~\bibnamefont {Morosan}}, \bibinfo {author} {\bibfnamefont
  {G.}~\bibnamefont {Chen}},\ and\ \bibinfo {author} {\bibfnamefont
  {P.}~\bibnamefont {Dai}},\ }\bibfield  {title} {\bibinfo {title}
  {Experimental signatures of a three-dimensional quantum spin liquid in
  effective spin-1/2 $\mathrm{Ce}_2\mathrm{Zr}_2\mathrm{O}_7$ pyrochlore},\
  }\href {https://doi.org/10.1038/s41567-019-0577-6} {\bibfield  {journal}
  {\bibinfo  {journal} {Nature Physics}\ }\textbf {\bibinfo {volume} {15}},\
  \bibinfo {pages} {1052} (\bibinfo {year} {2019})}\BibitemShut {NoStop}%
\bibitem [{\citenamefont {Chern}\ \emph {et~al.}(2022)\citenamefont {Chern},
  \citenamefont {Kim},\ and\ \citenamefont {Castelnovo}}]{Chern2022}%
  \BibitemOpen
  \bibfield  {author} {\bibinfo {author} {\bibfnamefont {L.~E.}\ \bibnamefont
  {Chern}}, \bibinfo {author} {\bibfnamefont {Y.~B.}\ \bibnamefont {Kim}},\
  and\ \bibinfo {author} {\bibfnamefont {C.}~\bibnamefont {Castelnovo}},\
  }\bibfield  {title} {\bibinfo {title} {Competing quantum spin liquids, gauge
  fluctuations, and anisotropic interactions in a breathing pyrochlore
  lattice},\ }\href {https://doi.org/10.1103/PhysRevB.106.134402} {\bibfield
  {journal} {\bibinfo  {journal} {Phys. Rev. B}\ }\textbf {\bibinfo {volume}
  {106}},\ \bibinfo {pages} {134402} (\bibinfo {year} {2022})}\BibitemShut
  {NoStop}%
\bibitem [{\citenamefont {Stavropoulos}\ \emph {et~al.}(2019)\citenamefont
  {Stavropoulos}, \citenamefont {Pereira},\ and\ \citenamefont
  {Kee}}]{Stavropoulos.P.Peter2019}%
  \BibitemOpen
  \bibfield  {author} {\bibinfo {author} {\bibfnamefont {P.~P.}\ \bibnamefont
  {Stavropoulos}}, \bibinfo {author} {\bibfnamefont {D.}~\bibnamefont
  {Pereira}},\ and\ \bibinfo {author} {\bibfnamefont {H.-Y.}\ \bibnamefont
  {Kee}},\ }\bibfield  {title} {\bibinfo {title} {Microscopic mechanism for a
  higher-spin {K}itaev model},\ }\href
  {https://doi.org/10.1103/PhysRevLett.123.037203} {\bibfield  {journal}
  {\bibinfo  {journal} {Phys. Rev. Lett.}\ }\textbf {\bibinfo {volume} {123}},\
  \bibinfo {pages} {037203} (\bibinfo {year} {2019})}\BibitemShut {NoStop}%
\bibitem [{\citenamefont {Koga}\ \emph {et~al.}(2018)\citenamefont {Koga},
  \citenamefont {Tomishige},\ and\ \citenamefont {Nasu}}]{Koga2018}%
  \BibitemOpen
  \bibfield  {author} {\bibinfo {author} {\bibfnamefont {A.}~\bibnamefont
  {Koga}}, \bibinfo {author} {\bibfnamefont {H.}~\bibnamefont {Tomishige}},\
  and\ \bibinfo {author} {\bibfnamefont {J.}~\bibnamefont {Nasu}},\ }\bibfield
  {title} {\bibinfo {title} {{Ground-state and Thermodynamic Properties of an S
  = 1 Kitaev Model}},\ }\href {https://doi.org/10.7566/JPSJ.87.063703}
  {\bibfield  {journal} {\bibinfo  {journal} {Journal of the Physical Society
  of Japan}\ }\textbf {\bibinfo {volume} {87}},\ \bibinfo {pages} {063703}
  (\bibinfo {year} {2018})}\BibitemShut {NoStop}%
\bibitem [{\citenamefont {Lee}\ \emph {et~al.}(2020)\citenamefont {Lee},
  \citenamefont {Kawashima},\ and\ \citenamefont {Kim}}]{Lee2020}%
  \BibitemOpen
  \bibfield  {author} {\bibinfo {author} {\bibfnamefont {H.-Y.}\ \bibnamefont
  {Lee}}, \bibinfo {author} {\bibfnamefont {N.}~\bibnamefont {Kawashima}},\
  and\ \bibinfo {author} {\bibfnamefont {Y.~B.}\ \bibnamefont {Kim}},\
  }\bibfield  {title} {\bibinfo {title} {{Tensor network wave function of S=1
  Kitaev spin liquids}},\ }\href
  {https://doi.org/10.1103/PhysRevResearch.2.033318} {\bibfield  {journal}
  {\bibinfo  {journal} {Phys. Rev. Res.}\ }\textbf {\bibinfo {volume} {2}},\
  \bibinfo {pages} {033318} (\bibinfo {year} {2020})}\BibitemShut {NoStop}%
\bibitem [{\citenamefont {Chen}\ \emph {et~al.}(2022)\citenamefont {Chen},
  \citenamefont {Genzor}, \citenamefont {Kim},\ and\ \citenamefont
  {Kao}}]{Chen2022}%
  \BibitemOpen
  \bibfield  {author} {\bibinfo {author} {\bibfnamefont {Y.-H.}\ \bibnamefont
  {Chen}}, \bibinfo {author} {\bibfnamefont {J.}~\bibnamefont {Genzor}},
  \bibinfo {author} {\bibfnamefont {Y.~B.}\ \bibnamefont {Kim}},\ and\ \bibinfo
  {author} {\bibfnamefont {Y.-J.}\ \bibnamefont {Kao}},\ }\bibfield  {title}
  {\bibinfo {title} {{Excitation spectrum of spin-1 Kitaev spin liquids}},\
  }\href {https://doi.org/10.1103/PhysRevB.105.L060403} {\bibfield  {journal}
  {\bibinfo  {journal} {Phys. Rev. B}\ }\textbf {\bibinfo {volume} {105}},\
  \bibinfo {pages} {L060403} (\bibinfo {year} {2022})}\BibitemShut {NoStop}%
\bibitem [{\citenamefont {Pohle}\ \emph {et~al.}(2023)\citenamefont {Pohle},
  \citenamefont {Shannon},\ and\ \citenamefont {Motome}}]{Pohle2023}%
  \BibitemOpen
  \bibfield  {author} {\bibinfo {author} {\bibfnamefont {R.}~\bibnamefont
  {Pohle}}, \bibinfo {author} {\bibfnamefont {N.}~\bibnamefont {Shannon}},\
  and\ \bibinfo {author} {\bibfnamefont {Y.}~\bibnamefont {Motome}},\
  }\bibfield  {title} {\bibinfo {title} {{Spin nematics meet spin liquids:
  Exotic quantum phases in the spin-1 bilinear-biquadratic model with Kitaev
  interactions}},\ }\href {https://doi.org/10.1103/PhysRevB.107.L140403}
  {\bibfield  {journal} {\bibinfo  {journal} {Phys. Rev. B}\ }\textbf {\bibinfo
  {volume} {107}},\ \bibinfo {pages} {L140403} (\bibinfo {year}
  {2023})}\BibitemShut {NoStop}%
\bibitem [{\citenamefont {Taddei}\ \emph {et~al.}(2023)\citenamefont {Taddei},
  \citenamefont {Garlea}, \citenamefont {Samarakoon}, \citenamefont {Sanjeewa},
  \citenamefont {Xing}, \citenamefont {Heitmann}, \citenamefont {dela Cruz},
  \citenamefont {Sefat},\ and\ \citenamefont {Parker}}]{Taddei2023}%
  \BibitemOpen
  \bibfield  {author} {\bibinfo {author} {\bibfnamefont {K.~M.}\ \bibnamefont
  {Taddei}}, \bibinfo {author} {\bibfnamefont {V.~O.}\ \bibnamefont {Garlea}},
  \bibinfo {author} {\bibfnamefont {A.~M.}\ \bibnamefont {Samarakoon}},
  \bibinfo {author} {\bibfnamefont {L.~D.}\ \bibnamefont {Sanjeewa}}, \bibinfo
  {author} {\bibfnamefont {J.}~\bibnamefont {Xing}}, \bibinfo {author}
  {\bibfnamefont {T.~W.}\ \bibnamefont {Heitmann}}, \bibinfo {author}
  {\bibfnamefont {C.}~\bibnamefont {dela Cruz}}, \bibinfo {author}
  {\bibfnamefont {A.~S.}\ \bibnamefont {Sefat}},\ and\ \bibinfo {author}
  {\bibfnamefont {D.}~\bibnamefont {Parker}},\ }\bibfield  {title} {\bibinfo
  {title} {{Zigzag magnetic order and possible Kitaev interactions in the
  spin-1 honeycomb lattice $\mathrm{KNiAsO}_{4}$}},\ }\href
  {https://doi.org/10.1103/PhysRevResearch.5.013022} {\bibfield  {journal}
  {\bibinfo  {journal} {Phys. Rev. Res.}\ }\textbf {\bibinfo {volume} {5}},\
  \bibinfo {pages} {013022} (\bibinfo {year} {2023})}\BibitemShut {NoStop}%
\bibitem [{\citenamefont {Mohapatra}\ and\ \citenamefont
  {Balram}(2023)}]{Mohapatra2023}%
  \BibitemOpen
  \bibfield  {author} {\bibinfo {author} {\bibfnamefont {S.}~\bibnamefont
  {Mohapatra}}\ and\ \bibinfo {author} {\bibfnamefont {A.~C.}\ \bibnamefont
  {Balram}},\ }\bibfield  {title} {\bibinfo {title} {Pronounced quantum
  many-body scars in the one-dimensional spin-1 {K}itaev model},\ }\href
  {https://doi.org/10.1103/PhysRevB.107.235121} {\bibfield  {journal} {\bibinfo
   {journal} {Phys. Rev. B}\ }\textbf {\bibinfo {volume} {107}},\ \bibinfo
  {pages} {235121} (\bibinfo {year} {2023})}\BibitemShut {NoStop}%
\bibitem [{\citenamefont {Xu}\ \emph {et~al.}(2020)\citenamefont {Xu},
  \citenamefont {Feng}, \citenamefont {Kawamura}, \citenamefont {Yamaji},
  \citenamefont {Nahas}, \citenamefont {Prokhorenko}, \citenamefont {Qi},
  \citenamefont {Xiang},\ and\ \citenamefont {Bellaiche}}]{Xu2020}%
  \BibitemOpen
  \bibfield  {author} {\bibinfo {author} {\bibfnamefont {C.}~\bibnamefont
  {Xu}}, \bibinfo {author} {\bibfnamefont {J.}~\bibnamefont {Feng}}, \bibinfo
  {author} {\bibfnamefont {M.}~\bibnamefont {Kawamura}}, \bibinfo {author}
  {\bibfnamefont {Y.}~\bibnamefont {Yamaji}}, \bibinfo {author} {\bibfnamefont
  {Y.}~\bibnamefont {Nahas}}, \bibinfo {author} {\bibfnamefont
  {S.}~\bibnamefont {Prokhorenko}}, \bibinfo {author} {\bibfnamefont
  {Y.}~\bibnamefont {Qi}}, \bibinfo {author} {\bibfnamefont {H.}~\bibnamefont
  {Xiang}},\ and\ \bibinfo {author} {\bibfnamefont {L.}~\bibnamefont
  {Bellaiche}},\ }\bibfield  {title} {\bibinfo {title} {{Possible Kitaev
  Quantum Spin Liquid State in 2D Materials with $S=3/2$}},\ }\href
  {https://doi.org/10.1103/PhysRevLett.124.087205} {\bibfield  {journal}
  {\bibinfo  {journal} {Phys. Rev. Lett.}\ }\textbf {\bibinfo {volume} {124}},\
  \bibinfo {pages} {087205} (\bibinfo {year} {2020})}\BibitemShut {NoStop}%
\bibitem [{\citenamefont {Jin}\ \emph {et~al.}(2022)\citenamefont {Jin},
  \citenamefont {Natori}, \citenamefont {Pollmann},\ and\ \citenamefont
  {Knolle}}]{Jin2022}%
  \BibitemOpen
  \bibfield  {author} {\bibinfo {author} {\bibfnamefont {H.-K.}\ \bibnamefont
  {Jin}}, \bibinfo {author} {\bibfnamefont {W.~M.~H.}\ \bibnamefont {Natori}},
  \bibinfo {author} {\bibfnamefont {F.}~\bibnamefont {Pollmann}},\ and\
  \bibinfo {author} {\bibfnamefont {J.}~\bibnamefont {Knolle}},\ }\bibfield
  {title} {\bibinfo {title} {{Unveiling the S=3/2 Kitaev honeycomb spin
  liquids}},\ }\href {https://doi.org/10.1038/s41467-022-31503-0} {\bibfield
  {journal} {\bibinfo  {journal} {Nature Communications}\ }\textbf {\bibinfo
  {volume} {13}},\ \bibinfo {pages} {3813} (\bibinfo {year}
  {2022})}\BibitemShut {NoStop}%
\bibitem [{\citenamefont {Natori}\ \emph {et~al.}()\citenamefont {Natori},
  \citenamefont {Jin},\ and\ \citenamefont {Knolle}}]{natori2023quantum}%
  \BibitemOpen
  \bibfield  {author} {\bibinfo {author} {\bibfnamefont {W.~M.~H.}\
  \bibnamefont {Natori}}, \bibinfo {author} {\bibfnamefont {H.-K.}\
  \bibnamefont {Jin}},\ and\ \bibinfo {author} {\bibfnamefont {J.}~\bibnamefont
  {Knolle}},\ }\href@noop {} {\bibinfo {title} {Quantum liquids of the {S}=3/2
  kitaev honeycomb and related kugel-khomskii models}},\ \Eprint
  {https://arxiv.org/abs/2304.13378 (2023)} {arXiv:2304.13378 (2023)}
  \BibitemShut {NoStop}%
\bibitem [{\citenamefont {Fukui}\ \emph {et~al.}(2022)\citenamefont {Fukui},
  \citenamefont {Kato}, \citenamefont {Nasu},\ and\ \citenamefont
  {Motome}}]{Fukui2022}%
  \BibitemOpen
  \bibfield  {author} {\bibinfo {author} {\bibfnamefont {K.}~\bibnamefont
  {Fukui}}, \bibinfo {author} {\bibfnamefont {Y.}~\bibnamefont {Kato}},
  \bibinfo {author} {\bibfnamefont {J.}~\bibnamefont {Nasu}},\ and\ \bibinfo
  {author} {\bibfnamefont {Y.}~\bibnamefont {Motome}},\ }\bibfield  {title}
  {\bibinfo {title} {{Ground-state phase diagram of spin-$S$ Kitaev-Heisenberg
  models}},\ }\href {https://doi.org/10.1103/PhysRevB.106.174416} {\bibfield
  {journal} {\bibinfo  {journal} {Phys. Rev. B}\ }\textbf {\bibinfo {volume}
  {106}},\ \bibinfo {pages} {174416} (\bibinfo {year} {2022})}\BibitemShut
  {NoStop}%
\bibitem [{\citenamefont {Xu}\ \emph {et~al.}(2018{\natexlab{b}})\citenamefont
  {Xu}, \citenamefont {Feng}, \citenamefont {Xiang},\ and\ \citenamefont
  {Bellaiche}}]{Xu2018}%
  \BibitemOpen
  \bibfield  {author} {\bibinfo {author} {\bibfnamefont {C.}~\bibnamefont
  {Xu}}, \bibinfo {author} {\bibfnamefont {J.}~\bibnamefont {Feng}}, \bibinfo
  {author} {\bibfnamefont {H.}~\bibnamefont {Xiang}},\ and\ \bibinfo {author}
  {\bibfnamefont {L.}~\bibnamefont {Bellaiche}},\ }\bibfield  {title} {\bibinfo
  {title} {Interplay between {K}itaev interaction and single ion anisotropy in
  ferromagnetic $\mathrm{CrI}_3$ and $\mathrm{CrGeTe}_3$ monolayers},\ }\href
  {https://doi.org/10.1038/s41524-018-0115-6} {\bibfield  {journal} {\bibinfo
  {journal} {npj Computational Materials}\ }\textbf {\bibinfo {volume} {4}},\
  \bibinfo {pages} {57} (\bibinfo {year} {2018}{\natexlab{b}})}\BibitemShut
  {NoStop}%
\bibitem [{\citenamefont {Bradley}\ and\ \citenamefont
  {Singh}(2022)}]{Bradley.Owen.2022}%
  \BibitemOpen
  \bibfield  {author} {\bibinfo {author} {\bibfnamefont {O.}~\bibnamefont
  {Bradley}}\ and\ \bibinfo {author} {\bibfnamefont {R.~R.~P.}\ \bibnamefont
  {Singh}},\ }\bibfield  {title} {\bibinfo {title} {Instabilities of spin-1
  {K}itaev spin liquid phase in presence of single-ion anisotropies},\ }\href
  {https://doi.org/10.1103/PhysRevB.105.L060405} {\bibfield  {journal}
  {\bibinfo  {journal} {Phys. Rev. B}\ }\textbf {\bibinfo {volume} {105}},\
  \bibinfo {pages} {L060405} (\bibinfo {year} {2022})}\BibitemShut {NoStop}%
\bibitem [{\citenamefont {S\o{}rensen}\ \emph {et~al.}(2023)\citenamefont
  {S\o{}rensen}, \citenamefont {Riddell},\ and\ \citenamefont
  {Kee}}]{Erik2023}%
  \BibitemOpen
  \bibfield  {author} {\bibinfo {author} {\bibfnamefont {E.~S.}\ \bibnamefont
  {S\o{}rensen}}, \bibinfo {author} {\bibfnamefont {J.}~\bibnamefont
  {Riddell}},\ and\ \bibinfo {author} {\bibfnamefont {H.-Y.}\ \bibnamefont
  {Kee}},\ }\bibfield  {title} {\bibinfo {title} {{Islands of chiral solitons
  in integer-spin Kitaev chains}},\ }\href
  {https://doi.org/10.1103/PhysRevResearch.5.013210} {\bibfield  {journal}
  {\bibinfo  {journal} {Phys. Rev. Res.}\ }\textbf {\bibinfo {volume} {5}},\
  \bibinfo {pages} {013210} (\bibinfo {year} {2023})}\BibitemShut {NoStop}%
\bibitem [{\citenamefont {Fishman}(2021)}]{Fishman.Randy.S.2021}%
  \BibitemOpen
  \bibfield  {author} {\bibinfo {author} {\bibfnamefont {R.~S.}\ \bibnamefont
  {Fishman}},\ }\bibfield  {title} {\bibinfo {title} {Single-ion anisotropy is
  necessary and appropriate to study the magnetic behavior of
  $\mathrm{Tb}^{3+}$ moments with ${J}_{\text{eff}}=\frac{1}{2}$ on the
  honeycomb lattice in $\mathrm{Tb}_{2}\mathrm{Ir}_{3}\mathrm{Ga}_{9}$},\
  }\href {https://doi.org/10.1103/PhysRevB.103.214440} {\bibfield  {journal}
  {\bibinfo  {journal} {Phys. Rev. B}\ }\textbf {\bibinfo {volume} {103}},\
  \bibinfo {pages} {214440} (\bibinfo {year} {2021})}\BibitemShut {NoStop}%
\bibitem [{\citenamefont {Zaletel}\ \emph {et~al.}(2015)\citenamefont
  {Zaletel}, \citenamefont {Mong}, \citenamefont {Karrasch}, \citenamefont
  {Moore},\ and\ \citenamefont {Pollmann}}]{Zaletel2015}%
  \BibitemOpen
  \bibfield  {author} {\bibinfo {author} {\bibfnamefont {M.~P.}\ \bibnamefont
  {Zaletel}}, \bibinfo {author} {\bibfnamefont {R.~S.~K.}\ \bibnamefont
  {Mong}}, \bibinfo {author} {\bibfnamefont {C.}~\bibnamefont {Karrasch}},
  \bibinfo {author} {\bibfnamefont {J.~E.}\ \bibnamefont {Moore}},\ and\
  \bibinfo {author} {\bibfnamefont {F.}~\bibnamefont {Pollmann}},\ }\bibfield
  {title} {\bibinfo {title} {Time-evolving a matrix product state with
  long-ranged interactions},\ }\href
  {https://doi.org/10.1103/PhysRevB.91.165112} {\bibfield  {journal} {\bibinfo
  {journal} {Phys. Rev. B}\ }\textbf {\bibinfo {volume} {91}},\ \bibinfo
  {pages} {165112} (\bibinfo {year} {2015})}\BibitemShut {NoStop}%
\bibitem [{\citenamefont {Fishman}\ \emph {et~al.}(2022)\citenamefont
  {Fishman}, \citenamefont {White},\ and\ \citenamefont
  {Stoudenmire}}]{ITensor}%
  \BibitemOpen
  \bibfield  {author} {\bibinfo {author} {\bibfnamefont {M.}~\bibnamefont
  {Fishman}}, \bibinfo {author} {\bibfnamefont {S.~R.}\ \bibnamefont {White}},\
  and\ \bibinfo {author} {\bibfnamefont {E.~M.}\ \bibnamefont {Stoudenmire}},\
  }\bibfield  {title} {\bibinfo {title} {{The ITensor Software Library for
  Tensor Network Calculations}},\ }\href
  {https://doi.org/10.21468/SciPostPhysCodeb.4} {\bibfield  {journal} {\bibinfo
   {journal} {SciPost Phys. Codebases}\ ,\ \bibinfo {pages} {4}} (\bibinfo
  {year} {2022})}\BibitemShut {NoStop}%
\bibitem [{\citenamefont {Vidal}(2007)}]{VidaliTEBD2007}%
  \BibitemOpen
  \bibfield  {author} {\bibinfo {author} {\bibfnamefont {G.}~\bibnamefont
  {Vidal}},\ }\bibfield  {title} {\bibinfo {title} {Classical simulation of
  infinite-size quantum lattice systems in one spatial dimension},\ }\href
  {https://doi.org/10.1103/PhysRevLett.98.070201} {\bibfield  {journal}
  {\bibinfo  {journal} {Phys. Rev. Lett.}\ }\textbf {\bibinfo {volume} {98}},\
  \bibinfo {pages} {070201} (\bibinfo {year} {2007})}\BibitemShut {NoStop}%
\bibitem [{\citenamefont {Sen}\ \emph {et~al.}(2010)\citenamefont {Sen},
  \citenamefont {Shankar}, \citenamefont {Dhar},\ and\ \citenamefont
  {Ramola}}]{Sen2010}%
  \BibitemOpen
  \bibfield  {author} {\bibinfo {author} {\bibfnamefont {D.}~\bibnamefont
  {Sen}}, \bibinfo {author} {\bibfnamefont {R.}~\bibnamefont {Shankar}},
  \bibinfo {author} {\bibfnamefont {D.}~\bibnamefont {Dhar}},\ and\ \bibinfo
  {author} {\bibfnamefont {K.}~\bibnamefont {Ramola}},\ }\bibfield  {title}
  {\bibinfo {title} {Spin-1 {K}itaev model in one dimension},\ }\href
  {https://doi.org/10.1103/PhysRevB.82.195435} {\bibfield  {journal} {\bibinfo
  {journal} {Phys. Rev. B}\ }\textbf {\bibinfo {volume} {82}},\ \bibinfo
  {pages} {195435} (\bibinfo {year} {2010})}\BibitemShut {NoStop}%
\bibitem [{\citenamefont {Sugita}\ \emph {et~al.}(2020)\citenamefont {Sugita},
  \citenamefont {Kato},\ and\ \citenamefont {Motome}}]{Sugita.Yusuke2020}%
  \BibitemOpen
  \bibfield  {author} {\bibinfo {author} {\bibfnamefont {Y.}~\bibnamefont
  {Sugita}}, \bibinfo {author} {\bibfnamefont {Y.}~\bibnamefont {Kato}},\ and\
  \bibinfo {author} {\bibfnamefont {Y.}~\bibnamefont {Motome}},\ }\bibfield
  {title} {\bibinfo {title} {Antiferromagnetic {K}itaev interactions in polar
  spin-orbit {M}ott insulators},\ }\href
  {https://doi.org/10.1103/PhysRevB.101.100410} {\bibfield  {journal} {\bibinfo
   {journal} {Phys. Rev. B}\ }\textbf {\bibinfo {volume} {101}},\ \bibinfo
  {pages} {100410} (\bibinfo {year} {2020})}\BibitemShut {NoStop}%
\bibitem [{\citenamefont {Sears}\ \emph {et~al.}(2020)\citenamefont {Sears},
  \citenamefont {Chern}, \citenamefont {Kim}, \citenamefont {Bereciartua},
  \citenamefont {Francoual}, \citenamefont {Kim},\ and\ \citenamefont
  {Kim}}]{Sears2020}%
  \BibitemOpen
  \bibfield  {author} {\bibinfo {author} {\bibfnamefont {J.~A.}\ \bibnamefont
  {Sears}}, \bibinfo {author} {\bibfnamefont {L.~E.}\ \bibnamefont {Chern}},
  \bibinfo {author} {\bibfnamefont {S.}~\bibnamefont {Kim}}, \bibinfo {author}
  {\bibfnamefont {P.~J.}\ \bibnamefont {Bereciartua}}, \bibinfo {author}
  {\bibfnamefont {S.}~\bibnamefont {Francoual}}, \bibinfo {author}
  {\bibfnamefont {Y.~B.}\ \bibnamefont {Kim}},\ and\ \bibinfo {author}
  {\bibfnamefont {Y.-J.}\ \bibnamefont {Kim}},\ }\bibfield  {title} {\bibinfo
  {title} {Ferromagnetic {K}itaev interaction and the origin of large magnetic
  anisotropy in $\alpha$-$\mathrm{RuCl}_3$},\ }\href
  {https://doi.org/10.1038/s41567-020-0874-0} {\bibfield  {journal} {\bibinfo
  {journal} {Nature Physics}\ }\textbf {\bibinfo {volume} {16}},\ \bibinfo
  {pages} {837} (\bibinfo {year} {2020})}\BibitemShut {NoStop}%
\bibitem [{\citenamefont {Moudgalya}\ \emph
  {et~al.}(2020{\natexlab{b}})\citenamefont {Moudgalya}, \citenamefont
  {Bernevig},\ and\ \citenamefont {Regnault}}]{Moudgalya2020_2}%
  \BibitemOpen
  \bibfield  {author} {\bibinfo {author} {\bibfnamefont {S.}~\bibnamefont
  {Moudgalya}}, \bibinfo {author} {\bibfnamefont {B.~A.}\ \bibnamefont
  {Bernevig}},\ and\ \bibinfo {author} {\bibfnamefont {N.}~\bibnamefont
  {Regnault}},\ }\bibfield  {title} {\bibinfo {title} {{Quantum many-body scars
  in a Landau level on a thin torus}},\ }\href
  {https://doi.org/10.1103/PhysRevB.102.195150} {\bibfield  {journal} {\bibinfo
   {journal} {Phys. Rev. B}\ }\textbf {\bibinfo {volume} {102}},\ \bibinfo
  {pages} {195150} (\bibinfo {year} {2020}{\natexlab{b}})}\BibitemShut
  {NoStop}%
\bibitem [{\citenamefont {Daniel}\ \emph {et~al.}()\citenamefont {Daniel},
  \citenamefont {Hallam}, \citenamefont {Desaules}, \citenamefont {Hudomal},
  \citenamefont {Su}, \citenamefont {Halimeh},\ and\ \citenamefont
  {Papić}}]{Daniel2023}%
  \BibitemOpen
  \bibfield  {author} {\bibinfo {author} {\bibfnamefont {A.}~\bibnamefont
  {Daniel}}, \bibinfo {author} {\bibfnamefont {A.}~\bibnamefont {Hallam}},
  \bibinfo {author} {\bibfnamefont {J.-Y.}\ \bibnamefont {Desaules}}, \bibinfo
  {author} {\bibfnamefont {A.}~\bibnamefont {Hudomal}}, \bibinfo {author}
  {\bibfnamefont {G.-X.}\ \bibnamefont {Su}}, \bibinfo {author} {\bibfnamefont
  {J.~C.}\ \bibnamefont {Halimeh}},\ and\ \bibinfo {author} {\bibfnamefont
  {Z.}~\bibnamefont {Papić}},\ }\href@noop {} {\bibinfo {title} {Bridging
  quantum criticality via many-body scarring}},\ \Eprint
  {https://arxiv.org/abs/2301.03631 (2023)} {arXiv:2301.03631 (2023)}
  \BibitemShut {NoStop}%
\bibitem [{myn()}]{mynote}%
  \BibitemOpen
  \href@noop {} {\bibinfo {title} {The data and the code that support the
  findings of this study are available from the corresponding authors upon
  reasonable request.}}\BibitemShut {Stop}%
\bibitem [{\citenamefont {Byrnes}\ \emph {et~al.}(2002)\citenamefont {Byrnes},
  \citenamefont {Sriganesh}, \citenamefont {Bursill},\ and\ \citenamefont
  {Hamer}}]{Byrnes2002}%
  \BibitemOpen
  \bibfield  {author} {\bibinfo {author} {\bibfnamefont {T.~M.~R.}\
  \bibnamefont {Byrnes}}, \bibinfo {author} {\bibfnamefont {P.}~\bibnamefont
  {Sriganesh}}, \bibinfo {author} {\bibfnamefont {R.~J.}\ \bibnamefont
  {Bursill}},\ and\ \bibinfo {author} {\bibfnamefont {C.~J.}\ \bibnamefont
  {Hamer}},\ }\bibfield  {title} {\bibinfo {title} {{Density matrix
  renormalization group approach to the massive Schwinger model}},\ }\href
  {https://doi.org/10.1103/PhysRevD.66.013002} {\bibfield  {journal} {\bibinfo
  {journal} {Phys. Rev. D}\ }\textbf {\bibinfo {volume} {66}},\ \bibinfo
  {pages} {013002} (\bibinfo {year} {2002})}\BibitemShut {NoStop}%
\bibitem [{\citenamefont {Rico}\ \emph {et~al.}(2014)\citenamefont {Rico},
  \citenamefont {Pichler}, \citenamefont {Dalmonte}, \citenamefont {Zoller},\
  and\ \citenamefont {Montangero}}]{Rico2014}%
  \BibitemOpen
  \bibfield  {author} {\bibinfo {author} {\bibfnamefont {E.}~\bibnamefont
  {Rico}}, \bibinfo {author} {\bibfnamefont {T.}~\bibnamefont {Pichler}},
  \bibinfo {author} {\bibfnamefont {M.}~\bibnamefont {Dalmonte}}, \bibinfo
  {author} {\bibfnamefont {P.}~\bibnamefont {Zoller}},\ and\ \bibinfo {author}
  {\bibfnamefont {S.}~\bibnamefont {Montangero}},\ }\bibfield  {title}
  {\bibinfo {title} {Tensor networks for lattice gauge theories and atomic
  quantum simulation},\ }\href {https://doi.org/10.1103/PhysRevLett.112.201601}
  {\bibfield  {journal} {\bibinfo  {journal} {Phys. Rev. Lett.}\ }\textbf
  {\bibinfo {volume} {112}},\ \bibinfo {pages} {201601} (\bibinfo {year}
  {2014})}\BibitemShut {NoStop}%
\bibitem [{\citenamefont {Yang}\ \emph {et~al.}(2020)\citenamefont {Yang},
  \citenamefont {Sun}, \citenamefont {Ott}, \citenamefont {Wang}, \citenamefont
  {Zache}, \citenamefont {Halimeh}, \citenamefont {Yuan}, \citenamefont
  {Hauke},\ and\ \citenamefont {Pan}}]{Yang2020}%
  \BibitemOpen
  \bibfield  {author} {\bibinfo {author} {\bibfnamefont {B.}~\bibnamefont
  {Yang}}, \bibinfo {author} {\bibfnamefont {H.}~\bibnamefont {Sun}}, \bibinfo
  {author} {\bibfnamefont {R.}~\bibnamefont {Ott}}, \bibinfo {author}
  {\bibfnamefont {H.-Y.}\ \bibnamefont {Wang}}, \bibinfo {author}
  {\bibfnamefont {T.~V.}\ \bibnamefont {Zache}}, \bibinfo {author}
  {\bibfnamefont {J.~C.}\ \bibnamefont {Halimeh}}, \bibinfo {author}
  {\bibfnamefont {Z.-S.}\ \bibnamefont {Yuan}}, \bibinfo {author}
  {\bibfnamefont {P.}~\bibnamefont {Hauke}},\ and\ \bibinfo {author}
  {\bibfnamefont {J.-W.}\ \bibnamefont {Pan}},\ }\bibfield  {title} {\bibinfo
  {title} {{Observation of gauge invariance in a 71-site Bose--Hubbard quantum
  simulator}},\ }\href {https://doi.org/10.1038/s41586-020-2910-8} {\bibfield
  {journal} {\bibinfo  {journal} {Nature}\ }\textbf {\bibinfo {volume} {587}},\
  \bibinfo {pages} {392} (\bibinfo {year} {2020})}\BibitemShut {NoStop}%
\bibitem [{\citenamefont {Damme}\ \emph {et~al.}()\citenamefont {Damme},
  \citenamefont {Halimeh},\ and\ \citenamefont {Hauke}}]{VanDamme2020}%
  \BibitemOpen
  \bibfield  {author} {\bibinfo {author} {\bibfnamefont {M.~V.}\ \bibnamefont
  {Damme}}, \bibinfo {author} {\bibfnamefont {J.~C.}\ \bibnamefont {Halimeh}},\
  and\ \bibinfo {author} {\bibfnamefont {P.}~\bibnamefont {Hauke}},\
  }\href@noop {} {\bibinfo {title} {Gauge-symmetry violation quantum phase
  transition in lattice gauge theories}},\ \Eprint
  {https://arxiv.org/abs/2010.07338 (2020)} {arXiv:2010.07338 (2020)}
  \BibitemShut {NoStop}%
\bibitem [{\citenamefont {Lin}\ \emph {et~al.}(2020{\natexlab{b}})\citenamefont
  {Lin}, \citenamefont {Chandran},\ and\ \citenamefont
  {Motrunich}}]{LinChengJu2020}%
  \BibitemOpen
  \bibfield  {author} {\bibinfo {author} {\bibfnamefont {C.-J.}\ \bibnamefont
  {Lin}}, \bibinfo {author} {\bibfnamefont {A.}~\bibnamefont {Chandran}},\ and\
  \bibinfo {author} {\bibfnamefont {O.~I.}\ \bibnamefont {Motrunich}},\
  }\bibfield  {title} {\bibinfo {title} {Slow thermalization of exact quantum
  many-body scar states under perturbations},\ }\href
  {https://doi.org/10.1103/PhysRevResearch.2.033044} {\bibfield  {journal}
  {\bibinfo  {journal} {Phys. Rev. Res.}\ }\textbf {\bibinfo {volume} {2}},\
  \bibinfo {pages} {033044} (\bibinfo {year} {2020}{\natexlab{b}})}\BibitemShut
  {NoStop}%
\bibitem [{\citenamefont {Langlett}\ \emph {et~al.}(2022)\citenamefont
  {Langlett}, \citenamefont {Yang}, \citenamefont {Wildeboer}, \citenamefont
  {Gorshkov}, \citenamefont {Iadecola},\ and\ \citenamefont
  {Xu}}]{Langlett2022}%
  \BibitemOpen
  \bibfield  {author} {\bibinfo {author} {\bibfnamefont {C.~M.}\ \bibnamefont
  {Langlett}}, \bibinfo {author} {\bibfnamefont {Z.-C.}\ \bibnamefont {Yang}},
  \bibinfo {author} {\bibfnamefont {J.}~\bibnamefont {Wildeboer}}, \bibinfo
  {author} {\bibfnamefont {A.~V.}\ \bibnamefont {Gorshkov}}, \bibinfo {author}
  {\bibfnamefont {T.}~\bibnamefont {Iadecola}},\ and\ \bibinfo {author}
  {\bibfnamefont {S.}~\bibnamefont {Xu}},\ }\bibfield  {title} {\bibinfo
  {title} {{Rainbow scars: From area to volume law}},\ }\href
  {https://doi.org/10.1103/PhysRevB.105.L060301} {\bibfield  {journal}
  {\bibinfo  {journal} {Phys. Rev. B}\ }\textbf {\bibinfo {volume} {105}},\
  \bibinfo {pages} {L060301} (\bibinfo {year} {2022})}\BibitemShut {NoStop}%
\bibitem [{\citenamefont {Gotta}\ \emph {et~al.}()\citenamefont {Gotta},
  \citenamefont {Moudgalya},\ and\ \citenamefont
  {Mazza}}]{gotta2023asymptotic}%
  \BibitemOpen
  \bibfield  {author} {\bibinfo {author} {\bibfnamefont {L.}~\bibnamefont
  {Gotta}}, \bibinfo {author} {\bibfnamefont {S.}~\bibnamefont {Moudgalya}},\
  and\ \bibinfo {author} {\bibfnamefont {L.}~\bibnamefont {Mazza}},\
  }\href@noop {} {\bibinfo {title} {Asymptotic quantum many-body scars}},\
  \Eprint {https://arxiv.org/abs/2303.05407 (2023)} {arXiv:2303.05407 (2023)}
  \BibitemShut {NoStop}%
\bibitem [{\citenamefont {den Nijs}\ and\ \citenamefont
  {Rommelse}(1989)}]{den.Nijs.Marcel1989}%
  \BibitemOpen
  \bibfield  {author} {\bibinfo {author} {\bibfnamefont {M.}~\bibnamefont {den
  Nijs}}\ and\ \bibinfo {author} {\bibfnamefont {K.}~\bibnamefont {Rommelse}},\
  }\bibfield  {title} {\bibinfo {title} {Preroughening transitions in crystal
  surfaces and valence-bond phases in quantum spin chains},\ }\href
  {https://doi.org/10.1103/PhysRevB.40.4709} {\bibfield  {journal} {\bibinfo
  {journal} {Phys. Rev. B}\ }\textbf {\bibinfo {volume} {40}},\ \bibinfo
  {pages} {4709} (\bibinfo {year} {1989})}\BibitemShut {NoStop}%
\bibitem [{\citenamefont {Tasaki}(1991)}]{Tasaki.Hal1991}%
  \BibitemOpen
  \bibfield  {author} {\bibinfo {author} {\bibfnamefont {H.}~\bibnamefont
  {Tasaki}},\ }\bibfield  {title} {\bibinfo {title} {Quantum liquid in
  antiferromagnetic chains: A stochastic geometric approach to the {H}aldane
  gap},\ }\href {https://doi.org/10.1103/PhysRevLett.66.798} {\bibfield
  {journal} {\bibinfo  {journal} {Phys. Rev. Lett.}\ }\textbf {\bibinfo
  {volume} {66}},\ \bibinfo {pages} {798} (\bibinfo {year} {1991})}\BibitemShut
  {NoStop}%
\bibitem [{\citenamefont {Liu}\ \emph {et~al.}(2015)\citenamefont {Liu},
  \citenamefont {Kong},\ and\ \citenamefont {You}}]{Liu2015}%
  \BibitemOpen
  \bibfield  {author} {\bibinfo {author} {\bibfnamefont {G.-H.}\ \bibnamefont
  {Liu}}, \bibinfo {author} {\bibfnamefont {L.-J.}\ \bibnamefont {Kong}},\ and\
  \bibinfo {author} {\bibfnamefont {W.-L.}\ \bibnamefont {You}},\ }\bibfield
  {title} {\bibinfo {title} {Quantum phase transitions in spin-1 compass
  chains},\ }\href {https://doi.org/10.1140/epjb/e2015-60247-6} {\bibfield
  {journal} {\bibinfo  {journal} {The European Physical Journal B}\ }\textbf
  {\bibinfo {volume} {88}},\ \bibinfo {pages} {284} (\bibinfo {year}
  {2015})}\BibitemShut {NoStop}%
\bibitem [{\citenamefont {Winter}\ \emph {et~al.}(2017)\citenamefont {Winter},
  \citenamefont {Tsirlin}, \citenamefont {Daghofer}, \citenamefont {van~den
  Brink}, \citenamefont {Singh}, \citenamefont {Gegenwart},\ and\ \citenamefont
  {Valent{\'{\i}}}}]{Winter_2017}%
  \BibitemOpen
  \bibfield  {author} {\bibinfo {author} {\bibfnamefont {S.~M.}\ \bibnamefont
  {Winter}}, \bibinfo {author} {\bibfnamefont {A.~A.}\ \bibnamefont {Tsirlin}},
  \bibinfo {author} {\bibfnamefont {M.}~\bibnamefont {Daghofer}}, \bibinfo
  {author} {\bibfnamefont {J.}~\bibnamefont {van~den Brink}}, \bibinfo {author}
  {\bibfnamefont {Y.}~\bibnamefont {Singh}}, \bibinfo {author} {\bibfnamefont
  {P.}~\bibnamefont {Gegenwart}},\ and\ \bibinfo {author} {\bibfnamefont
  {R.}~\bibnamefont {Valent{\'{\i}}}},\ }\bibfield  {title} {\bibinfo {title}
  {Models and materials for generalized {K}itaev magnetism},\ }\href
  {https://doi.org/10.1088/1361-648x/aa8cf5} {\bibfield  {journal} {\bibinfo
  {journal} {Journal of Physics: Condensed Matter}\ }\textbf {\bibinfo {volume}
  {29}},\ \bibinfo {pages} {493002} (\bibinfo {year} {2017})}\BibitemShut
  {NoStop}%
\bibitem [{\citenamefont {You}\ and\ \citenamefont {Tian}(2008)}]{W.L.You2008}%
  \BibitemOpen
  \bibfield  {author} {\bibinfo {author} {\bibfnamefont {W.-L.}\ \bibnamefont
  {You}}\ and\ \bibinfo {author} {\bibfnamefont {G.-S.}\ \bibnamefont {Tian}},\
  }\bibfield  {title} {\bibinfo {title} {Quantum phase transition in the
  one-dimensional compass model using the pseudospin approach},\ }\href
  {https://doi.org/10.1103/PhysRevB.78.184406} {\bibfield  {journal} {\bibinfo
  {journal} {Phys. Rev. B}\ }\textbf {\bibinfo {volume} {78}},\ \bibinfo
  {pages} {184406} (\bibinfo {year} {2008})}\BibitemShut {NoStop}%
\bibitem [{\citenamefont {Trousselet}\ \emph {et~al.}(2010)\citenamefont
  {Trousselet}, \citenamefont {Ole{\'{s}}},\ and\ \citenamefont
  {Horsch}}]{Trousselet_2010}%
  \BibitemOpen
  \bibfield  {author} {\bibinfo {author} {\bibfnamefont {F.}~\bibnamefont
  {Trousselet}}, \bibinfo {author} {\bibfnamefont {A.~M.}\ \bibnamefont
  {Ole{\'{s}}}},\ and\ \bibinfo {author} {\bibfnamefont {P.}~\bibnamefont
  {Horsch}},\ }\bibfield  {title} {\bibinfo {title} {Compass-{H}eisenberg model
  on the square lattice {\textemdash}spin order and elementary excitations},\
  }\href {https://doi.org/10.1209/0295-5075/91/40005} {\bibfield  {journal}
  {\bibinfo  {journal} {{EPL} (Europhysics Letters)}\ }\textbf {\bibinfo
  {volume} {91}},\ \bibinfo {pages} {40005} (\bibinfo {year}
  {2010})}\BibitemShut {NoStop}%
\bibitem [{\citenamefont {Trousselet}\ \emph {et~al.}(2012)\citenamefont
  {Trousselet}, \citenamefont {Ole\ifmmode~\acute{s}\else \'{s}\fi{}},\ and\
  \citenamefont {Horsch}}]{Trousselet2012}%
  \BibitemOpen
  \bibfield  {author} {\bibinfo {author} {\bibfnamefont {F.}~\bibnamefont
  {Trousselet}}, \bibinfo {author} {\bibfnamefont {A.~M.}\ \bibnamefont
  {Ole\ifmmode~\acute{s}\else \'{s}\fi{}}},\ and\ \bibinfo {author}
  {\bibfnamefont {P.}~\bibnamefont {Horsch}},\ }\bibfield  {title} {\bibinfo
  {title} {Magnetic properties of nanoscale compass-{H}eisenberg planar
  clusters},\ }\href {https://doi.org/10.1103/PhysRevB.86.134412} {\bibfield
  {journal} {\bibinfo  {journal} {Phys. Rev. B}\ }\textbf {\bibinfo {volume}
  {86}},\ \bibinfo {pages} {134412} (\bibinfo {year} {2012})}\BibitemShut
  {NoStop}%
\bibitem [{\citenamefont {Zvyagin}\ \emph {et~al.}(2023)\citenamefont
  {Zvyagin}, \citenamefont {Slavin},\ and\ \citenamefont
  {Zvyagina}}]{Zvyagin2023}%
  \BibitemOpen
  \bibfield  {author} {\bibinfo {author} {\bibfnamefont {A.~A.}\ \bibnamefont
  {Zvyagin}}, \bibinfo {author} {\bibfnamefont {V.~V.}\ \bibnamefont
  {Slavin}},\ and\ \bibinfo {author} {\bibfnamefont {G.~A.}\ \bibnamefont
  {Zvyagina}},\ }\bibfield  {title} {\bibinfo {title} {Manifestation of spin
  nematic ordering in the spin-1 chain system},\ }\href
  {https://doi.org/10.1103/PhysRevB.107.134421} {\bibfield  {journal} {\bibinfo
   {journal} {Phys. Rev. B}\ }\textbf {\bibinfo {volume} {107}},\ \bibinfo
  {pages} {134421} (\bibinfo {year} {2023})}\BibitemShut {NoStop}%
\bibitem [{\citenamefont {Yu}\ \emph {et~al.}(2017)\citenamefont {Yu},
  \citenamefont {Sun},\ and\ \citenamefont {Zhai}}]{Yu2017}%
  \BibitemOpen
  \bibfield  {author} {\bibinfo {author} {\bibfnamefont {J.}~\bibnamefont
  {Yu}}, \bibinfo {author} {\bibfnamefont {N.}~\bibnamefont {Sun}},\ and\
  \bibinfo {author} {\bibfnamefont {H.}~\bibnamefont {Zhai}},\ }\bibfield
  {title} {\bibinfo {title} {{Symmetry Protected Dynamical Symmetry in the
  Generalized Hubbard Models}},\ }\href
  {https://doi.org/10.1103/PhysRevLett.119.225302} {\bibfield  {journal}
  {\bibinfo  {journal} {Phys. Rev. Lett.}\ }\textbf {\bibinfo {volume} {119}},\
  \bibinfo {pages} {225302} (\bibinfo {year} {2017})}\BibitemShut {NoStop}%
\end{thebibliography}

%

\end{document}